\documentclass{article}
    \PassOptionsToPackage{numbers, compress}{natbib}
\usepackage[final]{neurips_2026}
\usepackage[utf8]{inputenc} 
\usepackage[T1]{fontenc}    
\usepackage{hyperref}       

\usepackage{url}            
\usepackage{array}          
\usepackage{booktabs}       
\usepackage{amsfonts}       
\usepackage{amssymb}        
\usepackage{nicefrac}       
\usepackage{microtype}      
\usepackage[table]{xcolor}  
\usepackage{graphicx}       
\usepackage{float}          
\usepackage{placeins}

\usepackage{wrapfig}        
\usepackage{subcaption}     
\usepackage{enumitem}
\usepackage{multirow}
\usepackage{makecell}
\usepackage{tikz}           
\usetikzlibrary{positioning, fit, calc, arrows.meta, shapes.misc, shadows}
\usepackage{pgfplots}       
\usepgfplotslibrary{groupplots,fillbetween}
\usepackage{tcolorbox}      
\usepackage{diagbox}        
\usepackage{fvextra}        
\DefineVerbatimEnvironment{VerbatimWrap}{Verbatim}{%
  breaklines=true, breakindent=0pt, breaksymbol={},
  fontsize=\small,%
}

\definecolor{poscolor}{HTML}{2CA02C}    
\definecolor{fncolor}{HTML}{999999}     
\definecolor{hncolor}{HTML}{E08000}     
\definecolor{linkcolor}{HTML}{1F4E99}   
\definecolor{pipepos}{HTML}{2CA02C}     
\definecolor{pipefn}{HTML}{D62728}
\definecolor{pipehn}{HTML}{E08000}
\definecolor{pipeaccent}{HTML}{1F4E99}
\definecolor{chocolate}{HTML}{7B3F00}
\definecolor{indigo}{HTML}{4B0082}
\definecolor{lightblue}{HTML}{3A8FCF}
\definecolor{cumcolor}{HTML}{6B8FB8}    
\definecolor{trafblue}{HTML}{1F77B4}    
\definecolor{shotorange}{HTML}{FF7F0E}  

\newif\ifcomments
\ifcomments
    \newcommand{\yichuan}[1]{{\color{red}[Yichuan: #1]}}
    \newcommand{\andyl}[1]{{\color{blue}[andyl: #1]}}
    \newcommand{\colin}[1]{{\color{teal}[Colin: #1]}}
    \newcommand{\zirui}[1]{{\color{cyan}[Zirui: #1]}}
    \newcommand{\paul}[1]{{\color{purple}[Paul: #1]}}
    \newcommand{\lesheng}[1]{{\color{orange}[Lesheng: #1]}}
    \newcommand{\matei}[1]{{\color{brown}[Matei: #1]}}
    \newcommand{\joey}[1]{{\color{olive}[Joe: #1]}}
    \newcommand{\sewon}[1]{{\color{magenta}[Sewon: #1]}}
\else
    \newcommand{\yichuan}[1]{}
    \newcommand{\andyl}[1]{}
    \newcommand{\colin}[1]{}
    \newcommand{\zirui}[1]{}
    \newcommand{\paul}[1]{}
    \newcommand{\lesheng}[1]{}
    \newcommand{\matei}[1]{}
    \newcommand{\joey}[1]{}
    \newcommand{\sewon}[1]{}
\fi

\newcommand{\sys}{\textsc{PixelRAG}}
\newcommand{\gc}[1]{{\color{gray}{#1}}}
\definecolor{citecolor}{HTML}{0b64c5}
\hypersetup{
 colorlinks=True,
 linkcolor=citecolor,
 citecolor=citecolor,
 urlcolor=citecolor}

\newcommand{\titlelogo}{\raisebox{-0.34\height}{\includegraphics[height=2.4em]{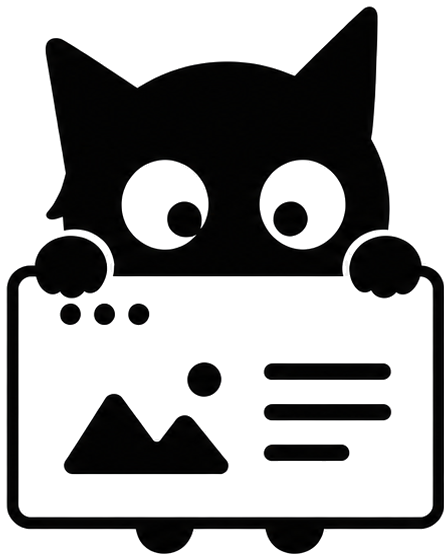}}}
\title{
\titlelogo\hspace{0.45em}\sys{}: Web Screenshots Beat Text for Retrieval-Augmented Generation
}

\author{%
  Yichuan Wang$^{*1}$ \quad
  Zhifei Li$^{*2,5}$ \quad
  Zirui Wang$^{1}$ \quad
  Paul Teiletche$^{3}$ \quad
  Lesheng Jin$^{4}$ \vspace{.3em} \\
  \textbf{Matei Zaharia}$^{\dagger 1}$ \quad
  \textbf{Joseph E.\ Gonzalez}$^{\dagger 1}$ \quad
  \textbf{Sewon Min}$^{\dagger 1}$ \vspace{.8em} \\
  {\normalfont $^{1}$UC Berkeley \quad
  $^{2}$Princeton University \quad
  $^{3}$EPFL \quad
  $^{4}$Databricks \quad
  $^{5}$Renmin University of China} \vspace{.8em} \\
  \texttt{yichuan\_wang@berkeley.edu} \quad \texttt{zhifei.li@princeton.edu}
}

\begin{document}

\maketitle

\renewcommand{\thefootnote}{\fnsymbol{footnote}}
\footnotetext[1]{Equal contribution.}
\footnotetext[2]{Equal advising.}
\renewcommand{\thefootnote}{\arabic{footnote}}

\begin{abstract}
Augmenting large language models (LLMs) with retrieved web text has become a dominant paradigm, yet the web is not natively textual: existing systems depend on complex parsing pipelines that linearize HTML and discard layout, visual structure, and formatting.
We introduce \textbf{\sys{}}, a new retrieval-augmented method that represents websites in their \emph{native visual form} and performs retrieval and reading entirely in pixel space, enabling an end-to-end architecture that eliminates text abstraction.
\sys{} is, to our knowledge, the first pipeline to operate over a full Wikipedia corpus in this form, scaling to a datastore of 30 million screenshot images with an efficient visual retrieval index. 
Built on an existing visual embedding model (i.e., Qwen3-VL-Embedding), \sys{} further fine-tunes this model on screenshot data with carefully curated contrastive training data.
Retrieved screenshots are then fed directly as pixel inputs to a VLM, without intermediate text conversion.
\sys{} consistently outperforms both no-retrieval and text-based RAG baselines, most surprisingly on widely studied text-centric tasks such as NQ and SimpleQA.
It also achieves strong gains on multimodal open-domain QA (e.g., MMSearch), benchmarks over noisy news corpora (e.g., LiveVQA), and agentic benchmarks (e.g., MoNaCo), improving accuracy by up to 18.1\% over text-based baselines.
Finally, pixel representations enable a new efficiency lever for RAG through image compression, achieving up to $3\times$ token cost reduction at lower resolutions while maintaining accuracy.
Our results challenge the necessity of text representations in web retrieval, suggesting that web RAG can operate directly in the web's native visual form while improving both performance and efficiency.
Our code is available at \url{https://github.com/StarTrail-org/PixelRAG}.

\end{abstract}

\section{Introduction}
\label{sec:introduction}

Retrieval-augmented generation (RAG) has become a dominant paradigm for grounding large language models (LLMs) in external knowledge, powering open-domain question answering, search-augmented agents, and deep-research systems~\cite{lewis2020retrieval,guu2020realm,karpukhin2020dpr,izacard2021fid,nakano2021webgpt,lyu2025frustratingly}. 
Among retrieval sources, the web is the largest and most diverse knowledge base: an increasing number of LLM applications rely on web retrieval to stay current and factually grounded~\cite{shao2024scaling,fang2025reusing,li2025websailor}.

\begin{figure}[t]
\centering
\includegraphics[width=\linewidth]{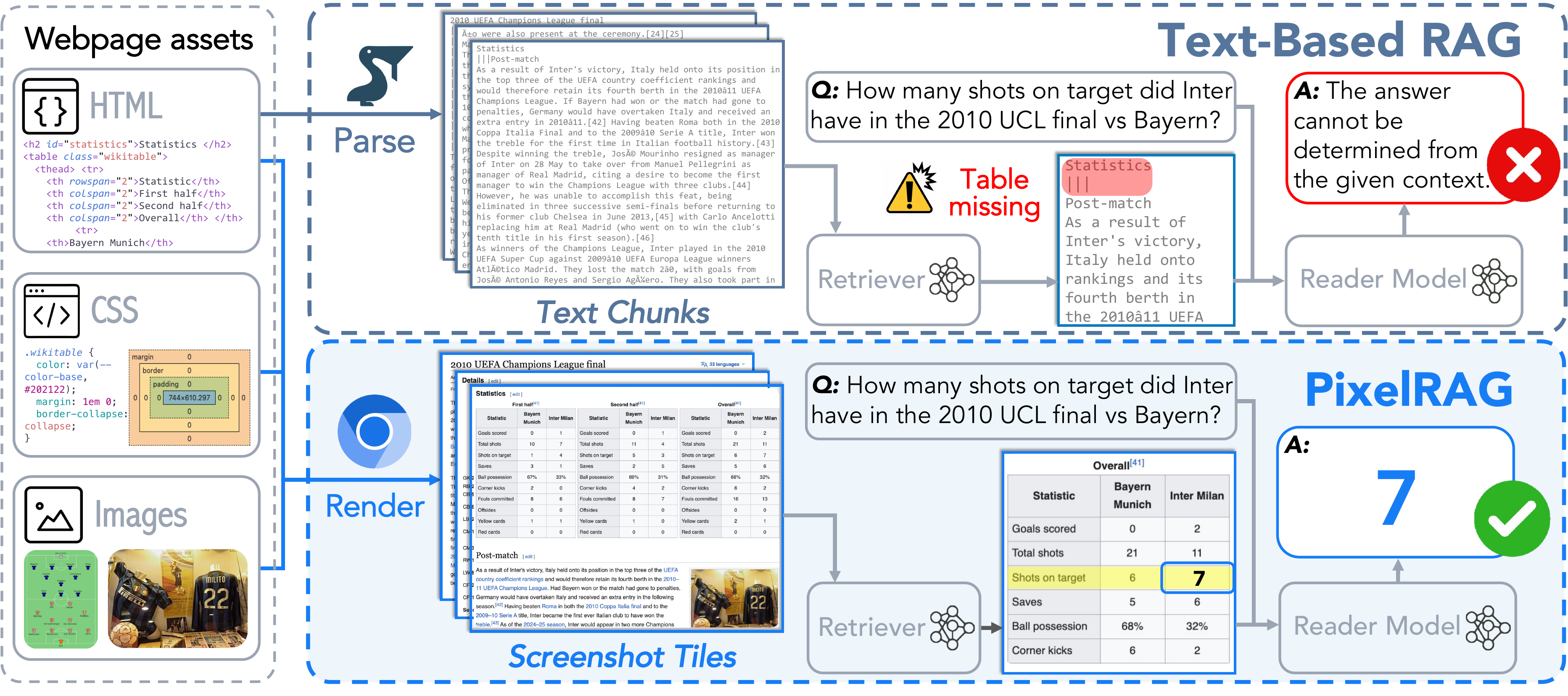}
\caption{\textbf{Overview of \sys{}.} Text-based RAG (top) parses HTML into a text index and retrieves text chunks for the reader model. \textbf{\sys{}} (bottom, ours) renders each webpage, builds a visual index, and retrieves screenshot tiles for the reader --- no parser required, fully visual.}
\label{fig:pipeline}
\end{figure}

Text-based RAG assumes access to clean text documents for retrieval and reading, but hinges on a critical, often overlooked step: HTML parsing. 
Clean text is not free: HTML-to-text extraction is complex, heavily engineered, and error-prone~\cite{li2026beyond,penedo2024fineweb,tan2025htmlrag}. 
Even state-of-the-art parsers~\cite{barbaresi-2021-trafilatura,wang2025readerlm,liu2025dripper} are brittle and inherently lossy, discarding visual cues (e.g., emphasis, layout) and structured content such as tables, charts, and images. These losses cascade: the retriever indexes a degraded view of the page, and the reader must reason over linearized, noisy text that may be difficult to comprehend.

In this work, we ask: can RAG operate directly in pixel space---on the web as users see it---bypassing complex and lossy HTML parsing? Concretely, this would index web screenshots, retrieve pages as images, and feed them to a VLM as context. While ambitious, recent advances in VLMs that significantly improved visual understanding, enabling capabilities such as OCR, document, and chart understanding, make this approach feasible~\cite{bai2025qwen3-vl-technical-report,openai2024gpt4o,geminiteam2025gemini25,anthropic2024claude3,zhu2025internvl3,wang2024charxiv}. 
As information-seeking increasingly involves structured and visual content, this approach also enables a unified representation without separate text and vision representations. Prior work has explored this paradigm in curated, small-scale domains such as PDFs~\cite{faysse2024colpali,dse,yu2024visrag,cho2024m3docrag,yan2026unlocking,huo2026causalembed}, but it remains largely unexplored for the web—the dominant source of open-domain knowledge.

We introduce \textbf{\sys{}}, to our knowledge, the first end-to-end screenshot-based RAG approach at web scale, operating over the full Wikipedia corpus (7M articles) and news corpora from CNN, AP News, and BBC. While scalable data collection is non-trivial, we build a pipeline that first collects rendering assets (HTML, CSS, images) and reconstructs pages locally to produce screenshots at scale.
Retrieval is powered by a Qwen3-VL-Embedding model~\cite{li2026qwen3-vl-embedding}, which we further adapt to screenshot data via efficient contrastive fine-tuning with synthetic QA pairs and dynamic hard-negative mining~\cite{Xiong2020ApproximateNN,karpukhin2020dpr}.
Retrieved images are then fed directly as pixel inputs to a VLM reader, bypassing intermediate text representations.

Our experiments show that \sys{} consistently outperforms no retrieval and text-based RAG baselines across benchmarks, most surprisingly on widely used text-centric Wikipedia QA tasks such as NQ~\cite{kwiatkowski2019natural} and SimpleQA~\cite{wei2024simpleqa}, where most, if not all, questions are answerable from text alone.
\sys{} also delivers strong gains on multimodal QA, benchmarks over noisy news corpora (LiveVQA~\cite{fu2025livevqa}), and agentic benchmarks~\cite{jiang2024mmsearch}.
Further analysis suggests that gains on text-centric tasks arise from leveraging 2D structural cues (e.g., tables, infoboxes) and avoiding information loss observed in HTML-to-text parsing.
Notably, these gains scale with VLM capability, suggesting further headroom as models continue to improve.

Finally, pixel representations introduce a new, underexplored dimension for improving token efficiency in RAG via image compression, in line with recent work~\cite{wei2025deepseek,cheng2025glyph}. Training the VLM reader to tolerate lower resolutions further improves the accuracy\textendash efficiency Pareto frontier, enabling up to $3\times$ token cost reduction while maintaining accuracy. To summarize, our contributions are fourfold:

\begin{enumerate}[leftmargin=14pt, topsep=1pt,itemsep=0pt]
    \item We introduce \sys{}, the first end-to-end pixel-space RAG system at web scale, operating directly over rendered screenshots and bypassing HTML-to-text parsing. This simplifies the pipeline, preserves visual cues, and avoids information loss.
    
    \item We develop a scalable pipeline for constructing screenshot-based corpora from the web, along with a visual embedding model trained on screenshot data via synthetic QA pairs and hard negatives. 
    
    \item We show that pixel-space retrieval on web data can consistently outperform text-based baselines across diverse benchmarks, including well-studied text-centric ones (e.g., NQ, SimpleQA).
    
    \item We identify a new efficiency dimension via image compression, achieving up to $3\times$ token cost reduction. These gains suggest an emerging shift toward pixel-space RAG as multimodal models continue to improve.
\end{enumerate}

\section{Related Work \& Motivation}
\label{sec:background_motivation}


\paragraph{Retrieval-Augmented Generation (RAG).}
RAG operates over large corpora of text documents, using a \emph{retrieval model} to identify relevant text and a \emph{reader model} to generate grounded answers~\cite{karpukhin2020dpr,lewis2020retrieval,guu2020realm}, powering search-augmented agents~\cite{nakano2021webgpt,jin2025search} and deep research systems~\cite{team2025tongyi}.
In open-domain settings, these corpora are predominantly web-derived, including knowledge sources such as Wikipedia.

Despite this reliance on web data, most prior work assumes that clean textual documents are readily available, largely abstracting away the upstream HTML-to-text parsing step that converts raw web pages into the textual representations ultimately consumed by retrieval and reader models.



\paragraph{Text Parsing from Web Data.}

Text-based RAG over web data, therefore, relies on HTML-to-text parsers as a standard preprocessing layer for constructing retrievable text corpora.
In large-scale data curation, reliable web parsing has long been recognized as a critical challenge~\cite{li2024datacomp,penedo2024fineweb}: web pages vary enormously in structure, interleaving natural language passages with tables, infoboxes, figures, charts, and dynamically rendered content.
Current pipelines typically rely on heuristic extractors such as \textit{trafilatura}~\cite{barbaresi-2021-trafilatura}, \textit{resiliparse}~\cite{bevendorff:2018}, and \textit{mwparserfromhell}~\cite{mwparserfromhell}, or learned parsers such as ReaderLM~\cite{wang2025readerlm} and Dripper~\cite{liu2025dripper}.

However, even state-of-the-art parsers remain brittle and inherently lossy, often discarding visual cues (e.g., emphasis and layout) and structured content such as tables, charts, and images. Recent work shows that parser choice alone can substantially affect downstream performance~\cite{li2024datacomp,penedo2024fineweb}: a single extractor may discard over 40\% of recoverable webpage text~\cite{li2026beyond}.
Similar effects have also been observed in RAG systems, both in prior work~\cite{fang2025reusing} and in our own experiments (\S\ref{sec:main_results}), where the gap between the two strongest parsing methods yields nearly a 10\% absolute difference on SimpleQA.

\paragraph{Vision-Language Models for Text-Rich Images.}
At the same time, recent vision-language models (VLMs) are becoming increasingly capable of understanding text rendered as pixels, narrowing the gap with text-input models~\cite{lu2024text,sun2026reading,lyu2025pixelworld,bai2025qwen3-vl-technical-report}.
Concurrently, systems such as DeepSeek-OCR~\cite{wei2025deepseek} and Glyph~\cite{cheng2025glyph} demonstrate that representing documents visually can sometimes be more token-efficient than fully textual pipelines.
Together, these advances suggest that directly modeling rendered web pages as images may be a promising alternative to HTML-to-text parsing.

\paragraph{Visual Document Retrieval.}
Motivated by these advances, a growing line of work explores retrieving documents directly from their rendered visual representations~\cite{faysse2024colpali,moreira2026nemotron,yu2024visrag}, primarily on small-scale, visually rich benchmarks built from PDFs and slides~\cite{mace2025vidore,zhu2024mmdocbench,shorten2026irpapers}.
Our work is inspired by this direction, but instead targets open-domain web retrieval---arguably the most dominant setting for modern RAG models. Compared to curated document collections, web corpora are vastly larger, more heterogeneous, and dominated by noisy, text-heavy pages rather than curated documents.

The closest prior work to ours are recent efforts to extend visual retrieval to Wikipedia~\cite{dse,cho2024m3docrag}; however, they remain limited to a small subset of Wikipedia and focus on retrieval rather than end-to-end RAG. In contrast, we study whether fully visual RAG can readily replace text-based RAG over large-scale Wikipedia, even on well-studied RAG benchmarks such as SimpleQA and NQ.

\paragraph{Summary: Why Pixel-Based RAG?}
\label{sec:motivation}
To summarize, three observations motivate a shift from text-based to pixel-based RAG.\begin{enumerate}[leftmargin=14pt, topsep=1pt,itemsep=0pt]
    \item \textbf{Parsing discards critical information.}
Even the best extractors strip images, charts, and rendered layout, and flatten or lose tables and other 2D structures---content that often contains the answer (Figure~\ref{fig:pipeline}; Appendix~\ref{app:visual_loss}).
Some content is only materialized during rendering, making it invisible to any HTML-level extractor.
    \item \textbf{Text loses visual structure}
    Even when all content is preserved, converting a two-dimensional layout into a one-dimensional token sequence discards spatial grouping, font hierarchy, and emphasis.
    Web pages are designed to be consumed as rendered artifacts---these visual cues help readers locate information, and stripping them makes it harder for retrieval to distinguish answer-bearing regions from surrounding text.
    As shown in \S\ref{sec:qualitative_analysis} and Appendix~\ref{app:retrieval_signal_loss}, text retrieval often surfaces topically relevant but uninformative passages, whereas pixel-based retrieval localizes the answer directly.
    \item \textbf{VLM advances increasingly favor pixel-based RAG.} Modern VLMs are becoming more capable and token-efficient, already outperforming text-only models on structurally rich content such as tables and infoboxes, making pixel-based retrieval increasingly practical and effective.
\end{enumerate}


\section{\sys{}}
\label{sec:vis_store}
We present \sys{}, an end-to-end retrieval-and-generation pipeline over web data that operates entirely in pixel space (Figure~\ref{fig:pipeline}).
\sys{} consists of three stages: \emph{data collection} (\S\ref{sec:data_collection}) renders webpages into screenshots and slices them into fixed-size tiles; \emph{index construction} (\S\ref{sec:retrieval}) encodes each tile into a visual embedding and builds an approximate nearest-neighbor index; and \emph{runtime retrieval and generation} (\S\ref{sec:generation}) retrieves the top-$K$ tiles for a query and feeds them to a vision-language reader that produces the answer directly from pixels.
We instantiate \sys{} on full Wikipedia as our running example.
Wikipedia is a canonical corpus of web knowledge and the basis of most widely used text-based RAG pipelines and benchmarks, so building the same datastore in pixel space enables direct comparison against text-based RAG over the same underlying corpus.
To our knowledge, \sys{} is the first system to operate over the full 7M-article Wikipedia corpus in pixel space.
The pipeline is datastore-agnostic: \S\ref{sec:main_results} applies the same stages to a 668K-article news corpus.

\subsection{Data Collection}
\label{sec:data_collection}

\paragraph{Challenges.}
In text-based retrieval, building a datastore is well-studied: download HTML files and parse them into plain text.
For pixel-based RAG, how to build a datastore is less clear.
A naive approach is to open each webpage in a browser instance and capture a screenshot; however, this does not scale to millions of pages (e.g., roughly 30 days for all 7M articles).
This is slow for two reasons: (a) online fetch of all webpage assets dominates wall-clock time (large network I/O), and (b) browsers are not designed for high-throughput rendering, limiting concurrency and causing frequent retries and stalls.
Instead, we decouple fetching from rendering: we first fetch all sources into a local mirror, then render and tile entirely offline.
This makes the pipeline scalable, reproducible, and fault-tolerant.

\paragraph{Fetching.}
Each corpus is first materialized into a local mirror.
For Wikipedia, all assets (HTML, CSS, and images) are extracted from a pre-built Kiwix ZIM archive; for the news corpus, an asynchronous crawler fetches them with parallel crawls over different domains.
Once cached, rendering proceeds entirely offline, making retries free and experiments reproducible.
Per-source details are in Appendix~\ref{app:datastore_fetch}.

\paragraph{Rendering.}
Each cached page is rendered with Playwright (headless Chromium).
We strip non-content elements (navigation bars, sidebars, surrounding whitespace; see Appendix~\ref{app:screenshot_comparison}) to produce clean, content-only screenshots.
Pages exceeding the browser's single-shot viewport are captured by scrolling, so coverage is independent of page length.

\paragraph{Tiling.}
We fix the viewport width to 875 pixels (Wikipedia’s default content width) and slice each full-page screenshot into non-overlapping $1024$-pixel-tall tiles (the last tile may be shorter).
This yields ${\sim}$30M tiles for Wikipedia (7M articles) and ${\sim}$3.6M for the news corpus (668K articles).

\subsection{Index Construction}
\label{sec:retrieval}

\paragraph{Embedding.}
Related work in PDF retrieval (\S\ref{sec:motivation}) largely relies on late-interaction multivector models such as ColPali, ColQwen~\cite{faysse2024colpali}, and Nemotron ColEmbed~\cite{moreira2026nemotron}, which were found to be critical for capturing fine-grained information in document images.
In our setting, however, there is a key difference: \emph{scale}.
Our datastore contains 30M tiles, far beyond the scale of prior visual-retrieval work (e.g., a few thousand pages~\cite{faysse2024colpali,shorten2026irpapers}), making multivector retrieval prohibitively expensive: each $875 \times 1024$-pixel tile emits ${\sim}875$ visual tokens\footnote{$875 \times 1024 / 16^2 / 4 \approx 875$; 16 is the ViT patch size, and the VLM projector merges every 4 spatially aligned patches~\cite{bai2025qwen3-vl-technical-report}.} at 128 dimensions each (${\sim}112$K dimensions per tile), which would inflate the 30M-tile index to ${\sim}6.5$\,TB in fp16, far beyond single-host RAM.
We therefore adopt single-vector retrieval: a single 2048-dimensional vector per tile keeps the full index at ${\sim}$120\,GB in fp16, manageable on one machine.
Our results (\S\ref{sec:evaluation}) show that single-vector retrieval with a state-of-the-art visual embedding model is already sufficiently competitive.
Here we use an off-the-shelf Qwen3-VL-Embedding-2B~\cite{li2026qwen3-vl-embedding} and its fine-tuned variant (\S\ref{sec:embedding_training}).

\paragraph{Index.}
At tens of millions of vectors, exact search is infeasible; we use a FAISS IVF index~\cite{faiss} for approximate nearest-neighbor search, which is fast to build and supports efficient addition and deletion of vectors, enabling incremental updates (e.g., ingesting new news articles daily or refreshing Wikipedia snapshots) without full re-indexing.

The full offline pipeline completes the 7M-article Wikipedia in ${\sim}$2 days on a single machine (128 CPU cores, 2\,TB RAM, 8 H100 GPUs).

\subsection{Runtime Retrieval and Generation}
\label{sec:generation}

At query time, the query is embedded with the same embedding model, the top-$K$ tiles are retrieved by inner-product similarity, and the vision-language reader produces the answer directly from the retrieved pixels and the query (prompt templates in Appendix~\ref{app:prompt_listings}).
The quality--cost trade-off is governed by:
(1)~\emph{Number of retrieved tiles $K$}: increasing $K$ supplies the reader with more evidence but costs more visual tokens per query.
(2)~\emph{Rendering resolution}: each tile is originally $875\times1024$ pixels, but can be downscaled (e.g., to $437\times512$) before being fed to the reader; since common open-source VLMs support dynamic resolution, visual-token count scales proportionally with pixel count, yielding a ${\sim}4\times$ token reduction~\cite{bai2025qwen3-vl-technical-report}.
We study both trade-offs in \S\ref{sec:evaluation}. 
\section{Embedding Model Contrastive Learning Pipeline}
\label{sec:embedding_training}
Given a screenshot datastore, we want to fine-tune a visual embedding model tailored to it, improving retrieval accuracy over the base model.
We describe a synthetic contrastive data generation recipe in \S\ref{sec:data_recipe} and the training procedure in \S\ref{sec:training_recipe}.

\subsection{Synthetic Contrastive Data Generation and Dynamic Hard-Negative Mining}
\label{sec:data_recipe}

Our goal is to curate a contrastive training set of $(q, p, \{n^-\})$ triples, where $q$ is a query grounded in the visual content of a page $p$, and each $n^-$ is a hard negative that shares page structure and topic with $p$ but does \emph{not} answer $q$.
We use no external labeled data: the entire training set is synthesized from the datastore, with an LLM as the sole annotator.
Note that the downstream benchmarks in \S\ref{sec:evaluation} are \emph{out-of-distribution}: the embedding model sees neither their queries nor their labels during training.

\paragraph{Stage 1: Synthetic query generation.}

We first filter out information-sparse pages such as listings and disambiguation pages, keeping only tiles that pass lightweight information-density heuristics (full method in Appendix~\ref{app:ki_sampling}).
Given a sampled tile $p$, we prompt an LLM to generate a natural-language query $q$ whose answer is present in $p$ (prompt in Appendix~\ref{app:query_gen_prompt}).
We then apply a two-part filter: (1)~a \emph{self-containedness} check discards queries that implicitly reference the source page (e.g., ``\textit{...listed in the screenshot?}''); (2)~an \emph{answerability} check re-prompts the LLM to answer $q$ from $p$ alone and keeps the pair only if correct (prompts in Appendix~\ref{app:selfcontained_prompt},~\ref{app:answerable_prompt}).

\paragraph{Stage 2: Dynamic hard-negative mining with false-negative filtering.}
Given a $(q, p)$ pair from Stage~1, we retrieve the top-$K$ tiles for $q$ using the base embedding model~\cite{Xiong2020ApproximateNN} and treat all except $p$ as hard-negative candidates.
However, the same knowledge often appears in more than one page; a top-$K$ neighbor may therefore also answer $q$, making it a false negative that penalizes a correct retrieval under the InfoNCE objective.

\begin{figure}[t]
  \centering
  \input{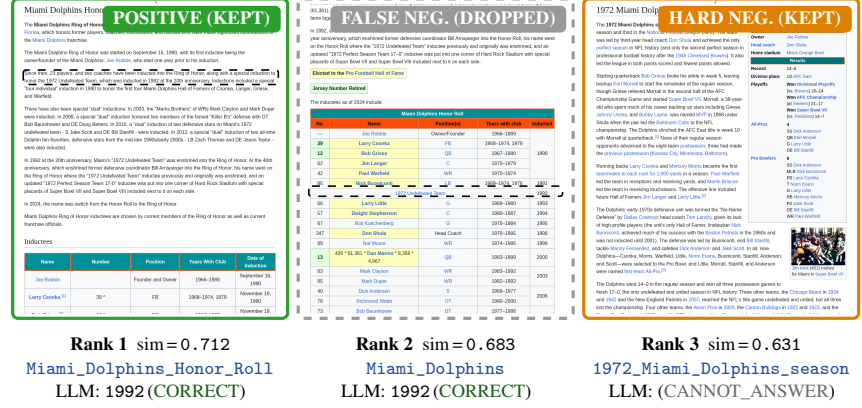}
  \caption{\textbf{Hard-negative mining with false-negative filtering.}
  The LLM correctly answers the query from both the positive tile (left, \textcolor{poscolor}{kept}) and a second tile (center, \textcolor{fncolor}{dropped as false negative}), but cannot answer from a topically adjacent page (right, \textcolor{hncolor}{kept as hard negative}).}
  \label{fig:hard_negative}

\end{figure}

We address this with an LLM-based false-negative filter.
For each top-$K$ candidate (excluding $p$), we prompt the LLM to answer $q$ from the candidate tile alone and judge the response as \texttt{CORRECT}, \texttt{WRONG}, or \texttt{CANNOT\_ANSWER} (prompts in Appendix~\ref{app:consistency_prompt}).
Candidates judged \texttt{CORRECT} are false negatives and dropped; the rest are kept as hard negatives.
We retain the first $M$ surviving candidates per query ($M{=}2$); Figure~\ref{fig:hard_negative} walks through a worked example.
This pipeline uses a strong LLM to distill clean supervision into the embedding model, improving retrieval without human labels.

\subsection{Training Recipe}
\label{sec:training_recipe}

\paragraph{Contrastive training loss.}
We fine-tune the visual embedding model with InfoNCE loss~\cite{oord2018representation} over mined triples and in-batch negatives.
For a mini-batch of $B$ triples $\{(q_i, p_i, \{n_{i,j}\}_{j=1}^{M})\}_{i=1}^{B}$,
\begin{equation}
\mathcal{L} = -\frac{1}{B}\sum_{i=1}^{B} \log
\frac{e^{\,\mathrm{sim}(q_i, p_i)/\tau}}
{e^{\,\mathrm{sim}(q_i, p_i)/\tau}
\;+\; \sum_{j=1}^{M} e^{\,\mathrm{sim}(q_i, n_{i,j})/\tau}
\;+\; \sum_{k\neq i} e^{\,\mathrm{sim}(q_i, p_k)/\tau}},
\label{eq:infonce}
\end{equation}
with $\mathrm{sim}(\cdot,\cdot)$ the cosine similarity and $\tau$ a temperature (full training configuration in Appendix~\ref{app:impl_details}).

\paragraph{LoRA fine-tuning with unfrozen ViT.}
Prior visual retriever training, such as ColPali~\cite{faysse2024colpali} and Swift~\cite{zhao2024swiftascalablelightweightinfrastructure}, applies LoRA only to the LLM backbone, as adding LoRA to the ViT degrades performance on document images.
We observe the opposite: we apply LoRA to both the LLM backbone and the ViT, and find it brings consistent gains on rendered webpage screenshots (Table~\ref{tab:ablation_recipe}).
We hypothesize that visually near-duplicate webpage tiles require stronger visual discrimination, especially in fine-grained rendered text details and table structure, which benefits from ViT adaptation.

The full training completes in under 3 hours on a single H100, adding modest cost.


\section{Evaluation}
\label{sec:evaluation}
\sys{} outperforms all text-based RAG baselines across six benchmarks (\S\ref{sec:main_results}).
We ablate key design choices (\S\ref{sec:ablation_modality}), extend to agentic search (\S\ref{sec:ablation_agent}), study image compression as a cost knob (\S\ref{sec:image_compression}), and show that performance scales with VLM capability (\S\ref{sec:vlm_progress}).
\subsection{Experimental Setup}
\label{sec:eval_setup}

\paragraph{Benchmarks.}
We evaluate on three task families across two corpora.
(1) \emph{Text-centric Wikipedia QA} includes NQ~\cite{kwiatkowski2019natural}, NQ-Tables~\cite{herzig2021nqtables}, and SimpleQA~\cite{wei2024simpleqa}, which are widely studied benchmarks whose questions are primarily answerable from textual content. 
(2) \emph{Multimodal Wikipedia QA} includes MMSearch~\cite{jiang2024mmsearch}\footnote{MMSearch queries span diverse web topics; we use our Wikipedia datastore as the retrieval backend for all queries.} and Encyclopedic VQA~\cite{mensink2023evqa}, consisting of image-grounded queries.
(3) \emph{News VQA} includes LiveVQA~\cite{fu2025livevqa}, which tests generalization beyond Wikipedia to a news corpus that is significantly noisier and more heterogeneous.
The first five benchmarks query the Wikipedia datastore (30M tiles, \S\ref{sec:data_collection}); LiveVQA queries a separate news datastore (3.6M tiles).
Dataset sizes and evaluation metrics are detailed in Appendix~\ref{app:benchmark_details}.
\paragraph{Baselines.}
We compare against no-retrieval and two text-based RAG baselines that differ only in the HTML-to-text parser: \emph{Trafilatura}~\cite{barbaresi-2021-trafilatura}, a general-purpose extractor ranked best in both prior work~\cite{penedo2024fineweb,liu2025dripper} and our own comparison against six alternatives (Appendix~\ref{app:parser_comparison}); and \emph{mwparserfromhell}~\cite{mwparserfromhell}, a widely used Wikipedia-specific parser~\cite{thrush2022pipeline,neuml2024wikipedia}.
Text indexes use 1024-token chunks, matching the ${\sim}875$ visual tokens per tile (\S\ref{sec:retrieval}), with the same embedding model~\cite{li2026qwen3-vl-embedding} and FAISS~\cite{faiss} index.
\paragraph{Configuration.}
Unless otherwise specified, the reader is Qwen3.5-4B~\cite{qwen2026qwen35} and receives $k{=}3$ retrieved items (text chunks or screenshot tiles).
\sys{} (base) retrieves with the pretrained Qwen3-VL-Embedding-2B~\cite{li2026qwen3-vl-embedding}; \sys{} adds contrastive fine-tuning from \S\ref{sec:embedding_training}.
Recall@$k$ measures whether any of the top-$k$ retrieved items comes from the gold article (matched by URL).
\subsection{End-to-End QA Results}
\label{sec:main_results}

\begin{table}[t]
\centering
\caption{Recall@3 and end-to-end QA accuracy. The vertical line separates benchmarks using the Wikipedia corpus (left) and news corpus (right). \emph{mwparserfromhell} is Wikipedia-specific; MMSearch lacks gold articles, so recall is marked {--}.}
\label{tab:main_results}
\vspace{4pt}
\footnotesize
\begin{tabular*}{\textwidth}{@{\extracolsep{\fill}} l r@{\hskip 6pt}r r@{\hskip 6pt}r r@{\hskip 6pt}r r@{\hskip 6pt}r r@{\hskip 6pt}r !{\vrule} r@{\hskip 6pt}r @{}}
\toprule
 & \multicolumn{2}{c}{NQ} & \multicolumn{2}{c}{NQ-Tables} & \multicolumn{2}{c}{SimpleQA} & \multicolumn{2}{c}{MMSearch} & \multicolumn{2}{c}{EVQA} & \multicolumn{2}{c}{LiveVQA}  \\
\cmidrule(lr){2-3} \cmidrule(lr){4-5} \cmidrule(lr){6-7} \cmidrule(lr){8-9} \cmidrule(lr){10-11}  \cmidrule(lr){12-13}
Method & Recall & Acc & Recall & Acc & Recall & Acc & Recall & Acc & Recall & Acc  & Recall & Acc \\
\midrule
No retrieval             && 30.4 && 24.5 && 7.0 && 12.7 && 27.2 && 63.6 \\
\midrule
\emph{Text-based retrieval} &&&&&&&&&&&& \\
\quad mwparserfromhell        & 48.6 & 54.2 & 34.8 & 35.9 & 74.2 & 60.7 &-- & 25.3 & 6.5 & 31.5 && \\
\quad Trafilatura  & 45.8 & 55.9 & 37.2 & 42.5 & 77.4 & 71.6 &-- & 24.7 & 6.4 & 29.6 & 16.2 &59.0  \\
\midrule
\emph{Pixel-based retrieval} &&&&&&&&&&&& \\
\quad \sys{} (base) & 53.5 & 57.9 & 45.5 & 47.0 & 80.8 & 73.8 &-- & \textbf{28.3} &27.1 & 40.7 & \textbf{38.9} & \textbf{70.3}   \\
\quad \sys{} & \textbf{58.8} & \textbf{58.7} & \textbf{51.1} & \textbf{48.8} & \textbf{83.8} & \textbf{78.8} &-- & \textbf{28.3} & \textbf{35.0} & \textbf{45.1} & 33.3 & 70.0 \\
\bottomrule
\end{tabular*}
\vspace{-8pt}
\end{table}

\sys{} improves end-to-end QA accuracy across all six benchmarks (Table~\ref{tab:main_results}), most notably on text-centric tasks where questions are answerable from text alone.
Even \sys{} (base) outperforms both text-based baselines on every task, with gains of up to 8.3\% in recall and 4.5\% in accuracy. Fine-tuning yields an additional 5.3\% recall improvement and 5.0\% accuracy gain. The largest improvements appear on NQ-Tables, where answering questions requires structured content such as tables and infoboxes.
The gap widens further on multimodal VQA benchmarks.
On EVQA, text retrieval provides little benefit over no retrieval, suggesting that retrieved text chunks are often off-target. In contrast, \sys{} improves QA accuracy by up to 18\% while increasing retrieval recall on EVQA by over 5$\times$. These gains suggest that preserving images and layout allows the embedding model to directly align visual queries with article content.
This advantage extends beyond Wikipedia: on LiveVQA over our news corpus, \sys{} (base) outperforms the text baseline by 11.3\% and more than doubles recall without domain-specific training.
Wikipedia-trained fine-tuning does not transfer well to news domains; scaling training across domains remains future work.

\begin{figure}[t]
\centering
\includegraphics[width=\linewidth]{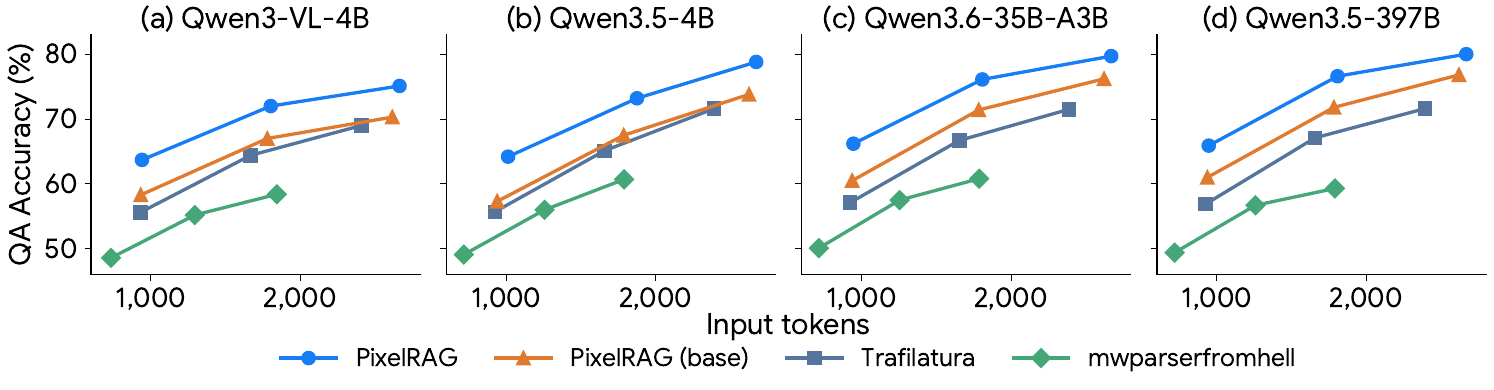}
\caption{SimpleQA accuracy versus average input tokens across four reader models ($k \in \{1,2,3\}$).}
\label{fig:token_efficiency}
\end{figure}

\paragraph{Varying $k$ and reader model.}
\label{sec:token_efficiency}
We vary $k \in \{1,2,3\}$ and the reader model on SimpleQA (Figure~\ref{fig:token_efficiency}).
\sys{} consistently outperforms text retrieval across all configurations: even \sys{} (base) beats both text baselines, and fine-tuning further widens the gap.
At comparable token budgets, two \sys{} tiles match or exceed three text chunks while consuming fewer tokens, as a single tile packs denser evidence than a 1024-token chunk~\cite{wei2025deepseek}.
The advantage grows with stronger readers: at $k{=}2$, \sys{} leads Trafilatura by 9.4\% with Qwen3.6-35B-A3B vs.\ 8.1\% with Qwen3.5-4B.

\begin{table}[t]
\centering
\caption{SimpleQA accuracy and Evidence Recall@3 by evidence type. Abbreviations: T. = Trafilatura, M. = mwparserfromhell.}
\label{tab:failure_by_evidence_type}
\footnotesize
\newcommand{\dlt}[1]{\makebox[3.2em][l]{\gc{\,(#1)}}}
\begin{tabular*}{\textwidth}{@{\extracolsep{\fill}}lr rrr rrr@{}}
\toprule
& & \multicolumn{3}{c}{Accuracy (\%)} & \multicolumn{3}{c}{Evidence Recall@3 (\%)} \\
\cmidrule(lr){3-5} \cmidrule(lr){6-8}
Type & $n$ & \sys{} & T. & M. & \sys{} & T. & M. \\
\midrule
Table     & 282 & \textbf{75.8}\dlt{+9.1}  & 66.7 & 48.2 & \textbf{34.8}\dlt{+11.0}  & 23.8 & 5.0  \\
List      & 290 & 77.4\dlt{+0.5}           & \textbf{76.9} & 62.8 & \textbf{36.9}\dlt{+8.3}   & 28.6 & 23.8 \\
Infobox   & 503 & \textbf{83.3}\dlt{+4.6}  & 78.7 & 65.2 & 63.0\dlt{$-$2.2}          & \textbf{65.2} & 54.9 \\
Paragraph & 571 & \textbf{79.4}\dlt{+7.9}  & 71.5 & 64.1 & \textbf{63.9}\dlt{+19.8}  & 44.1 & 38.7 \\
\midrule
Overall & 1000\footnotemark & \textbf{78.8}\dlt{+7.1} & 71.6 & 60.7 & \textbf{83.8}\dlt{+6.4} & 77.4 & 74.2 \\
\bottomrule
\end{tabular*}
\footnotetext{A question may appear in several rows when its answer spans multiple content types (e.g., both a table and a paragraph); $n$ values therefore do not sum to 1{,}000. Such multi-type questions tend to be easier, raising per-type averages above the overall.}
\end{table}

\paragraph{Qualitative analysis: SimpleQA accuracy breakdown by answer location.}
\label{sec:qualitative_analysis}

We examine what each method actually retrieves using \emph{evidence Recall@3}: whether the answer-bearing tile or chunk appears in the reader's top-3, broken down by where the answer lives on the page (Table~\ref{tab:failure_by_evidence_type}; definition in Appendix~\ref{app:failure_detailed}).
Overall, \sys{} retrieves answer-bearing evidence 6\% more often than Trafilatura (83.8\% vs.\ 77.4\%), explaining most of the end-to-end accuracy gap.

The advantage concentrates on two evidence types.
For \emph{tables}, parsers frequently lose structured content during linearization, dropping evidence Recall@3 to 23.8\% (Trafilatura) and 5.0\% (mwparserfromhell) versus 34.8\% for \sys{}.
For \emph{paragraphs}, the gap is even larger (+19.8\%): once linearized, keyword-dense infobox text overlaps with nearly any factual query about the article, displacing the answer-bearing paragraph from top-3; the visual embedding model is immune because infoboxes have a visually distinct bordered-sidebar layout.
Switching to mwparserfromhell widens the overall accuracy gap by a further 10.9\%, confirming that the advantage is not parser-specific.

Text-based retrieval exhibits three distinct failure modes (Appendix~\ref{app:failure_detailed}).
\emph{Parser loss} (36.6\%): linearization destroys 2D structures such as tables, dropping the answer from the text corpus entirely; e.g., a match-statistics table is flattened to empty delimiter characters, leaving no chunk that preserves its cell values (Fig.~\ref{fig:fm1_inter}).
\emph{Rank loss} (55.2\%): answer-bearing text exists in the corpus but ranks outside the reader's top-3.
A common pattern in text retrieval is infobox displacement: the linearized infobox produces keyword-dense but answer-irrelevant text that matches nearly any factual query about the entity, occupying rank~1 while the answer paragraph falls to rank~60+.
\sys{} is less susceptible because its visual encoder distinguishes the infobox's bordered-sidebar layout from body text (Figs.~\ref{fig:fm2_dali}--\ref{fig:fm2_shepard}).
\emph{Reader loss} (8.2\%): even when evidence reaches top-3, linearizing tables and lists collapses the row and hierarchy grouping the reader needs; a multi-year honors list, for example, loses its year headings once flattened, causing the reader to attribute an entry to the wrong year (Fig.~\ref{fig:fm3_morgan_prize}).

\subsection{Ablation Studies}
\label{sec:ablation_modality}

\begin{table}[t]
\centering
\begin{minipage}[t]{0.46\textwidth}
\vspace{0pt}
\centering
\caption{Retrieval--reader modality ablation.}
\label{tab:ablation_modality}
\footnotesize
\begin{tabular*}{\textwidth}{@{\extracolsep{\fill}}llrr@{}}
\toprule
Retrieval & Reader & SimpleQA & LiveVQA \\
\midrule
Screenshot & Screenshot & \textbf{73.8} & \textbf{70.3} \\
\midrule
Screenshot & OCR text & 72.2 & 64.5 \\
\multirow{2}{*}{Text} & Rendered & \multirow{2}{*}{67.4} & \multirow{2}{*}{56.6} \\
     & image                    &                       &                       \\
Text & Text & 71.6 & 59.0 \\
Text & HTML & 59.8 & 56.6 \\
\bottomrule
\end{tabular*}
\end{minipage}%
\hfill
\begin{minipage}[t]{0.50\textwidth}
\vspace{0pt}
\centering
\caption{Embedding training recipe ablation. We use dynamic hard negatives with ViT unfreezing.}
\label{tab:ablation_recipe}
\footnotesize
\begin{tabular}{@{}lcc@{}}
\toprule
Recipe & ViT & QA \\
\midrule
Base model & -- & 0.725 \\
In-batch negatives & Frozen & 0.710 \\
Naive hard negatives & Frozen & 0.723 \\
Dynamic hard negatives & Frozen & 0.770 \\
\textbf{Dynamic hard negatives (ours)} & \textbf{Unfrozen} & \textbf{0.793} \\
\bottomrule
\end{tabular}
\end{minipage}
\end{table}

\paragraph{Retrieval and reader input modality ablation.}

To disentangle the contributions of retrieval modality and reader input format, we fix the embedding model to Qwen3-VL-Embedding-2B, then vary only these two factors across five configurations (Table~\ref{tab:ablation_modality}):
(1)~\textbf{Screenshot $\to$ Screenshot} (\sys{} (base));
(2)~\textbf{Screenshot $\to$ OCR};
(3)~\textbf{Text $\to$ Rendered image};
(4)~\textbf{Text $\to$ Text} (standard text-based RAG); and
(5)~\textbf{Text $\to$ HTML}.
Both retrieval and reader input format contribute, but retrieval is the larger factor (Table~\ref{tab:ablation_modality}).
Screenshot retrieval is consistently better: the first two rows always outperform the text-retrieval variants, and Screenshot $\to$ OCR outperforms Text $\to$ Text on both benchmarks, showing that visual retrieval improves downstream accuracy even when the reader sees only text.
We also test feeding raw HTML to the reader to preserve table and list structure (Text $\to$ HTML), but this \emph{underperforms} flat text on both benchmarks: HTML markup inflates context by $3.8\times$, consuming the reader's token budget on tags rather than evidence (Appendix~\ref{app:html_dom_lookup}).
Directly indexing and reading raw HTML fares no better, with accuracy dropping by up to 29\% (Appendix~\ref{app:html_rag_full}).

\paragraph{Retriever training recipe ablation.}
We ablate the visual embedding training recipe (\S\ref{sec:embedding_training}) by varying the negative curation strategy and the ViT training mode (Table~\ref{tab:ablation_recipe}; setup in Appendix~\ref{app:mini_datastore}).
In-batch negatives alone slightly underperform the base model.
Naive hard negatives (top-ranked non-positive passages from the base embedding model) improve over in-batch negatives but introduce noisy supervision from false negatives; dynamic hard-negative mining (\S\ref{sec:embedding_training}), which filters these, accounts for the largest single jump (+4.7\%).
Finally, unlike prior work on document retrieval~\cite{faysse2024colpali}, unfreezing the ViT through LoRA further improves results: rendered web pages are more visually homogeneous than PDFs or slides, requiring finer visual discrimination than a frozen backbone provides.

\subsection{Agentic Search with \sys{}}
\label{sec:ablation_agent}
\begin{figure}[t]
\begin{minipage}[t]{0.34\textwidth}
  \centering
  \includegraphics[height=4.3cm]{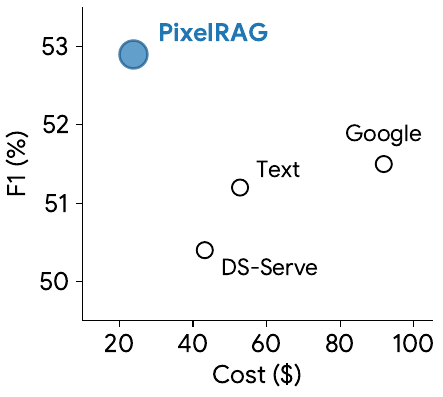}

  \captionof{figure}{Agentic multi-hop QA on MoNaCo: F1 vs.\ total cost.}
  \label{fig:ablation_agent}
\end{minipage}%
\hfill
\begin{minipage}[t]{0.64\textwidth}
  \centering
  \includegraphics[height=4.3cm]{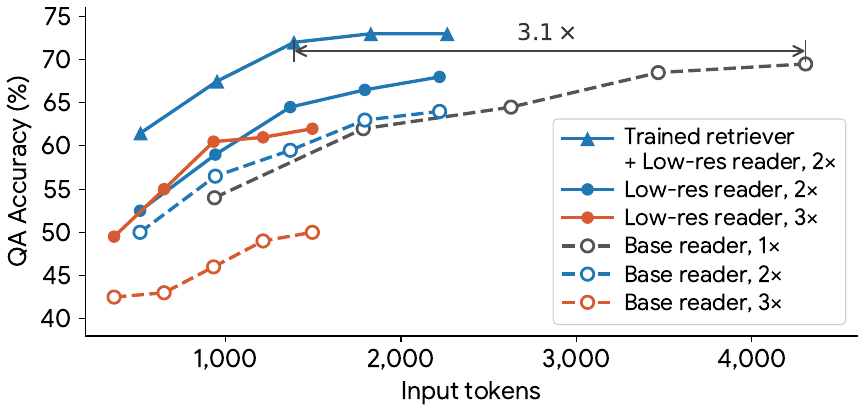}
  \captionof{figure}{SimpleQA accuracy vs.\ input tokens under image compression ($k \in \{1,\dots,5\}$). $c\times$ denotes the ratio.}
  \label{fig:sft_compression_curve}
\end{minipage}
\vspace{-8pt}
\end{figure}

We evaluate \sys{} as the search backend for a GPT-5 ReAct agent on MoNaCo~\cite{wolfson2025monaco}, a multi-hop Wikipedia QA benchmark (setup in Appendix~\ref{app:benchmark_details}).
\sys{} achieves the highest F1 at the lowest cost (Figure~\ref{fig:ablation_agent}), outperforming text-chunk retrieval, Google~\cite{serpapi2025}, and DS-Serve~\cite{liu2026ds} while costing $2$--$4\times$ less.
We report token-level F1, the official MoNaCo metric, as answers are often multi-value lists where partial credit is appropriate.
Because screenshot tiles pack more information per retrieval step than text chunks, the agent requires fewer searches and accumulates less conversation history: 3.6M prompt tokens versus 37.5M for text retrieval ($10\times$ reduction).

\subsection{Image Compression to Save Visual Tokens}
\label{sec:image_compression}

In RAG systems, prefilling retrieved context dominates inference cost, and reducing context length directly lowers latency and expense.
Pixel inputs expose a compression knob unavailable to text: image resolution.
Downsampling tiles via Lanczos resampling~\cite{lanczos_resampling} reduces token count proportionally through VLM dynamic resolution~\cite{bai2025qwen3-vl-technical-report}, without changing the datastore or retriever: a $c\times$ compression scales each side by $1/\sqrt{c}$.
To maintain accuracy at lower resolution, we fine-tune the reader (Qwen3-VL-4B) on compressed tiles.
Training data is constructed at zero additional annotation cost by reusing the $(q, p, a)$ triples from the embedding pipeline (\S\ref{sec:data_recipe}), augmented with retrieved distractor tiles to simulate the inference setting (training details in Appendix~\ref{app:reader_sft}).
At $2\times$ compression, the fine-tuned reader almost matches native-resolution accuracy while halving tokens (Figure~\ref{fig:sft_compression_curve}).
Combining the trained retriever with the $2\times$-compressed reader reaches $72\%$ at $k{=}3$, surpassing the uncompressed $k{=}5$ ceiling at one third of the token cost.

\subsection{\sys{} Performance Scales with VLM Capability}
\label{sec:vlm_progress}

Progress in VLM reading ability is a key driver of \sys{}'s performance (Table~\ref{tab:ablation_reader_scale_full}).
Early VLMs struggle with screenshots: small models in the Llama-3.2-Vision and Qwen2-VL families trail text retrieval by over 12.5\%.
From Qwen3-VL-8B ($\blacktriangle$) onward, \sys{} matches or beats text, and the advantage reaches up to 5.9\%, consistent with VLMs becoming increasingly capable at understanding text in pixels~\cite{lyu2025pixelworld}.
Table~\ref{tab:ablation_reader_scale_full} pinpoints the crossover where pixel retrieval first matches text at Qwen3-VL-4B ($\bigstar$, 70.5\% vs.\ 69.0\%): below this capability threshold models cannot reliably read text rendered in pixels and text retrieval dominates, while above it every model we test favors pixel retrieval, the gap widening to $+4.6$\,pp for Qwen3.6-35B-A3B ($\blacklozenge$).
Concurrently, successive VLM generations encode the same tile into fewer input tokens, so the cost of pixel retrieval decreases with each model generation without any pipeline change (Table~\ref{tab:ablation_reader_scale_full}).
These trends suggest that \sys{}'s pixel-native approach will continue to benefit from VLM progress, with the reader as the only component that needs upgrading---the retrieval pipeline and datastore remain unchanged.

\definecolor{hmapblue}{HTML}{1A73E8}
\definecolor{hmaporange}{HTML}{E8740A}
\def\hAcc#1{%
  \pgfmathtruncatemacro{\hpctA}{%
    (#1>=0.750)*65 + and(#1>=0.733, #1<0.750)*56 + and(#1>=0.716, #1<0.733)*48 +%
    and(#1>=0.710, #1<0.716)*40 + and(#1>=0.691, #1<0.710)*32 +%
    and(#1>=0.667, #1<0.691)*24 + and(#1>=0.593, #1<0.667)*16 +%
    and(#1>=0.05, #1<0.593)*8}%
  \edef\doheatA{\noexpand\cellcolor{hmapblue!\hpctA!white}}\doheatA #1%
}
\def\hTok#1{%
  \pgfmathtruncatemacro{\hpctT}{min(65, max(0, (#1 - 1800) / (6800 - 1800) * 65))}%
  \edef\doheatT{\noexpand\cellcolor{hmaporange!\hpctT!white}}\doheatT #1%
}

\begin{table}[htbp]
\centering
\caption{Reader-model sweep on SimpleQA. Shading: \colorbox{hmapblue!45!white}{\strut}\,accuracy, \colorbox{hmaporange!35!white}{\strut}\,input tokens.}
\label{tab:ablation_reader_scale_full}
\vspace{4pt}
\footnotesize
\begin{tabular*}{\textwidth}{@{\extracolsep{\fill}}lrrrrrr@{}}
\toprule
& \multicolumn{3}{c}{Pixel retrieval} & \multicolumn{3}{c}{Text retrieval} \\
\cmidrule(lr){2-4} \cmidrule(lr){5-7}
Reader model & Acc & in tok & out tok & Acc & in tok & out tok \\
\midrule
LLaVA-1.5-7B~\cite{liu2023llava15} & \hAcc{0.092} & \hTok{1817} &   32 & \hAcc{0.504} & \hTok{2537} &   33 \\
Llama-3.2-11B-Vision~\cite{meta2024llama32}\textsuperscript{$\dagger$} & \hAcc{0.534} & \hTok{6485} &   30 & \hAcc{0.700} & \hTok{2252} &   26 \\
Llama-3.2-90B-Vision~\cite{meta2024llama32}\textsuperscript{$\dagger$} & \hAcc{0.574} & \hTok{6485} &   32 & \hAcc{0.670} & \hTok{2252} &   39 \\
Llama-4-Scout-17B-16E~\cite{meta2026llama4} & \hAcc{0.618} & \hTok{5613} &   93 & \hAcc{0.667} & \hTok{2175} &   60 \\
Llama-4-Maverick-17B-128E~\cite{meta2026llama4} & \hAcc{0.756} & \hTok{5613} &   57 & \hAcc{0.719} & \hTok{2175} &   73 \\
Qwen2-VL-2B~\cite{wang2024qwen2vl}           & \hAcc{0.439} & \hTok{3446} &   27 & \hAcc{0.564} & \hTok{2410} &   41 \\
Qwen2-VL-7B~\cite{wang2024qwen2vl}           & \hAcc{0.628} & \hTok{3446} &   15 & \hAcc{0.663} & \hTok{2410} &   26 \\
Qwen2-VL-72B~\cite{wang2024qwen2vl}          & \hAcc{0.745} & \hTok{3446} &   23 & \hAcc{0.718} & \hTok{2410} &   14 \\
Qwen2.5-VL-3B~\cite{bai2025qwen25vl}         & \hAcc{0.460} & \hTok{3446} &   20 & \hAcc{0.661} & \hTok{2410} &   17 \\
Qwen2.5-VL-7B~\cite{bai2025qwen25vl}         & \hAcc{0.713} & \hTok{3446} &   16 & \hAcc{0.678} & \hTok{2410} &   21 \\
Qwen2.5-VL-32B~\cite{bai2025qwen25vl}        & \hAcc{0.717} & \hTok{3446} &   29 & \hAcc{0.691} & \hTok{2410} &   56 \\
Qwen2.5-VL-72B~\cite{bai2025qwen25vl}        & \hAcc{0.745} & \hTok{3446} &   17 & \hAcc{0.694} & \hTok{2346} &   26 \\
Qwen3-VL-2B~\cite{bai2025qwen3-vl-technical-report} & \hAcc{0.613} & \hTok{2615} &  229 & \hAcc{0.676} & \hTok{2408} &   34 \\
\llap{$\bigstar$\;\,}Qwen3-VL-4B~\cite{bai2025qwen3-vl-technical-report} & \hAcc{0.705} & \hTok{2615} &   70 & \hAcc{0.690} & \hTok{2408} &   62 \\
\llap{$\blacktriangle\;\mkern4mu$}Qwen3-VL-8B~\cite{bai2025qwen3-vl-technical-report} & \hAcc{0.726} & \hTok{2615} &   95 & \hAcc{0.697} & \hTok{2408} &  100 \\
Qwen3-VL-30B-A3B~\cite{bai2025qwen3-vl-technical-report} & \hAcc{0.733} & \hTok{2615} &  109 & \hAcc{0.715} & \hTok{2344} &   37 \\
Qwen3-VL-32B~\cite{bai2025qwen3-vl-technical-report} & \hAcc{0.742} & \hTok{2615} &   52 & \hAcc{0.710} & \hTok{2344} &   43 \\
Qwen3-VL-235B-A22B~\cite{bai2025qwen3-vl-technical-report} & \hAcc{0.759} & \hTok{2615} &   52 & \hAcc{0.720} & \hTok{2408} &   42 \\
Qwen3.5-0.8B~\cite{qwen2026qwen35}           & \hAcc{0.602} & \hTok{2624} &   73 & \hAcc{0.593} & \hTok{2390} &   50 \\
Qwen3.5-2B~\cite{qwen2026qwen35}             & \hAcc{0.667} & \hTok{2624} &  116 & \hAcc{0.667} & \hTok{2390} &  104 \\
Qwen3.5-4B~\cite{qwen2026qwen35}             & \hAcc{0.738} & \hTok{2624} &  118 & \hAcc{0.716} & \hTok{2459} &   95 \\
Qwen3.5-9B~\cite{qwen2026qwen35}             & \hAcc{0.750} & \hTok{2624} &  115 & \hAcc{0.701} & \hTok{2390} &   83 \\
Qwen3.5-27B~\cite{qwen2026qwen35}            & \hAcc{0.759} & \hTok{2624} &   88 & \hAcc{0.711} & \hTok{2390} &   69 \\
Qwen3.5-35B-A3B~\cite{qwen2026qwen35}        & \hAcc{0.737} & \hTok{2622} &  148 & \hAcc{0.713} & \hTok{2390} &   89 \\
Qwen3.6-27B~\cite{qwen2026qwen36}            & \hAcc{0.754} & \hTok{2624} &   73 & \hAcc{0.715} & \hTok{2390} &   65 \\
\llap{$\blacklozenge\;\mkern4mu$}Qwen3.6-35B-A3B~\cite{qwen2026qwen36} & \hAcc{0.751} & \hTok{2624} &  110 & \hAcc{0.705} & \hTok{2390} &   50 \\
\midrule
Qwen3.5-4B (reasoning)~\cite{qwen2026qwen35} & \hAcc{0.745} & \hTok{2622} & 1311 & \hAcc{0.697} & \hTok{2457} & 1597 \\
Qwen3.5-397B-A17B (reasoning)~\cite{qwen2026qwen35} & \hAcc{0.769} & \hTok{2622} & 1081 & \hAcc{0.710} & \hTok{2461} & 1188 \\
Qwen3.6-35B-A3B (reasoning)~\cite{qwen2026qwen36} & \hAcc{0.763} & \hTok{2622} & 1254 & \hAcc{0.717} & \hTok{2457} & 1005 \\
\bottomrule
\end{tabular*}\\[2pt]
{\footnotesize $\dagger$ Llama-3.2-Vision accepts only one image; pixel column uses $k{=}1$.}
\end{table}

\section{Conclusion}
\label{sec:conclusion}
RAG over web data is critical for grounding LLMs, yet extracting clean text from diverse webpages remains a persistent challenge.
\sys{} sidesteps this problem by performing retrieval and generation directly over rendered screenshots, with no HTML parsing or text extraction in the loop.
We scale this approach to over 30M tiles covering all of Wikipedia, and show that this pixel-native pipeline outperforms text-based RAG even on text-centric benchmarks where no visual reasoning is needed.
The advantage grows with VLM progress: newer models achieve higher accuracy with fewer tokens, and image compression provides up to $3\times$ token reduction, making \sys{} increasingly practical with each generation.

Limitations, broader impact, and future directions are discussed in Appendix~\ref{app:limitations}.

\section*{Acknowledgement}
We thank Akari Asai for sharing knowledge about the MoNaCo dataset, Zhiying Xu for support during the early stages of the project, Lisa Dunlap for inspiring discussions on benchmark selection, Rulin Shao for insightful discussions throughout the project and support during the release phase, and Xueguang Ma for providing a naive version of screenshot capture code that helped bootstrap the project. We also thank members of Sky Lab and the SM group for helpful discussions and proofreading.
This work was supported in part by the Gemini Academic Program and the NVIDIA Academic Grant Program. This research was also supported by gifts from Accenture, Amazon, AMD, Anyscale, Broadcom Inc., Google, IBM, Intel, Intesa Sanpaolo, Lambda, Mibura Inc, Samsung SDS, and SAP.

\setcitestyle{numbers,sort&compress}
\bibliographystyle{unsrt}
\bibliography{references}

@misc{serpapi2025,
  title={{SerpApi}: Google Search API},
  author={{SerpApi}},
  year={2025},
  howpublished={\url{https://serpapi.com}}
}

@inproceedings{weyand2020gldv2,
  title={{Google Landmarks Dataset v2} --- A Large-Scale Benchmark for Instance-Level Recognition and Retrieval},
  author={Weyand, Tobias and Araujo, Andr{\'e} and Cao, Bingyi and Sim, Jack},
  booktitle={CVPR},
  year={2020}
}

@article{wang2024charxiv,
  title={Charxiv: Charting gaps in realistic chart understanding in multimodal llms},
  author={Wang, Zirui and Xia, Mengzhou and He, Luxi and Chen, Howard and Liu, Yitao and Zhu, Richard and Liang, Kaiqu and Wu, Xindi and Liu, Haotian and Malladi, Sadhika and others},
  journal={Advances in Neural Information Processing Systems},
  volume={37},
  pages={113569--113697},
  year={2024}
}

@misc{mwparserfromhell,
  author = {Ben Kurtovic and contributors},
  title = {mwparserfromhell: A Python parser for MediaWiki wikicode},
  howpublished = {\url{https://github.com/earwig/mwparserfromhell}},
  year = {2026}
}

@article{oord2018representation,
  title={Representation learning with contrastive predictive coding},
  author={Oord, Aaron van den and Li, Yazhe and Vinyals, Oriol},
  journal={arXiv preprint arXiv:1807.03748},
  year={2018}
}

@misc{openai2025gpt41,
  author={{OpenAI}},
  title={Introducing {GPT-4.1} in the {API}},
  year={2025},
  month=apr,
  url={https://openai.com/index/gpt-4-1/},
  note={Blog post}
}

@inproceedings{liu2026ds,
  title={{DS SERVE}: A Framework for Efficient and Scalable Neural Retrieval},
  author={Liu, Jinjian and Wang, Yichuan and Lyu, Xinxi and Shao, Rulin and Gonzalez, Joseph E. and Zaharia, Matei and Min, Sewon},
  booktitle={Fortieth AAAI Conference on Artificial Intelligence (AAAI)},
  pages={41631--41633},
  year={2026},
  doi={10.1609/aaai.v40i48.42363}
}

@article{li2025websailor,
  title={Websailor: Navigating super-human reasoning for web agent},
  author={Li, Kuan and Zhang, Zhongwang and Yin, Huifeng and Zhang, Liwen and Ou, Litu and Wu, Jialong and Yin, Wenbiao and Li, Baixuan and Tao, Zhengwei and Wang, Xinyu and others},
  journal={arXiv preprint arXiv:2507.02592},
  year={2025}
}

@inproceedings{Xiong2020ApproximateNN,
  title={Approximate Nearest Neighbor Negative Contrastive Learning for Dense Text Retrieval},
  author={Xiong, Lee and Xiong, Chenyan and Li, Ye and Tang, Kwok-Fung and Liu, Jialin and Bennett, Paul N. and Ahmed, Junaid and Overwijk, Arnold},
  booktitle={International Conference on Learning Representations},
  year={2021}
}

@inproceedings{barbaresi-2021-trafilatura,
    title = "Trafilatura: {A} Web Scraping Library and Command-Line Tool for Text Discovery and Extraction",
    author = "Barbaresi, Adrien",
    editor = "Ji, Heng  and
      Park, Jong C.  and
      Xia, Rui",
    booktitle = "Proceedings of the 59th Annual Meeting of the Association for Computational Linguistics and the 11th International Joint Conference on Natural Language Processing: System Demonstrations",
    month = aug,
    year = "2021",
    address = "Online",
    publisher = "Association for Computational Linguistics",
    url = "https://aclanthology.org/2021.acl-demo.15/",
    doi = "10.18653/v1/2021.acl-demo.15",
    pages = "122--131"
}

@InProceedings{bevendorff:2018,
  address =             {Berlin Heidelberg New York},
  author =              {Janek Bevendorff and Benno Stein and Matthias Hagen and Martin Potthast},
  booktitle =           {Advances in Information Retrieval. 40th European Conference on IR Research (ECIR 2018)},
  editor =              {Leif Azzopardi and Allan Hanbury and Gabriella Pasi and Benjamin Piwowarski},
  month =               mar,
  publisher =           {Springer},
  series =              {Lecture Notes in Computer Science},
  site =                {Grenoble, France},
  title =               {{Elastic ChatNoir: Search Engine for the ClueWeb and the Common Crawl}},
  year =                2018
}

@misc{zhao2024swiftascalablelightweightinfrastructure,
      title={SWIFT:A Scalable lightWeight Infrastructure for Fine-Tuning},
      author={Yuze Zhao and Jintao Huang and Jinghan Hu and Xingjun Wang and Yunlin Mao and Daoze Zhang and Hong Zhang and Zeyinzi Jiang and Zhikai Wu and Baole Ai and Ang Wang and Wenmeng Zhou and Yingda Chen},
      year={2024},
      eprint={2408.05517},
      archivePrefix={arXiv},
      primaryClass={cs.CL},
      url={https://arxiv.org/abs/2408.05517},
}

@article{lyu2025frustratingly,
  title={Frustratingly simple retrieval improves challenging, reasoning-intensive benchmarks},
  author={Lyu, Xinxi and Duan, Michael and Shao, Rulin and Koh, Pang Wei and Min, Sewon},
  journal={arXiv preprint arXiv:2507.01297},
  year={2025}
}

@inproceedings{faysse2024colpali,
  title={Colpali: Efficient document retrieval with vision language models},
  author={Faysse, Manuel and Sibille, Hugues and Wu, Tony and Omrani, Bilel and Viaud, Gautier and Hudelot, C{\'e}line and Colombo, Pierre},
  booktitle={International Conference on Learning Representations},
  year={2025}
}

@article{moreira2026nemotron,
  title={Nemotron ColEmbed V2: Top-Performing Late Interaction Embedding Models for Visual Document Retrieval},
  author={Moreira, Gabriel de Souza P and Ak, Ronay and Xu, Mengyao and Holworthy, Oliver and Schifferer, Benedikt and Yu, Zhiding and Babakhin, Yauhen and Osmulski, Radek and Cai, Jiarui and Chesler, Ryan and others},
  journal={arXiv preprint arXiv:2602.03992},
  year={2026}
}

@article{faiss,
  author    = {Jeff Johnson and Matthijs Douze and Herv\'e J\'egou},
  title     = {{Billion-Scale Similarity Search with GPUs}},
  journal   = {IEEE Transactions on Big Data},
  volume    = {7},
  number    = {3},
  pages     = {535--547},
  year      = {2021},
  doi       = {10.1109/TBDATA.2019.2921572},
  url       = {https://doi.org/10.1109/TBDATA.2019.2921572}
}

@article{bai2025qwen3-vl-technical-report,
  title={Qwen3-vl technical report},
  author={Bai, Shuai and Cai, Yuxuan and Chen, Ruizhe and Chen, Keqin and Chen, Xionghui and Cheng, Zesen and Deng, Lianghao and Ding, Wei and Gao, Chang and Ge, Chunjiang and others},
  journal={arXiv preprint arXiv:2511.21631},
  year={2025}
}

@article{shorten2026irpapers,
  title={Irpapers: A visual document benchmark for scientific retrieval and question answering},
  author={Shorten, Connor and Skaburskas, Augustas and Jones, Daniel M and Pierse, Charles and Esposito, Roberto and Trengrove, John and Dilocker, Etienne and van Luijt, Bob},
  journal={arXiv preprint arXiv:2602.17687},
  year={2026}
}

@article{li2026qwen3-vl-embedding,
  title={Qwen3-VL-Embedding and Qwen3-VL-Reranker: A Unified Framework for State-of-the-Art Multimodal Retrieval and Ranking},
  author={Li, Mingxin and Zhang, Yanzhao and Long, Dingkun and Chen, Keqin and Song, Sibo and Bai, Shuai and Yang, Zhibo and Xie, Pengjun and Yang, An and Liu, Dayiheng and others},
  journal={arXiv preprint arXiv:2601.04720},
  year={2026}
}

@article{wei2025deepseek,
  title={DeepSeek-OCR: Contexts Optical Compression},
  author={Wei, Haoran and Sun, Yaofeng and Li, Yukun},
  journal={arXiv preprint arXiv:2510.18234},
  year={2025}
}

@article{cheng2025glyph,
  title={Glyph: Scaling context windows via visual-text compression},
  author={Cheng, Jiale and Liu, Yusen and Zhang, Xinyu and Fei, Yulin and Hong, Wenyi and Lyu, Ruiliang and Wang, Weihan and Su, Zhe and Gu, Xiaotao and Liu, Xiao and others},
  journal={arXiv preprint arXiv:2510.17800},
  year={2025}
}

@article{lyu2025pixelworld,
  title={PixelWorld: How Far Are We from Perceiving Everything as Pixels?},
  author={Lyu, Zhiheng and Ma, Xueguang and Chen, Wenhu},
  journal={arXiv preprint arXiv:2501.19339},
  year={2025}
}

@inproceedings{dse,
  title={Unifying multimodal retrieval via document screenshot embedding},
  author={Ma, Xueguang and Lin, Sheng-Chieh and Li, Minghan and Chen, Wenhu and Lin, Jimmy},
  booktitle={Proceedings of the 2024 Conference on Empirical Methods in Natural Language Processing},
  pages={6492--6505},
  year={2024}
}

@article{shao2024scaling,
  title={Scaling retrieval-based language models with a trillion-token datastore},
  author={Shao, Rulin and He, Jacqueline and Asai, Akari and Shi, Weijia and Dettmers, Tim and Min, Sewon and Zettlemoyer, Luke and Koh, Pang W},
  journal={Advances in Neural Information Processing Systems},
  volume={37},
  pages={91260--91299},
  year={2024}
}

@article{fang2025reusing,
  title={Reusing Pre-Training Data at Test Time is a Compute Multiplier},
  author={Fang, Alex and Voice, Thomas and Pang, Ruoming and Schmidt, Ludwig and Gunter, Tom},
  journal={arXiv preprint arXiv:2511.04234},
  year={2025}
}

@article{li2024datacomp,
  title={Datacomp-lm: In search of the next generation of training sets for language models},
  author={Li, Jeffrey and Fang, Alex and Smyrnis, Georgios and Ivgi, Maor and Jordan, Matt and Gadre, Samir and Bansal, Hritik and Guha, Etash and Keh, Sedrick and Arora, Kushal and others},
  journal={Advances in Neural Information Processing Systems},
  volume={37},
  pages={14200--14282},
  year={2024}
}

@article{penedo2024fineweb,
  title={The fineweb datasets: Decanting the web for the finest text data at scale},
  author={Penedo, Guilherme and Kydl{\'\i}{\v{c}}ek, Hynek and Ben Allal, Loubna and Lozhkov, Anton and Mitchell, Margaret and Raffel, Colin and Von Werra, Leandro and Wolf, Thomas},
  journal={Advances in Neural Information Processing Systems},
  volume={37},
  pages={30811--30849},
  year={2024}
}

@article{wolfson2025monaco,
  title={{MoNaCo}: More Natural and Complex Questions for Reasoning Across Dozens of Documents},
  author={Wolfson, Tomer and Trivedi, Harsh and Geva, Mor and Goldberg, Yoav and Roth, Dan and Khot, Tushar and Sabharwal, Ashish and Tsarfaty, Reut},
  journal={Transactions of the Association for Computational Linguistics},
  year={2025},
  eprint={2508.11133},
  archivePrefix={arXiv}
}

@misc{lanczos_resampling,
  title={Lanczos resampling},
  author={{Wikipedia contributors}},
  year={2025},
  howpublished={\url{https://en.wikipedia.org/wiki/Lanczos_resampling}},
  note={Accessed: 2026-04-29}
}

@article{lewis2020retrieval,
  title={Retrieval-augmented generation for knowledge-intensive nlp tasks},
  author={Lewis, Patrick and Perez, Ethan and Piktus, Aleksandra and Petroni, Fabio and Karpukhin, Vladimir and Goyal, Naman and K{\"u}ttler, Heinrich and Lewis, Mike and Yih, Wen-tau and Rockt{\"a}schel, Tim and others},
  journal={Advances in neural information processing systems},
  volume={33},
  pages={9459--9474},
  year={2020}
}

@inproceedings{guu2020realm,
  title={{REALM}: Retrieval-augmented language model pre-training},
  author={Guu, Kelvin and Lee, Kenton and Tung, Zora and Pasupat, Panupong and Chang, Mingwei},
  booktitle={International Conference on Machine Learning},
  pages={3929--3938},
  year={2020}
}

@inproceedings{karpukhin2020dpr,
  title={Dense passage retrieval for open-domain question answering},
  author={Karpukhin, Vladimir and O{\u{g}}uz, Barlas and Min, Sewon and Lewis, Patrick and Wu, Ledell and Edunov, Sergey and Chen, Danqi and Yih, Wen-tau},
  booktitle={Proceedings of the 2020 Conference on Empirical Methods in Natural Language Processing},
  pages={6769--6781},
  year={2020}
}

@inproceedings{izacard2021fid,
  title={Leveraging passage retrieval with generative models for open domain question answering},
  author={Izacard, Gautier and Grave, Edouard},
  booktitle={Proceedings of the 16th Conference of the European Chapter of the Association for Computational Linguistics},
  pages={874--880},
  year={2021}
}

@article{nakano2021webgpt,
  title={Web{GPT}: Browser-assisted question-answering with human feedback},
  author={Nakano, Reiichiro and Hilton, Jacob and Balaji, Suchir and Wu, Jeff and Ouyang, Long and Kim, Christina and Hesse, Christopher and Jain, Shantanu and Kosaraju, Vineet and Saunders, William and others},
  journal={arXiv preprint arXiv:2112.09332},
  year={2021}
}

@article{jin2025search,
  title={Search-r1: Training llms to reason and leverage search engines with reinforcement learning},
  author={Jin, Bowen and Zeng, Hansi and Yue, Zhenrui and Yoon, Jinsung and Arik, Sercan and Wang, Dong and Zamani, Hamed and Han, Jiawei},
  journal={arXiv preprint arXiv:2503.09516},
  year={2025}
}

@article{kwiatkowski2019natural,
  title={Natural Questions: A Benchmark for Question Answering Research},
  author={Kwiatkowski, Tom and Palomaki, Jennimaria and Redfield, Olivia and Collins, Michael and Parikh, Ankur and Alberti, Chris and Epstein, Danielle and Polosukhin, Illia and Devlin, Jacob and Lee, Kenton and others},
  journal={Transactions of the Association for Computational Linguistics},
  volume={7},
  pages={453--466},
  year={2019}
}

@inproceedings{herzig2021nqtables,
  title={Open Domain Question Answering over Tables via Dense Retrieval},
  author={Herzig, Jonathan and M{\"u}ller, Thomas and Krichene, Syrine and Eisenschlos, Julian Martin},
  booktitle={Proceedings of the 2021 Conference of the North American Chapter of the Association for Computational Linguistics},
  pages={512--519},
  year={2021}
}

@article{wei2024simpleqa,
  title={Measuring short-form factuality in large language models},
  author={Wei, Jason and Karina, Nguyen and Chung, Hyung Won and Jiao, Yunxin Joy and Papay, Spencer and Glaese, Amelia and Schulman, John and Fedus, William},
  journal={arXiv preprint arXiv:2411.04368},
  year={2024}
}

@article{jiang2024mmsearch,
  title={{MMSearch}: Benchmarking the Potential of Large Models as Multi-modal Search Engines},
  author={Jiang, Dongzhi and Zhang, Renrui and Guo, Ziyu and Wu, Yanmin and Lei, Jiayi and Qiu, Pengshuo and Lu, Pan and Chen, Zehui and Fu, Chaoyou and Song, Guanglu and Gao, Peng and Liu, Yu and Li, Chunyuan and Li, Hongsheng},
  journal={arXiv preprint arXiv:2409.12959},
  year={2024}
}

@inproceedings{mensink2023evqa,
  title={Encyclopedic {VQA}: Visual questions about detailed properties of fine-grained categories},
  author={Mensink, Thomas and Uijlings, Jasper and Castrejon, Lluis and Goel, Arushi and Cadar, Felipe and Zhou, Howard and Sha, Fei and Araujo, Andr{\'e} and Ferrari, Vittorio},
  booktitle={Proceedings of the IEEE/CVF International Conference on Computer Vision},
  pages={3082--3092},
  year={2023}
}

@article{fu2025livevqa,
  title={Seeking and Updating with Live Visual Knowledge},
  author={Fu, Mingyang and Peng, Yuyang and Chen, Dongping and Zhou, Zetong and Liu, Benlin and Wan, Yao and Zhao, Zhou and Yu, Philip S. and Krishna, Ranjay},
  journal={arXiv preprint arXiv:2504.05288},
  year={2025}
}

@misc{qwen2026qwen35,
  title={{Qwen3.5}: Towards Native Multimodal Agents},
  author={{Qwen Team}},
  year={2026},
  howpublished={\url{https://qwen.ai/blog?id=qwen3.5}}
}

@misc{qwen2026qwen36,
  title={{Qwen3.6}: Towards Real World Agents},
  author={{Qwen Team}},
  year={2026},
  howpublished={\url{https://qwen.ai/blog?id=qwen3.6/}}
}

@inproceedings{li2026beyond,
  title={Beyond a Single Extractor: Re-thinking HTML-to-Text Extraction for LLM Pre-training},
  author={Li, Jeffrey and Gardner, Joshua P and Kang, Doug and Shi, Fangping and Singh, Karanjeet and Li, Chun-Liang and Shandilya, Herumb and Hall, David Leo Wright and Tuzel, Oncel and Liang, Percy and others},
  booktitle={Findings of the Association for Computational Linguistics: EACL 2026},
  pages={5836--5861},
  year={2026}
}

@article{wang2025readerlm,
  title={{ReaderLM-v2}: Small Language Model for {HTML} to Markdown and {JSON}},
  author={Wang, Feng and Shi, Zesheng and Wang, Bo and Wang, Nan and Xiao, Han},
  journal={arXiv preprint arXiv:2503.01151},
  year={2025}
}

@article{liu2025dripper,
  title={Dripper: Token-Efficient Main {HTML} Extraction with a Lightweight {LM}},
  author={Liu, Mengjie and Peng, Jiahui and Ning, Wenchang and others},
  journal={arXiv preprint arXiv:2511.23119},
  year={2025}
}

@inproceedings{yu2024visrag,
  title={{VisRAG}: Vision-based Retrieval-augmented Generation on Multi-modality Documents},
  author={Yu, Shi and Tang, Chaoyue and Xu, Bokai and Cui, Junbo and Ran, Junhao and Yan, Yukun and Liu, Zhenghao and Wang, Shuo and Han, Xu and Liu, Zhiyuan and Sun, Maosong},
  booktitle={The Thirteenth International Conference on Learning Representations (ICLR)},
  year={2025}
}

@misc{neuml2024wikipedia,
  title={NeuML/wikipedia: Wikipedia text dataset},
  author={{NeuML}},
  year={2024},
  howpublished={\url{https://huggingface.co/datasets/NeuML/wikipedia}},
  note={Text extracted from Wikipedia XML dumps via \texttt{mwparserfromhell}}
}

@article{cho2024m3docrag,
  title={M3docrag: Multi-modal retrieval is what you need for multi-page multi-document understanding},
  author={Cho, Jaemin and Mahata, Debanjan and Irsoy, Ozan and He, Yujie and Bansal, Mohit},
  journal={arXiv preprint arXiv:2411.04952},
  year={2024}
}

@article{zhu2024mmdocbench,
  title={Mmdocbench: Benchmarking large vision-language models for fine-grained visual document understanding},
  author={Zhu, Fengbin and Liu, Ziyang and Ng, Xiang Yao and Wu, Haohui and Wang, Wenjie and Feng, Fuli and Wang, Chao and Luan, Huanbo and Chua, Tat Seng},
  journal={arXiv preprint arXiv:2410.21311},
  year={2024}
}

@article{mace2025vidore,
  title={Vidore benchmark v2: Raising the bar for visual retrieval},
  author={Mac{\'e}, Quentin and Loison, Ant{\'o}nio and Faysse, Manuel},
  journal={arXiv preprint arXiv:2505.17166},
  year={2025}
}

@article{wang2024qwen2vl,
  title={Qwen2-{VL}: Enhancing Vision-Language Model's Perception of the World at Any Resolution},
  author={Wang, Peng and Bai, Shuai and Tan, Sinan and Wang, Shijie and Fan, Zhihao and Bai, Jinze and Chen, Keqin and Liu, Xuejing and Wang, Jialin and Ge, Wenbin and others},
  journal={arXiv preprint arXiv:2409.12191},
  year={2024}
}

@misc{meta2024llama32,
  title={Llama 3.2: Lightweight Text and Multimodal Models},
  author={{Meta AI}},
  year={2024},
  howpublished={\url{https://ai.meta.com/blog/llama-3-2-connect-2024-vision-edge-mobile-devices/}}
}

@article{openai2024gpt4o,
  title={{GPT-4o} System Card},
  author={{OpenAI}},
  journal={arXiv preprint arXiv:2410.21276},
  year={2024}
}

@techreport{anthropic2024claude3,
  title={The {Claude} 3 Model Family: {Opus}, {Sonnet}, {Haiku}},
  author={{Anthropic}},
  year={2024},
  url={https://www-cdn.anthropic.com/de8ba9b01c9ab7cbabf5c33b80b7bbc618857627/Model_Card_Claude_3.pdf},
  note={Technical Report}
}

@article{geminiteam2025gemini25,
  title={Gemini 2.5: Pushing the Frontier with Advanced Reasoning, Multimodality, Long Context, and Next Generation Agentic Capabilities},
  author={{Gemini Team}},
  journal={arXiv preprint arXiv:2507.06261},
  year={2025}
}

@article{zhu2025internvl3,
  title={{InternVL3}: Exploring Advanced Training and Test-Time Recipes for Open-Source Multimodal Models},
  author={Zhu, Jinguo and Wang, Weiyun and Chen, Zhe and others},
  journal={arXiv preprint arXiv:2504.10479},
  year={2025}
}

@inproceedings{tan2025htmlrag,
  title={{HtmlRAG}: {HTML} is Better Than Plain Text for Modeling Retrieved Knowledge in {RAG} Systems},
  author={Tan, Jiejun and Dou, Zhicheng and Wang, Wen and Wang, Mang and Chen, Weipeng and Wen, Ji-Rong},
  booktitle={Proceedings of the ACM Web Conference 2025 (WWW)},
  year={2025}
}

@article{yan2026unlocking,
  title={Unlocking Multimodal Document Intelligence: From Current Triumphs to Future Frontiers of Visual Document Retrieval},
  author={Yan, Yibo and Huo, Jiahao and Feng, Guanbo and Ou, Mingdong and Cao, Yi and Zou, Xin and Liu, Shuliang and Lyu, Yuanhuiyi and Huang, Yu and Li, Jungang and others},
  journal={arXiv preprint arXiv:2602.19961},
  year={2026}
}

@article{huo2026causalembed,
  title={{CausalEmbed}: Auto-Regressive Multi-Vector Generation in Latent Space for Visual Document Embedding},
  author={Huo, Jiahao and Huang, Yu and Yan, Yibo and Pan, Ye and Zheng, Kening and Huang, Wei-Chieh and Cao, Yi and Ou, Mingdong and Yu, Philip S. and Hu, Xuming},
  journal={arXiv preprint arXiv:2601.21262},
  year={2026}
}

@article{team2025tongyi,
  title={Tongyi deepresearch technical report},
  author={{Tongyi DeepResearch Team} and Li, Baixuan and Zhang, Bo and Zhang, Dingchu and Huang, Fei and Li, Guangyu and Chen, Guoxin and Yin, Huifeng and Wu, Jialong and Zhou, Jingren and others},
  journal={arXiv preprint arXiv:2510.24701},
  year={2025}
}

@article{liu2023llava15,
  title={Improved Baselines with Visual Instruction Tuning},
  author={Liu, Haotian and Li, Chunyuan and Li, Yuheng and Lee, Yong Jae},
  journal={arXiv preprint arXiv:2310.03744},
  year={2023}
}

@misc{meta2026llama4,
  title={The {Llama} 4 herd: The beginning of a new era of natively multimodal {AI} innovation},
  author={{Meta AI}},
  year={2026},
  howpublished={\url{https://ai.meta.com/blog/llama-4-multimodal-intelligence/}},
  note={Blog post}
}

@article{bai2025qwen25vl,
  title={Qwen2.5-{VL} Technical Report},
  author={Bai, Shuai and Chen, Keqin and Liu, Xuejing and Wang, Jialin and Ge, Wenbin and Song, Sibo and Dang, Kai and Wang, Peng and Wang, Shijie and Tang, Jun and others},
  journal={arXiv preprint arXiv:2502.13923},
  year={2025}
}

@misc{thrush2022pipeline,
    title={Online Language Modelling Data Pipeline},
    author={Tristan Thrush and Helen Ngo and Nathan Lambert and Douwe Kiela},
    year={2022},
    howpublished={\url{https://github.com/huggingface/olm-datasets}}
}

@misc{kiwixzim,
  author={{Kiwix Association}},
  title={Kiwix --- Offline Reader for Web Content},
  year={2007},
  howpublished={\url{https://kiwix.org}},
  note={Open-source offline browser using the {ZIM} archive format}
}

@article{sun2026reading,
  title={Reading, Not Thinking: Understanding and Bridging the Modality Gap When Text Becomes Pixels in Multimodal LLMs},
  author={Sun, Kaiser and Yuan, Xiaochuang and Liu, Hongjun and Zhao, Chen and Zhang, Cheng and Dredze, Mark and Bai, Fan},
  journal={arXiv preprint arXiv:2603.09095},
  year={2026}
}

@article{lu2024text,
  title={From text to pixel: Advancing long-context understanding in mllms},
  author={Lu, Yujie and Li, Xiujun and Fu, Tsu-Jui and Eckstein, Miguel and Wang, William Yang},
  journal={arXiv preprint arXiv:2405.14213},
  year={2024}
}

@article{wang2025leann,
  title={LEANN: A Low-Storage Vector Index},
  author={Wang, Yichuan and Li, Zhifei and Liu, Shu and Wu, Yongji and Mao, Ziming and Zhao, Yilong and Yan, Xiao and Xu, Zhiying and Zhou, Yang and Stoica, Ion and others},
  journal={arXiv preprint arXiv:2506.08276},
  year={2025}
}

\appendix
\clearpage
\section*{Technical appendices and supplementary material}

\vspace{4pt}
\noindent\textbf{Contents}
\vspace{2pt}

\noindent\begin{tabular}{@{}p{0.85\textwidth}r@{}}
\multicolumn{2}{@{}l}{\emph{A\quad System \& Implementation Details}} \\
\quad A.1\; Rendering Pipeline & \pageref{app:screenshot_comparison} \\
\quad A.2\; Datastore Fetching & \pageref{app:datastore_fetch} \\
\quad A.3\; Embedding Training: Data Recipe Details and Prompts & \pageref{app:embedding_training_details} \\
\quad A.4\; Reader Fine-tuning: SFT Results across Compression Ratios & \pageref{app:reader_sft} \\[3pt]
\multicolumn{2}{@{}l}{\emph{B\quad Evaluation Protocol}} \\
\quad B.1\; Benchmark Details & \pageref{app:benchmark_details} \\
\quad B.2\; Grading Protocol & \pageref{app:grading_protocol} \\[3pt]
\multicolumn{2}{@{}l}{\emph{C\quad Additional Results}} \\
\quad C.1\; Results with Alternative Readers & \pageref{app:alt_reader_results} \\
\quad C.2\; Full Results of Scaling VLM Performance & \pageref{app:reader_scale_full} \\[3pt]
\multicolumn{2}{@{}l}{\emph{D\quad Analysis}} \\
\quad D.1\; Wikipedia Text Extractor Comparison & \pageref{app:parser_comparison} \\
\quad D.2\; Visual Information Loss During HTML Parsing & \pageref{app:visual_loss} \\
\quad D.3\; Retrieval Signal Loss Under Text Linearization & \pageref{app:retrieval_signal_loss} \\
\quad D.4\; Detailed Failure Decomposition & \pageref{app:failure_detailed} \\
\quad D.5\; HTML DOM Lookup Baseline: Setup and Analysis & \pageref{app:html_dom_lookup} \\
\quad D.6\; Directly RAG on Raw HTML Data & \pageref{app:html_rag_full} \\[3pt]
\multicolumn{2}{@{}l}{\emph{E\quad Limitations, Broader Impact, and Future Work}} \\
\quad & \pageref{app:limitations} \\[3pt]
\multicolumn{2}{@{}l}{\emph{F\quad Prompt Listings}} \\
\quad & \pageref{app:prompt_listings} \\
\end{tabular}

\vspace{6pt}


\section{System \& Implementation Details}

\subsection{Rendering Pipeline}
\label{app:screenshot_comparison}

\begin{figure}[htbp]
  \centering
  \includegraphics[width=\linewidth]{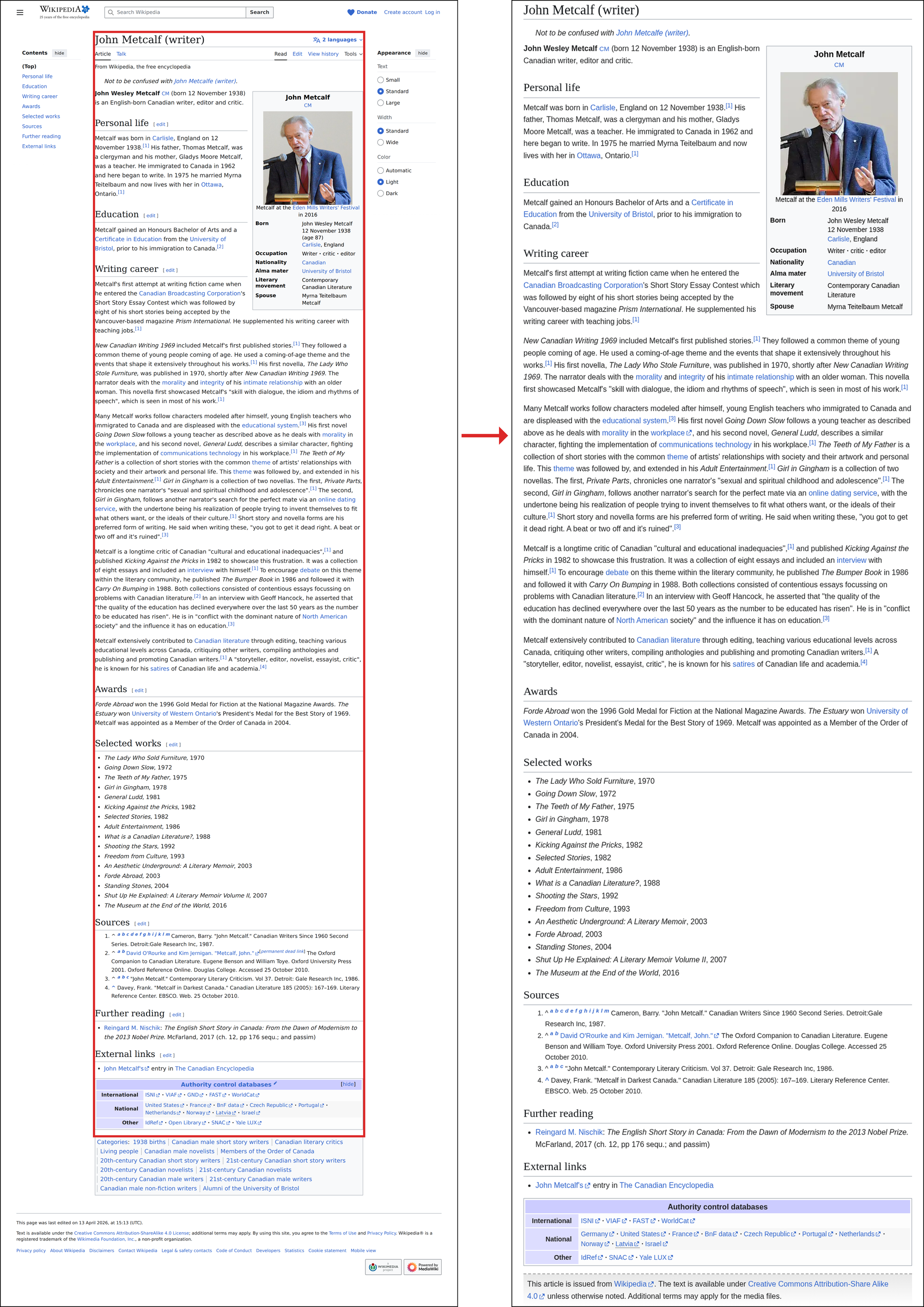}
  \caption{A Wikipedia page as it appears online in a browser (left) and after our rendering pipeline (right). We strip surrounding UI elements and whitespace, keeping only the article content, which is then tiled for indexing.}
  \label{fig:screenshot_comparison}
\end{figure}

We render each Wikipedia page using Playwright in a headless Chromium browser at a fixed viewport width.
After capture, we strip browser UI elements (navigation bars, sidebars, toolbars) and crop away surrounding whitespace, retaining only the article content area.
This produces a clean, content-only screenshot that is tiled and indexed into \sys{}.
Figure~\ref{fig:screenshot_comparison} shows an example before and after this processing step.

\paragraph{Storage considerations.}
The rendered tile images can be sizable (${\sim}5.6$\,TB for Wikipedia and ${\sim}469$\,GB for news).
In principle, the tile images need not be stored persistently: after embedding, one can retain only the vector index and re-render the top-$K$ pages on the fly from their original HTML, CSS, and JavaScript sources at query time, eliminating the image storage cost entirely.
The vector index itself can be further compressed with low-storage indexing techniques such as LEANN~\cite{wang2025leann}, reducing the embedding storage cost as well.

\subsection{Datastore Fetching}
\label{app:datastore_fetch}

\paragraph{Wikipedia.}
We populate the local cache from a Kiwix-serve instance backed by a ZIM archive (dump date: 2025-08).
Kiwix exposes the full Wikipedia corpus as a static snapshot, which avoids network bottlenecks entirely during rendering and sidesteps potential legal concerns associated with large-scale crawling of live pages.
After excluding redirects, we render 7,134,778 articles (100\% coverage of content pages), yielding approximately 30M tiles at ${\sim}4.2$ tiles per article.

\paragraph{News corpus.}
We collect articles from three major English-language news outlets: BBC (356,358 articles), AP~News (261,241 articles), and CNN (49,924 articles), totaling 667,523 articles and 3.6M screenshot tiles.

\subsection{Embedding Training: Data Recipe Details and Prompts}
\label{app:embedding_training_details}

This appendix collects the full prompts, heuristics, and implementation details behind the embedding-training pipeline of \S\ref{sec:embedding_training}.
Each sub-subsection is referenced from the exact point of choice in the main text.

\subsubsection{Knowledge-intensive tile sampling}
\label{app:ki_sampling}

Referenced from \S\ref{sec:data_recipe}, Stage~1.
Not every Wikipedia page, or every tile within a page, is suitable for synthetic query generation.
We apply three successive filters to select a pool of knowledge-intensive tiles.

\paragraph{Render-quality gate.}
We drop any page whose offline render is too small or failed: \texttt{page\_height} $< 3{,}000$\,px.
This removes empty pages, redirects, and partial renders before they are ever sampled.

\paragraph{Title/URL regex blocklist.}
We exclude meta and namespace pages by title and URL pattern matching: disambiguation pages, pages in the \texttt{Category:}, \texttt{Template:}, \texttt{Wikipedia:}, \texttt{Portal:}, \texttt{File:}, \texttt{Help:}, \texttt{Talk:}, \texttt{Module:}, and \texttt{Draft:} namespaces, as well as \texttt{*\_deaths} and \texttt{*\_births} list pages.

\paragraph{Chunk-position filter.}
Within a page that passes both filters, we sample only from the first 70\% of tiles, so that footers, reference lists, and ``See also'' boilerplate never become positive evidence.

\subsubsection{Synthetic query generation prompt}
\label{app:query_gen_prompt}

Referenced from \S\ref{sec:data_recipe}, Stage~1.
We prompt \texttt{gemini-3.1-flash-lite-preview} (Vertex AI) with the rendered tile and the prompt in Figure~\ref{fig:prompt-query-gen}, decoding at \texttt{temperature=0.7} and \texttt{max\_output\_tokens=1024}.
The model returns either the five-line \texttt{Q\,/\,A\,/\,S\,/\,T\,/\,C} block or the literal string \texttt{SKIP}.

\begin{figure}[htbp]
\centering
\begin{minipage}[t]{0.38\linewidth}
\centering
\includegraphics[width=\linewidth]{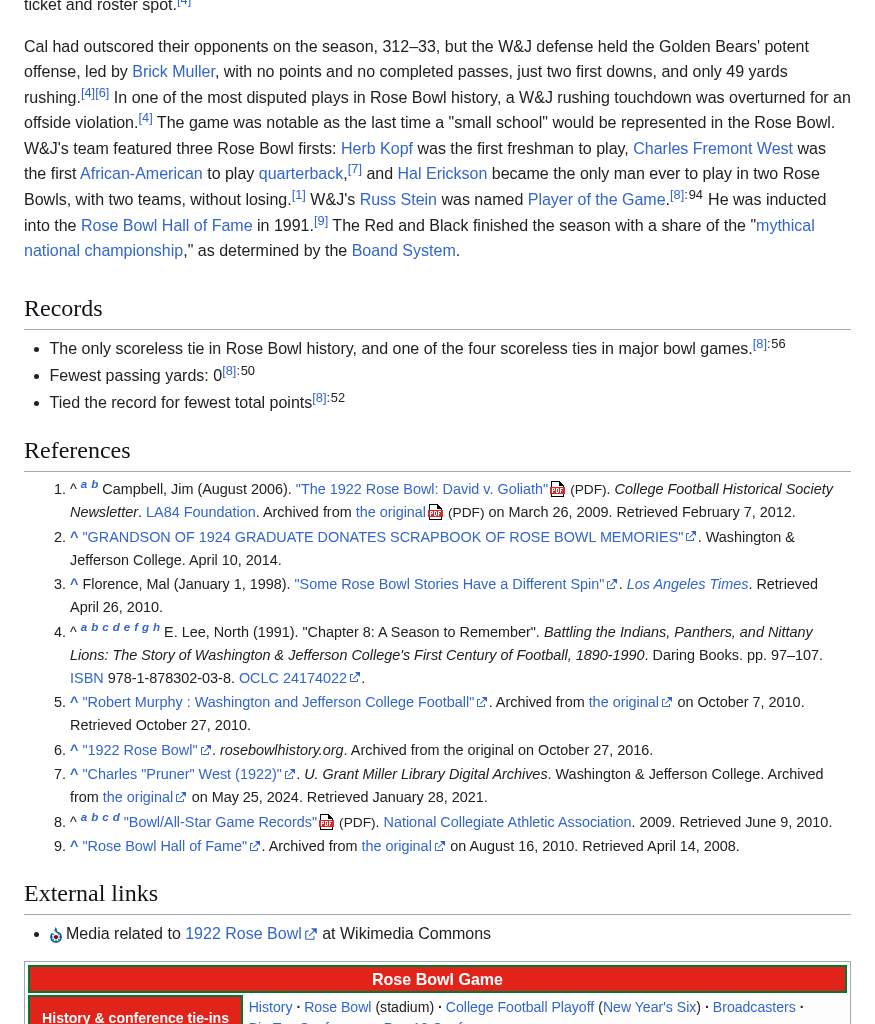}
\small\textit{Input: rendered Wikipedia tile}
\end{minipage}
\hfill
\begin{minipage}[t]{0.58\linewidth}
\begin{tcolorbox}[colback=gray!5!white,colframe=gray!50!black,
                  title=Model output $\to$ parsed record,fontupper=\small\ttfamily,
                  left=4pt,right=4pt,top=4pt,bottom=4pt]
\begin{verbatim}
{
  "query": "Who was the first African-
    American to play quarterback in
    the Rose Bowl?",
  "answer": "Charles Fremont West",
  "source_sentence": "Charles Fremont
    West was the first African-American
    to play quarterback, and Hal
    Erickson became the only man ever
    to play in two Rose Bowls, with
    two teams, without losing.",
  "source_type": "prose",
  "subject": "sports",
  "title": "1922 Rose Bowl",
  "url": "https://en.wikipedia.org/
    wiki/1922_Rose_Bowl"
}
\end{verbatim}
\end{tcolorbox}
\end{minipage}
\caption{Example of synthetic query generation (Stage~1). The rendered tile (left) is sent together with the prompt in Figure~\ref{fig:prompt-query-gen}; the structured five-line output is parsed into the JSON record shown (right).}
\label{fig:query-gen-example}
\end{figure}

Figure~\ref{fig:query-gen-example} shows a concrete example; the full prompt is in Figure~\ref{fig:prompt-query-gen} (Appendix~\ref{app:prompt_listings}).

\subsubsection{Self-contained-query filter prompt}
\label{app:selfcontained_prompt}

Referenced from \S\ref{sec:data_recipe}, Stage~1 (first false-positive filter).
After generation, we ask \texttt{gpt-4o} (OpenAI) at \texttt{temperature=0.0} to label each query as self-contained (YES) or not (NO), batching 50 queries per request.
On our canonical training set this filter drops $15.1\%$ of candidate pairs ($195{,}079 \to 165{,}537$).
The prompt is given in Figure~\ref{fig:prompt-selfcontained} (Appendix~\ref{app:prompt_listings}).

\subsubsection{Hard-negative consistency-judge prompt}
\label{app:consistency_prompt}

Referenced from \S\ref{sec:data_recipe}, Stage~2.
For each $(q, p)$ pair, we retrieve the top-$K$ candidates of $q$ from the base embedding model ($K{=}20$ in our pipeline), skip the positive, and walk the remaining candidates in rank order.
For each candidate, we run a two-stage filter that takes the candidate tile as the only visual context.
\textbf{Stage~A} (\emph{answer}, Figure~\ref{fig:prompt-hn-answer}) shows the candidate tile and the query to GPT-4o and asks for a short answer or the literal string \texttt{CANNOT\_ANSWER}.
\textbf{Stage~B} (\emph{judge}, Figure~\ref{fig:prompt-hn-judge}) shows the same tile, the query, and Stage~A's answer to GPT-4o, and classifies the candidate answer as exactly one of \texttt{CORRECT}, \texttt{WRONG}, or \texttt{CANNOT\_ANSWER}.
A \texttt{CORRECT} verdict means the candidate truly answers the query and is a false negative; we drop it.
\texttt{WRONG} or \texttt{CANNOT\_ANSWER} means the candidate is visually or topically related but does not actually answer the query, which is the hard-negative signal we want; we keep it.
We retain the first $M$ candidates that pass both stages; if the top-$K$ pool yields fewer than $M$ surviving hard negatives, the entire $(q, p)$ example is dropped from the training set.
Full prompts are in Figures~\ref{fig:prompt-hn-answer}--\ref{fig:prompt-hn-judge} (Appendix~\ref{app:prompt_listings}).

\subsubsection{Answerability filter (Stage~1, second false-positive filter)}
\label{app:answerable_prompt}

Referenced from \S\ref{sec:data_recipe}, Stage~1 (second false-positive filter).
The answerability filter reuses the same two prompts as the consistency judge above (Figures~\ref{fig:prompt-hn-answer} and~\ref{fig:prompt-hn-judge}); the only difference is that the visual context is the \emph{positive} tile $p$ rather than a candidate tile.
Concretely, we run Stage~A on $(q, p)$ to obtain a candidate answer $a^+$.
If $a^+$ equals the literal string \texttt{CANNOT\_ANSWER}, the verdict short-circuits to \texttt{CANNOT\_ANSWER} (Stage~B is not invoked); otherwise, we run Stage~B on $(q, p, a^+)$ to obtain a verdict in \{\texttt{CORRECT}, \texttt{WRONG}, \texttt{CANNOT\_ANSWER}\}.
We keep the $(q, p)$ pair iff the verdict is \texttt{CORRECT}, dropping it otherwise.

\subsubsection{Training implementation details}
\label{app:impl_details}

Referenced from \S\ref{sec:training_recipe}.
We implement contrastive fine-tuning with GradCache to decouple the effective batch size from GPU memory.
Training uses a batch size of 64 with a grad-cache chunk size of 4, 2 hard negatives per query, a peak learning rate of $7\!\times\!10^{-6}$ with 20 warmup steps and cosine decay.
The training set contains approximately 40K synthetic query--tile pairs after all filtering stages described above.
Training completes in approximately 3 hours on a single H100 GPU.

\subsection{Mini-Datastore for Embedding Ablations}
\label{app:mini_datastore}

Evaluating every embedding checkpoint against the full 30M-tile Wikipedia datastore is prohibitively expensive.
We instead construct a compact mini-datastore for rapid iteration.
We sample 400 queries from the evaluation set.
For each query, the mini-datastore contains all tiles from the gold article plus the top-100 tiles retrieved by the base embedding model, producing a per-query pool that mixes relevant and irrelevant candidates.
The resulting mini-datastore contains 400 questions and 7{,}426 tiles, approximating the difficulty of full-scale retrieval in a compact form suitable for rapid evaluation.
All accuracy numbers reported in the embedding training recipe ablation (Table~\ref{tab:ablation_recipe}) are evaluated on this mini-datastore.
We will publicly release this mini-datastore as a lightweight benchmark for evaluating visual retrieval over noisy webpage screenshots.

\subsection{Reader Fine-tuning: Method and Results}
\label{app:reader_sft}

\paragraph{Data.}
We reuse data from the embedding training pipeline (\S\ref{sec:data_recipe}): each Stage-1 $(q, p)$ pair already comes with a verified answer $a$ from the answerability filter (Appendix~\ref{app:query_gen_prompt}), giving us SFT triples $(q, p, a)$ at no additional annotation cost.
To simulate the retrieval setting, we use the embedding model from \S\ref{sec:training_recipe} to retrieve $k\!\in\!\{1,\dots,6\}$ tiles for $q$ over the datastore, and add $p$ to the set if it is not already present.
All tiles are downsampled via Lanczos resampling~\cite{lanczos_resampling} to a target low resolution, where a $c\times$ compression scales each side by $1/\sqrt{c}$ (e.g., $4\times$ halves both width and height).
By construction, this data format teaches the reader to (i)~read low-resolution images and (ii)~ignore distractor tiles.

\paragraph{SFT loss.}
We fine-tune the reader $\phi$ with standard token-level cross-entropy over answer tokens: $\mathcal{L}_{\mathrm{SFT}} = -\mathbb{E}_{(q,\mathcal{T},a)}\sum_{t} \log P_\phi(a_t \mid \mathcal{T}, q, a_{<t})$, where $\mathcal{T}$ is the retrieved tile set.

\paragraph{Evaluation.}
Table~\ref{tab:reader_sft_compression} reports LLM-judge accuracy (GPT-4.1) on a held-out 500-example test set, sweeping the compression factor $c$ and the number of retrieved tiles $k$.
At each $c$ we compare the no-SFT base reader applied directly to compressed tiles (\emph{compression-only}) against the SFT-trained reader, with the uncompressed base reader at $1\times$ as the ceiling reference.

\paragraph{Training observations.}
\emph{(i) Compression alone hurts.}
Applying $2\times$ compression to the base reader without retraining drops average accuracy from $0.905$ to $0.854$ ($-5.1$pp), and $3\times$ drops it to $0.738$ ($-16.7$pp).
\emph{(ii) SFT recovers the loss and meets or exceeds the uncompressed ceiling.}
SFT at $2\times$ reaches an average of $0.947$, $+9.3$pp above the compression-only baseline at the same $c$ and $+4.2$pp above the uncompressed ceiling itself; SFT at $3\times$ reaches $0.910$, $+17.2$pp above its compression-only baseline and matching the uncompressed ceiling on average ($+0.5$pp) while using one third of the pixel budget.
\emph{(iii) SFT improves robustness to distractor tiles as $k$ grows.}
The base reader at $1\times$ degrades from $0.958$ at $k\!=\!1$ to $0.856$ at $k\!=\!4$ ($-10.2$pp) as more distractors are added; SFT at $2\times$ stays within $1.4$pp of its $k\!=\!2$ peak across the full $k$ sweep, consistent with the ``ignore distractor / focus on gold tile'' training goal described above.

\begin{table}[htbp]
\centering
\caption{Reader SFT training evaluation across compression ratios (GPT-4.1 judge, 500-example held-out set).}
\label{tab:reader_sft_compression}
\vspace{4pt}
\footnotesize
\begin{tabular*}{\textwidth}{@{\extracolsep{\fill}}lccccc@{}}
\toprule
Setting & $k{=}1$ & $k{=}2$ & $k{=}3$ & $k{=}4$ & avg \\
\midrule
Base @ $1\times$ (no SFT, ceiling)             & 0.958 & 0.912 & 0.892 & 0.856 & 0.905 \\
\midrule
Base @ $2\times$ (compression only, no SFT)    & 0.908 & 0.862 & 0.852 & 0.794 & 0.854 \\
SFT  @ $2\times$                                & \textbf{0.946} & \textbf{0.950} & \textbf{0.954} & \textbf{0.936} & \textbf{0.947} \\
\midrule
Base @ $3\times$ (compression only, no SFT)    & 0.830 & 0.726 & 0.710 & 0.684 & 0.738 \\
SFT  @ $3\times$                                & 0.904 & 0.918 & 0.932 & 0.884 & 0.910 \\
\bottomrule
\end{tabular*}
\end{table}


\section{Evaluation Protocol}

\subsection{Benchmark Details}
\label{app:benchmark_details}

Table~\ref{tab:benchmark_details} summarizes the evaluation benchmarks used in this paper.

\begin{table}[htbp]
\centering
\small
\caption{Benchmark details. All Wikipedia benchmarks query the same 30M-page datastore (\S\ref{sec:data_collection}); LiveVQA queries a separate news datastore (Appendix~\ref{app:datastore_fetch}).}
\label{tab:benchmark_details}
\begin{tabular*}{\textwidth}{@{\extracolsep{\fill}}llcll@{}}
\toprule
Benchmark & Task family & $n$ & Metric & Source \\
\midrule
Natural Questions & Text-centric Wiki QA & 1{,}000 & GPT-4.1 judge & \cite{kwiatkowski2019natural} \\
NQ-Tables & Text-centric Wiki QA & 1{,}000 & GPT-4.1 judge & \cite{herzig2021nqtables} \\
SimpleQA & Text-centric Wiki QA & 1{,}000 & GPT-4.1 judge & \cite{wei2024simpleqa} \\
MMSearch & Multimodal Wiki QA & 300 & GPT-4.1 judge & \cite{jiang2024mmsearch} \\
Encyclopedic VQA & Multimodal Wiki QA & 749 & GPT-4.1 judge & \cite{mensink2023evqa} \\
LiveVQA-2025 & News VQA & 6{,}632\textsuperscript{$\star$} & Accuracy & \cite{fu2025livevqa} \\
\bottomrule
\end{tabular*}
\vspace{2pt}
{\footnotesize $\star$ CNN/BBC/AP~News subset; the full LiveVQA dataset has 26{,}888 QA pairs --- we exclude Forbes and Variety due to anti-bot protections that prevent reliable page capture.}
\end{table}

\paragraph{Encyclopedic VQA subset.}
We evaluate on the \emph{landmarks} subset (Google Landmarks v2~\cite{weyand2020gldv2}) with \emph{automatic} questions only ($n{=}749$).
The iNaturalist subset is excluded due to missing query images in the official dataset release; the automatic question type is used as it is the largest category and contains naturally phrased questions generated from Wikipedia sections.

\paragraph{MMSearch subset.}
We evaluate on all 300 end-to-end examples.
Of these, 171 include a query image and 129 are text-only.

\paragraph{LiveVQA setup.}
Each query consists of an editorial photo and a multiple-choice question; we jointly embed the photo and question text to retrieve from the news tile index (Appendix~\ref{app:datastore_fetch}).
We evaluate on CNN, BBC, and AP~News ($n{=}6{,}632$ QA pairs), excluding Forbes and Variety due to anti-bot protections that prevent reliable page capture.

\paragraph{MoNaCo setup.}
We evaluate on the full MoNaCo benchmark~\cite{wolfson2025monaco} (1{,}315 multi-hop Wikipedia QA questions) using a GPT-5 ReAct agent with a single \texttt{search} tool.
We use GPT-5 as the agent controller because it exhibits stronger and more stable agentic behavior than current open-source alternatives, allowing us to better isolate the impact of the retrieval backend.
Only the search backend varies across conditions; the agent loop, prompt, and reader (GPT-5) are identical.
We compare four backends: \sys{} (pixel retrieval), Trafilatura text retrieval, Google via SerpApi~\cite{serpapi2025}, and DS-Serve~\cite{liu2026ds} (open-source neural retrieval endpoint).
The agent issues up to 20 search queries per question with top-$k{=}5$ results per query.

\subsection{Grading Protocol}
\label{app:grading_protocol}

All Wikipedia QA benchmarks (NQ, NQ-Tables, SimpleQA, MMSearch, Encyclopedic VQA) use LLM-as-judge grading with GPT-4.1~\cite{openai2025gpt41} at temperature~0, seed~42, and \texttt{max\_tokens=1000}.
Following~\cite{wei2024simpleqa}, the grader classifies each prediction as CORRECT, INCORRECT, or NOT\_ATTEMPTED; we score CORRECT as 1.0 and the rest as 0.0.
For NQ and NQ-Tables, up to 10 gold answer aliases are joined with ``OR'' so any match counts as correct.
LiveVQA is a 5-option multiple-choice task graded by exact letter match (no LLM grader).
Reader prompt templates are listed in Appendix~\ref{app:prompt_listings}.


\section{Additional Results}

\subsection{Results with Alternative Readers}
\label{app:alt_reader_results}

The main text reports results with the Qwen3.5-4B reader.
Tables~\ref{tab:main_results_vl4b}--\ref{tab:main_results_top1} repeat the main results at top-$k{=}3$ and top-$k{=}1$ with the Qwen3-VL-4B reader.
All directional findings are identical across reader variants.
The tuned retriever (LoRA) consistently improves over the base checkpoint: on SimpleQA with the VL-4B reader, accuracy rises from 70.3\% to 75.1\% at top-$3$ and from 58.3\% to 63.7\% at top-$1$.

\begin{table}[htbp]
\centering
\caption{End-to-end results with Qwen3-VL-4B reader (cf.\ Table~\ref{tab:main_results}).}
\label{tab:main_results_vl4b}
\vspace{4pt}
\footnotesize
\begin{tabular*}{\textwidth}{@{\extracolsep{\fill}} l r@{\hskip 6pt}r r@{\hskip 6pt}r r@{\hskip 6pt}r r r@{\hskip 6pt}r !{\vrule} r@{\hskip 6pt}r @{}}
\toprule
 & \multicolumn{2}{c}{NQ} & \multicolumn{2}{c}{NQ-Tables} & \multicolumn{2}{c}{SimpleQA} & MMSearch & \multicolumn{2}{c}{EVQA} & \multicolumn{2}{c}{LiveVQA}  \\
\cmidrule(lr){2-3} \cmidrule(lr){4-5} \cmidrule(lr){6-7} \cmidrule(lr){8-8} \cmidrule(lr){9-10}  \cmidrule(lr){11-12}
Method & Recall & Acc & Recall & Acc & Recall & Acc & Acc & Recall & Acc  & Recall & Acc \\
\midrule
No retrieval             && 13.1 && 10.9 && 6.4 & 13.5 && 27.3 && 54.2 \\
\midrule
\multicolumn{12}{@{}l}{\quad\emph{Text-based retrieval}} \\
mwparserfromhell        & 48.6 & 27.7 & 35.0 & 18.3 & 72.9 & 58.4 & 12.9 & 3.2 & 30.7  \\
Trafilatura  & 45.8 & 29.4 & 37.7 & 21.9 & 79.3 & 69.0 & 13.5 & 3.2 & 29.2 &16.2 &55.9  \\
\midrule
\multicolumn{12}{@{}l}{\quad\emph{Pixel-based retrieval}} \\
\sys{} (base) & 53.4 & 31.0 & 44.4 & 24.3 & 79.9 & 70.3 & 19.9 &17.0 & 39.4 & \textbf{38.9} & \textbf{66.1}   \\
\sys{} & \textbf{58.9} & \textbf{31.1} & \textbf{50.7} & \textbf{25.8} & \textbf{83.1} & \textbf{75.1} & \textbf{21.6} & \textbf{20.2} & \textbf{40.8} & 33.3 & 65.3\\
\bottomrule
\end{tabular*}
\end{table}

\begin{table}[htbp]
\centering
\caption{End-to-end results at top-$k{=}1$ with Qwen3-VL-4B reader (cf.\ Table~\ref{tab:main_results}).}
\label{tab:main_results_top1}
\footnotesize
\begin{tabular*}{\textwidth}{@{\extracolsep{\fill}} l r@{\hskip 6pt}r r@{\hskip 6pt}r r@{\hskip 6pt}r r r@{\hskip 6pt}r !{\vrule} r@{\hskip 6pt}r @{}}
\toprule
 & \multicolumn{2}{c}{NQ} & \multicolumn{2}{c}{NQ-Tables} & \multicolumn{2}{c}{SimpleQA} & MMSearch & \multicolumn{2}{c}{EVQA} & \multicolumn{2}{c}{LiveVQA}  \\
\cmidrule(lr){2-3} \cmidrule(lr){4-5} \cmidrule(lr){6-7} \cmidrule(lr){8-8} \cmidrule(lr){9-10}  \cmidrule(lr){11-12}
 & Recall & Acc & Recall & Acc & Recall & Acc & Acc & Recall & Acc  & Recall & Acc \\
\midrule
No retrieval             && 13.1 && 10.9 && 6.4 & 12.7 && 27.3 && 54.2 \\
\midrule
\multicolumn{12}{@{}l}{\quad\emph{Text-based retrieval}} \\
mwparserfromhell        & 28.1 & 23.1 & 17.8 & 14.9 & 58.8 & 48.6 & 18.1 & 1.77 & 29.2  \\
Trafilatura  & 26.3 & 23.6 & 19.3 & 15.2 & 62.6 & 55.6 & 19.0 & 1.7 & 28.0 & 9.8 & 53.0  \\
\midrule
\multicolumn{12}{@{}l}{\quad\emph{Pixel-based retrieval}} \\
\sys{} (base)   & 36.3 & 25.7 & 28.2 & 20.0 & 65.7 & 58.3 & 20.3 & 9.8 & 37.0 & \textbf{26.8} & \textbf{61.2}   \\
\sys{} & \textbf{40.3} & \textbf{27.1} & \textbf{33.7} & \textbf{21.7} & \textbf{71.2} & \textbf{63.7} & \textbf{22.7} & \textbf{12.9} & \textbf{39.4} & 22.4 & 60.8 \\
\bottomrule
\end{tabular*}
\end{table}

\begin{table}[htbp]
\centering
\caption{Retrieval--reader modality ablation with Qwen3-VL-4B reader (cf.\ Table~\ref{tab:ablation_modality}).}
\label{tab:ablation_modality_vl4b}
\vspace{4pt}
\footnotesize
\begin{tabular*}{\textwidth}{@{\extracolsep{\fill}}llrr@{}}
\toprule
Retrieval & Reader & SimpleQA & LiveVQA \\
\midrule
Screenshot & Screenshot        & \textbf{70.3} & \textbf{66.1} \\
\midrule
Screenshot & OCR text          & 69.8          & 63.9          \\
Text       & Rendered img      & 63.1          & 56.8          \\
Text       & Text              & 69.0          & 55.9          \\
\bottomrule
\end{tabular*}
\end{table}

\subsection{Full Results of Scaling VLM Performance}
\label{app:reader_scale_full}

\begin{figure}[t]
\centering
\begin{subfigure}[t]{0.54\textwidth}
    \centering
    \includegraphics[width=\linewidth]{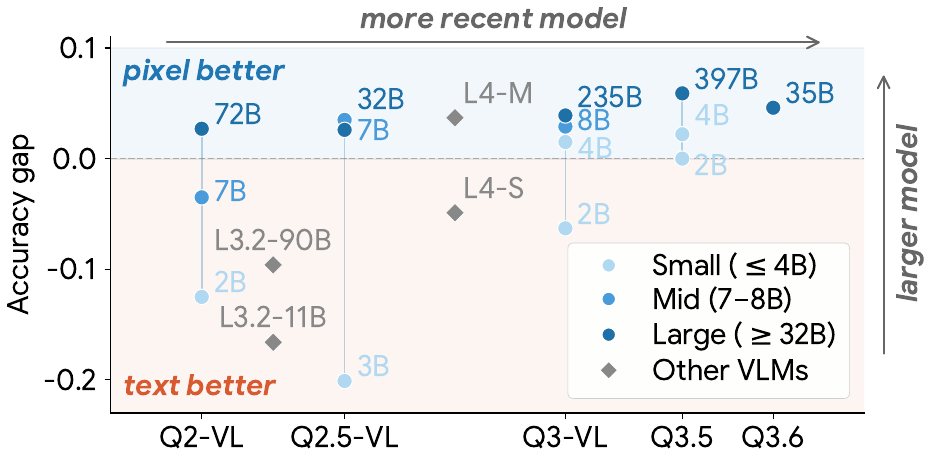}
    \caption{Accuracy gain over text retrieval}
    \label{fig:reader_scale_a}
\end{subfigure}
\hfill
\begin{subfigure}[t]{0.44\textwidth}
    \centering
    \includegraphics[width=\linewidth]{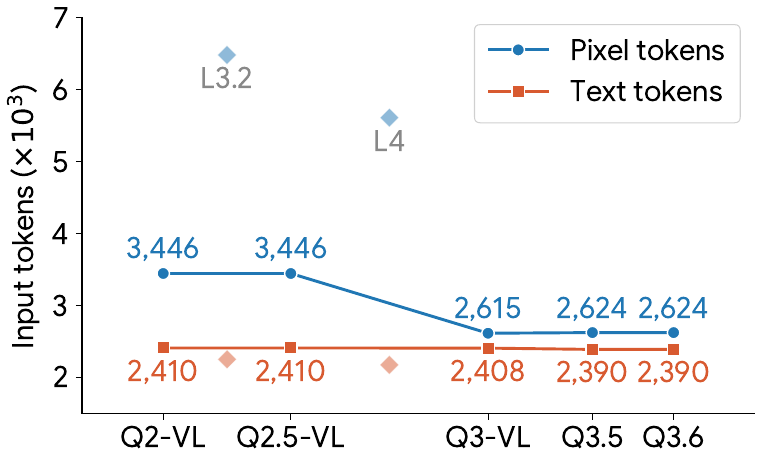}
    \caption{Input token cost}
    \label{fig:reader_scale_b}
\end{subfigure}
\caption{VLM reading ability across model generations on SimpleQA. Q = Qwen; L3.2 = Llama-3.2-Vision; L4 = Llama-4. Markers denote parameter scale.}
\vspace{-2ex}
\label{fig:reader_scale}
\vspace{-8pt}
\end{figure}

Figure~\ref{fig:reader_scale} visualizes the accuracy and token-cost trends from the full 31-model sweep in Table~\ref{tab:ablation_reader_scale_full} (main text).
Reasoning-mode models (bottom rows of the table) show a similar pattern but with higher output token cost due to chain-of-thought generation; the pixel advantage persists (+4.8\,pp for Qwen3.5-4B reasoning, +4.6\,pp for Qwen3.6-35B-A3B reasoning).

\section{Analysis}

\subsection{Wikipedia Text Extractor Comparison}
\label{app:parser_comparison}

To validate our choice of HTML-to-text extractor, we compare seven mainstream Wikipedia text parsers using GPT-4.1 as a judge on a six-dimension rubric covering QA-usability, completeness, cleanliness, tabular data, hierarchy, and overall quality, evaluated over 50 diverse Wikipedia pages (Table~\ref{tab:parser_comparison}).
\texttt{trafilatura} with \texttt{output\_format='markdown'} produces the highest-quality output overall (6.40/10, winning 36 out of 50 pages), followed closely by \texttt{resiliparse} (6.34/10, 31 wins).
Our pipeline uses \texttt{trafilatura} as the text-side parser for all text-retrieval baselines, \textbf{ensuring that the text backend we compare against is the strongest available}.

Critically, \emph{even the best parser scores only 4.38/10 on tabular data}, and no parser exceeds 4.38 on this dimension.
This ceiling is structural: linearization cannot preserve multi-row, multi-column alignment, merged cells, or the spatial grouping that makes table content interpretable.
This observation directly motivates our pixel-space approach: for table-bearing content, bypassing the parser entirely and operating on rendered screenshots avoids the information loss that every text extractor introduces.

\begin{table}[htbp]
\centering
\small
\caption{Wikipedia text extractor comparison. GPT-4.1 judges seven parsers on six dimensions (scale 1--10) over 50 diverse pages; ``Wins'' counts pages where the parser ranked first overall. Bold marks column-best values. Even the top parser scores below 4.4/10 on tabular data, motivating pixel-space retrieval for table-bearing content.}
\label{tab:parser_comparison}
\begin{tabular*}{\textwidth}{@{\extracolsep{\fill}}clccccccc@{}}
\toprule
Rank & Parser & Overall & QA-usab. & Compl. & Clean. & Tables & Hier. & Wins \\
\midrule
1 & \texttt{trafilatura} (md) & \textbf{6.40} & 6.56 & 7.46 & 6.24 & \textbf{4.38} & \textbf{8.04} & \textbf{36} \\
2 & \texttt{resiliparse}      & 6.34 & \textbf{6.64} & \textbf{7.60} & 6.02 & 4.02 & 7.16 & 31 \\
3 & BeautifulSoup              & 6.06 & 6.26 & 7.20 & 6.88 & 3.86 & 5.88 & 29 \\
4 & \texttt{mwparserfromhell} & 5.56 & 5.60 & 6.52 & 6.24 & 3.16 & 6.60 & 15 \\
5 & \texttt{wikiextractor}    & 5.20 & 5.26 & 6.10 & \textbf{7.54} & 2.82 & 5.26 & 8 \\
6 & raw wikitext              & 4.60 & 4.92 & 6.12 & 3.62 & 2.76 & 5.86 & 3 \\
7 & \texttt{jina\_reader}     & 2.30 & 1.78 & 1.84 & 2.22 & 1.80 & 4.90 & 2 \\
\bottomrule
\end{tabular*}
\end{table}

\subsection{Visual Information Loss During HTML Parsing}
\label{app:visual_loss}

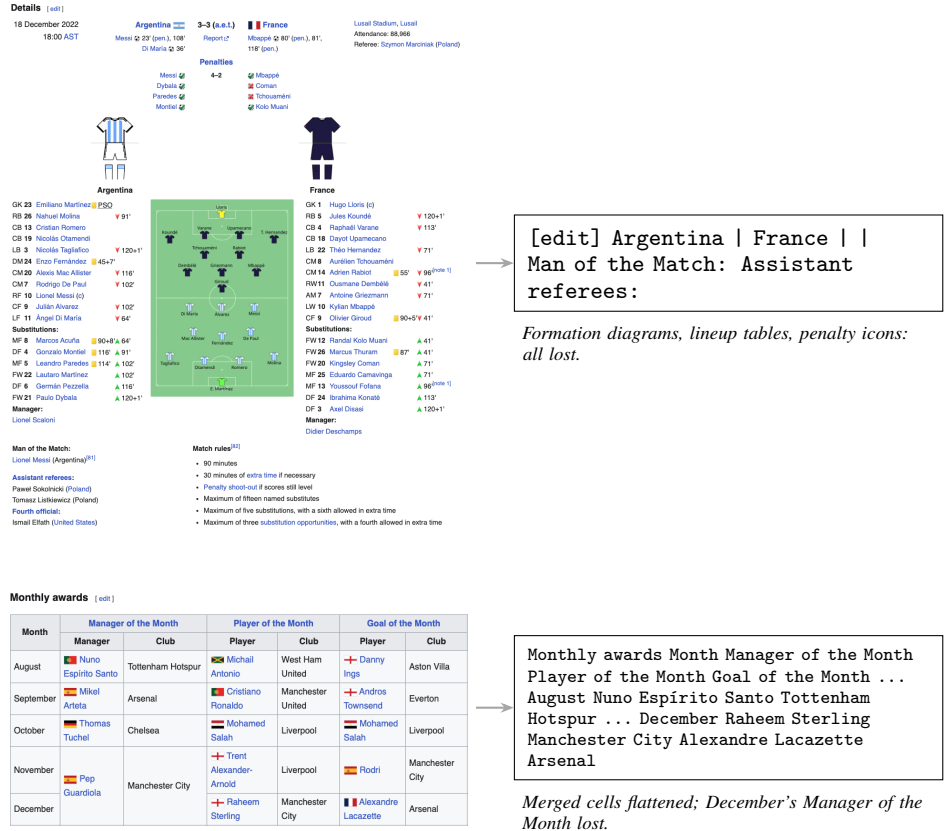
\begin{figure}[htbp]
\centering
\begin{tikzpicture}[
    imgnode/.style={inner sep=0pt},
    txtnode/.style={draw, text width=0.38\textwidth, font=\small\ttfamily, inner sep=5pt, align=left},
    arrowlabel/.style={font=\large},
    annot/.style={font=\scriptsize\itshape, text width=0.38\textwidth, align=left},
  ]
  \node[imgnode] (img1) {\includegraphics[width=0.44\textwidth]{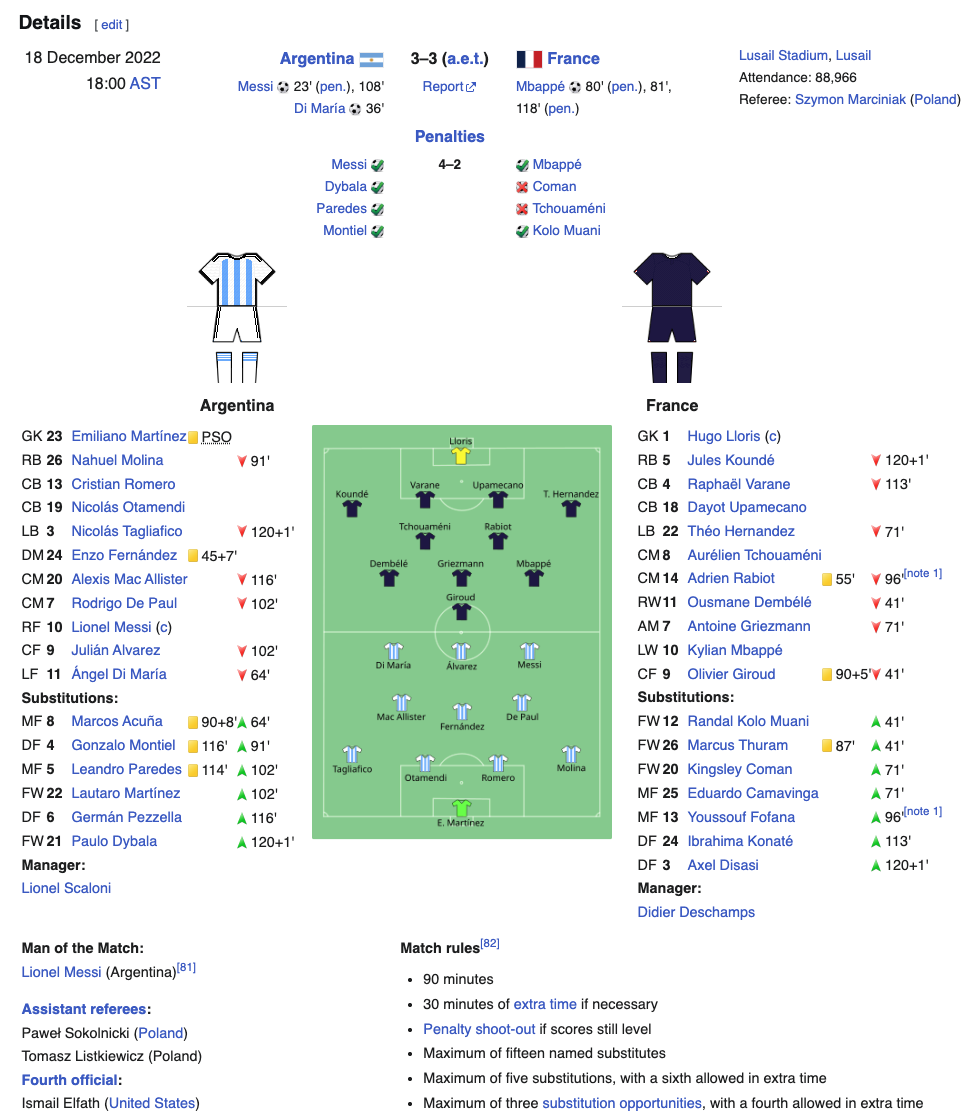}};
  \node[txtnode, right=0.6cm of img1.east, anchor=west, yshift=0cm]
    (txt1) {[edit] Argentina | France | | Man of the Match: Assistant referees:};
  \node[annot, below=2pt of txt1.south west, anchor=north west]
    (ann1) {Formation diagrams, lineup tables, penalty icons: all lost.};
  \draw[-{Stealth[length=5pt]}, thick, gray!70] (img1.east) ++(0.1,0) -- (txt1.west);

  \node[imgnode, below=0.8cm of img1.south, anchor=north] (img2)
    {\includegraphics[width=0.44\textwidth]{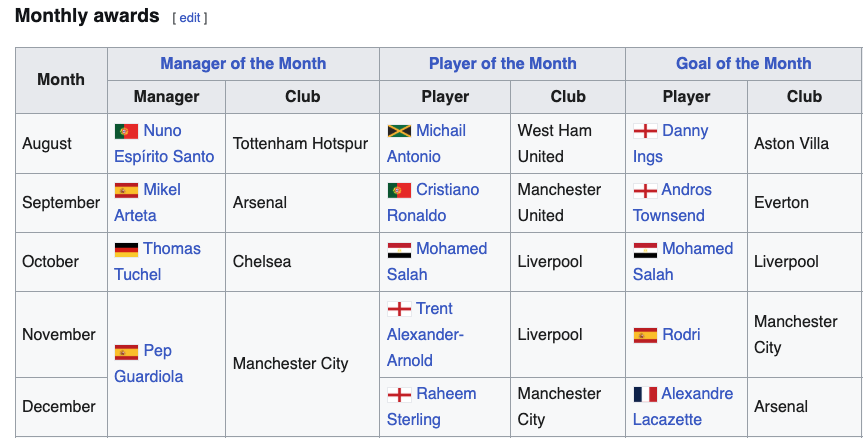}};
  \node[txtnode, right=0.6cm of img2.east, anchor=west, font=\scriptsize\ttfamily]
    (txt2) {Monthly awards Month Manager of the Month Player of the Month Goal of the Month \ldots\ August Nuno Esp\'irito Santo Tottenham Hotspur \ldots\ December Raheem Sterling Manchester City Alexandre Lacazette Arsenal};
  \node[annot, below=2pt of txt2.south west, anchor=north west]
    (ann2) {Merged cells flattened; December's Manager of the Month lost.};
  \draw[-{Stealth[length=5pt]}, thick, gray!70] (img2.east) ++(0.1,0) -- (txt2.west);
\end{tikzpicture}
\caption{Visual information loss during HTML parsing. Each row shows a rendered Wikipedia page (left) and the text extracted by \texttt{trafilatura} (right). Top: a match page with rich visual structure is reduced to a near-empty string. Bottom: a table with merged cells is linearized, making certain cell associations unrecoverable.}
\label{fig:visual_loss}
\end{figure}

Figure~\ref{fig:visual_loss} shows two concrete examples of visual information loss when HTML is parsed into plain text.
In the first example, a match page containing formation diagrams, penalty shoot-out icons, and lineup tables is reduced to a near-empty string after parsing.
In the second, a multi-row awards table with merged cells is flattened into a wall of text; for instance, December's ``Manager of the Month'' is unrecoverable from the linearized output because the merged cell spans two rows.
An additional end-to-end example from our evaluation appears in Figure~\ref{fig:fm1_inter} (\S\ref{app:failure_detailed}): the 2010 Champions League Final article's match-statistics table is destroyed by linearization, so no text chunk in the corpus contains the answer, whereas the pixel retriever surfaces the rendered table directly.

\subsection{Retrieval Signal Loss Under Text Linearization}
\label{app:retrieval_signal_loss}

\begin{figure}[htbp]
\centering
\begin{minipage}[t]{0.47\textwidth}
\centering
\textbf{Text ranking} {\footnotesize(1024-token chunks)}\\[4pt]
\fbox{\parbox{0.93\textwidth}{\raggedright\footnotesize
\textbf{rank 1, score 0.715}\\
\texttt{Indigenous peoples in Ecuador; Paraguay; Costa Rica; Argentina; Peru; Canada; Colombia; Brazil; \ldots}
}}

\vspace{4pt}
\fbox{\parbox{0.93\textwidth}{\raggedright\footnotesize
\textbf{rank 2, score 0.647}\\
\texttt{Portal: Indigenous peoples of the Americas / box-footer}
}}

\vspace{4pt}
\fbox{\parbox{0.93\textwidth}{\raggedright\footnotesize
\textbf{rank 3, score 0.639}\\
\texttt{| Honduras | 49\% | 19\% | 10\% | \ldots\ | Mexico | 58\% | 9\% | 21\% | \ldots}
}}

\vspace{4pt}
\centering $\vdots$
\vspace{4pt}

\fbox{\parbox{0.93\textwidth}{\raggedright\footnotesize
\textbf{rank 38, score 0.580}\\
\texttt{\ldots\ INEGI, Mexico \ldots\ United States Census Bureau \ldots}
}}
\end{minipage}
\hfill
\begin{minipage}[t]{0.47\textwidth}
\centering
\textbf{Pixel ranking} {\footnotesize($875\times1024$ tile, 880 tokens)}\\[4pt]
\fbox{\parbox{0.93\textwidth}{\centering\footnotesize
\textbf{rank 2, score 0.616}\\[2pt]
\begin{tikzpicture}
\node[inner sep=0pt] (img) {\includegraphics[width=0.9\textwidth]{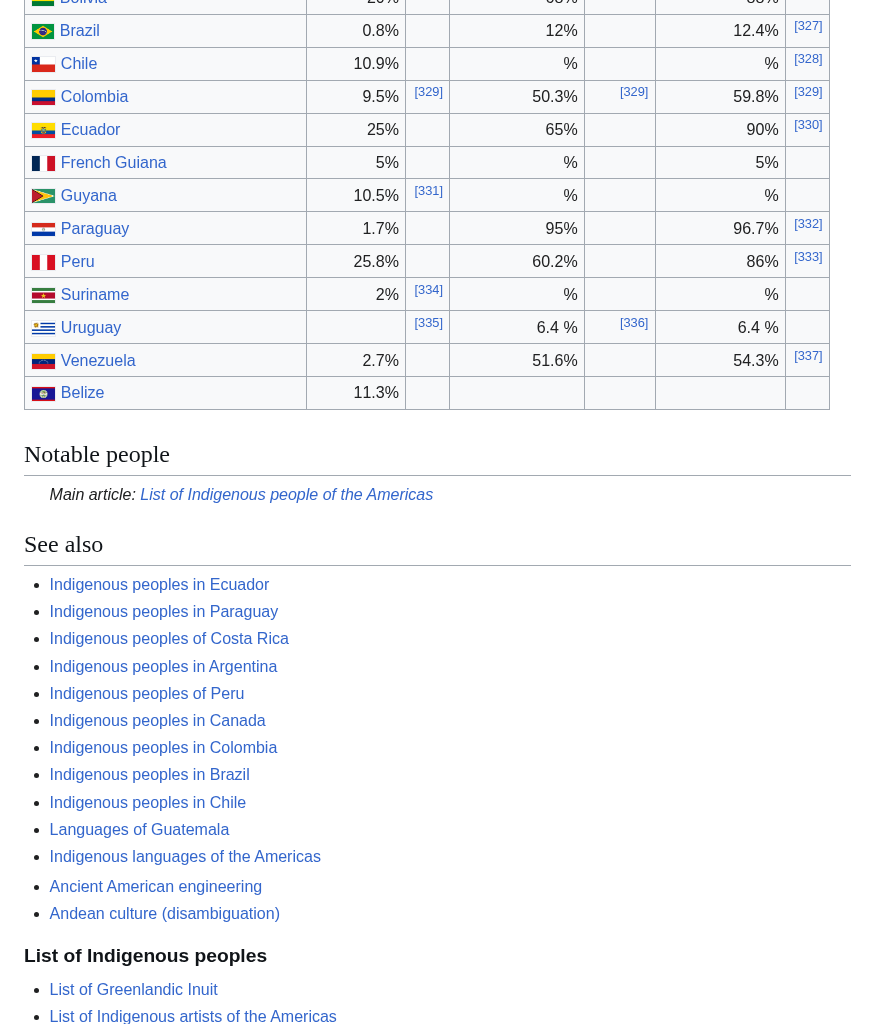}};
\draw[red, line width=1.2pt]
  ($(img.north west)+(0.005\textwidth,-0.005\textwidth)$)
  rectangle
  ($(img.north east)+(-0.005\textwidth,-0.42\textwidth)$);
\end{tikzpicture}
}}
\end{minipage}
\caption{Retrieval signal loss under text linearization. The left panel shows ranked text chunks for the table-formulated query, with ``\ldots'' marking omitted text. Across 12 text queries, the answer-bearing percentage chunks never entered the top-5, whereas the pixel path retrieved the rendered table tile containing the answer evidence.}
\label{fig:retrieval_signal_loss}
\end{figure}

Figure~\ref{fig:retrieval_signal_loss} shows a representative MoNaCo retrieval trace for the question: ``What is the current percentage of Indigenous peoples in each country in America?''
The text retriever repeatedly surfaced passages from the correct topic area (an article lead, see-also links, and portal/navigation fragments) but none contained the country-by-country percentage table.
The pixel retriever instead surfaced a mid-page table tile where the country rows and percentage values are visually grouped.
This is not a case where extraction deleted the answer; rather, linearization made the answer-bearing table a weaker retrieval target than topic-matching prose and navigation text.
In the logged run, no answer-bearing percentage chunk entered the text top-5 across the agent's 12 queries.
For the closest table-formulated text query, the corresponding same-article table/reference chunk appeared only at rank 38 (score 0.580), while the pixel query retrieved the rendered table tile at rank 2 (score 0.616).
Two additional evaluation examples illustrate the same rank-loss pattern on SimpleQA: in Figure~\ref{fig:fm2_dali}, text retrieval ranks the Dal\'i infobox above the answer-bearing body paragraph (rank 12 vs.\ rank 3 for \sys{}); in Figure~\ref{fig:fm2_shepard}, the gap is even more extreme, with the answer paragraph falling to rank 66 under text retrieval while \sys{} recovers it at rank 3.

\subsection{Detailed Failure Decomposition}
\label{app:failure_detailed}

\paragraph{Evidence verification.}
To distinguish genuine retrieval failures from borderline misses, we deliberately use a strong verifier (GPT-5.1) over a generous retrieval scope (top-100): if no answer-bearing item appears anywhere in the top-100, the miss is structural rather than a ranking threshold artifact.
Concretely, for each SimpleQA question we retrieve the top-100 text chunks and screenshot tiles from their respective indexes and ask GPT-5.1 whether each item contains enough information to answer the question (screenshot items use multimodal input; text items use text-only input).
An item is \emph{valid evidence} if GPT-5.1 answers correctly, verified by GPT-4.1 grading against the ground-truth answer.
This two-stage verify-then-grade protocol mirrors the consistency-judge pipeline used for hard-negative mining during training (Appendix~\ref{app:consistency_prompt}): a capable model attempts to answer from the candidate, and a separate grader confirms correctness.

\paragraph{Evidence type classification.}
Each valid evidence \emph{tile} is classified by the DOM structure of the chunk it occupies.
We re-render the gold article in Playwright at the same 875px viewport used for tile capture, then for each 1024px vertical strip, query all DOM elements via \texttt{getBoundingClientRect} and assign the tile's type as the element with the largest pixel coverage: \texttt{<table class="infobox">} $\to$ infobox, \texttt{<table>} $\to$ table, \texttt{<ul>/<ol>} $\to$ list, \texttt{<p>} $\to$ paragraph.
A question can have valid evidence tiles of multiple types; rather than projecting it onto a single bucket, we report each per-type analysis on the conditional sub-population of questions for which the corresponding type-$\tau$ valid evidence exists.
The sub-populations therefore overlap, and per-type sample sizes do not sum to $N$.

\paragraph{Failure decomposition.}
Table~\ref{tab:failure_decomposition} pairs each question's text and screenshot outcomes.
A question exhibits \emph{parser loss} when \sys{} retrieval surfaces valid evidence in top-3 but no valid text chunk exists anywhere in the retrieved top-100, meaning the text corpus cannot answer the question.
\emph{Rank loss} applies when valid text evidence exists but falls outside the reader's top-3 window.
The rank-loss columns compare the two modalities on the rank-loss subset only: text fails by definition (rank $>$ 3), while the \sys{} column shows where it succeeds on the same questions (rank $\leq$ 3 by definition).

\begin{table}[htbp]
\centering
\caption{Paired failure decomposition on SimpleQA by evidence type, restricted to questions where \sys{} succeeds but text fails. Parser and Rank give the counts of parser-loss and rank-loss; the rank-loss columns give the mean evidence rank per modality on the rank-loss subset. Rows overlap (column sums $>$ Overall).}
\label{tab:failure_decomposition}
\footnotesize
\begin{tabular*}{\textwidth}{@{\extracolsep{\fill}}lc cc cc cc@{}}
\toprule
& & \multicolumn{2}{c}{Text fail$^\dagger$} & \multicolumn{2}{c}{Rank-loss rank} & & \\
\cmidrule(lr){3-4} \cmidrule(lr){5-6}
Type & $n$ & Parser & Rank & \sys{} & Trafilatura & \sys{} fail & Both fail \\
\midrule
Table     & 282 & 21 & 33 & 1.2 & 17.7 & 17 & 36 \\
List      & 290 & 16 & 29 & 1.6 & 16.3 & 22 & 25 \\
Infobox   & 503 & 29 & 41 & 1.4 & 16.6 & 24 & 36 \\
Paragraph & 571 & 34 & 73 & 1.6 & \textbf{22.5} & 30 & 58 \\
\midrule
Overall   & 946 & 67 & 91 & 1.5 & 20.1 & 55 & 99 \\
\bottomrule
\end{tabular*}

\vspace{2pt}
{\scriptsize $^\dagger$\sys{} succeeds (rank $\leq 3$) but text does not.}
\end{table}

Table~\ref{tab:failure_examples} shows representative SimpleQA examples for each failure mode.

\begin{table}[htbp]
\centering
\caption{Representative SimpleQA examples for each failure mode.}
\label{tab:failure_examples}
\footnotesize
\begin{tabular*}{\textwidth}{@{\extracolsep{\fill}}p{0.08\textwidth}p{0.07\textwidth}p{0.41\textwidth}p{0.14\textwidth}cc@{}}
\toprule
Mode & Ev.\ type & Question (abbreviated) & Answer & \sys{} & Trafilatura \\
\midrule
Parser & table & How many shots did Inter attempt on target in the CL Final \ldots? & 7 & 1 & {$>$100} \\
Parser & para. & What disease was Elizabeth Esteve-Coll diagnosed with \ldots? & multiple sclerosis & 1 & {$>$100} \\
Parser & list & Which award did Reza Aslan receive in 2014? & Intersections Int'l Award & 2 & {$>$100} \\
\midrule
Rank & table & On what day was Javier Zanetti's first daughter born? & 11 June 2005 & 2 & 89 \\
Rank & para. & What was the rate of climb of the Grumman F4F-3 Wildcat \ldots? & 11.70 m/s & 2 & 91 \\
Rank & list & Who sketched the Taddei Tondo following its arrival at the RA \ldots? & John Constable & 1 & 62 \\
Rank & infobox & What team finished with 38 pts in the 2021--22 PL season? & Leeds United & 1 & 68 \\
\bottomrule
\end{tabular*}
\end{table}

\paragraph{Reader loss.}
Reader loss is computed by cross-referencing the evidence annotations with actual reader outputs: questions where valid evidence appears in the reader's top-3 but the reader still answers incorrectly.
With the Qwen3.5-4B reader, reader loss is 7.0\% (56 of 799 questions with valid \sys{} evidence in top-3).

\begin{figure}[htbp]
\centering
\small
\fbox{\parbox{\dimexpr\textwidth-2\fboxsep-2\fboxrule}{\small
\emph{Q:} ``Who received an honorable mention at the 1996 Frank and Brennie Morgan Prize \ldots?'' \hfill \emph{A:} \textbf{Lenhard Ng}
}}

\vspace{4pt}
\begin{minipage}[t]{0.50\textwidth}
\textbf{Trafilatura text top-3}\\[2pt]
\fbox{\parbox{\dimexpr\textwidth-2\fboxsep-2\fboxrule}{\raggedright\scriptsize
\textbf{rank 1} --- \textit{Morgan Prize} (chunk 0)\\[2pt]
\texttt{\ldots Previous winners}\\
\texttt{- 1995}\\
\texttt{- Winner: Kannan Soundararajan \ldots}\\
\texttt{- Honorable mention: \textbf{Kiran Kedlaya} (Harvard)}\\
\texttt{- 1996}\\
\texttt{- Winner: Manjul Bhargava \ldots}\\
\texttt{- Honorable mention: \textbf{Lenhard Ng} (Harvard)}\\
\texttt{- 1997}\\
\texttt{- Winner: Jade Vinson \ldots}\\
\texttt{\phantom{x}\scriptsize [\ldots 20 more year--name entries \ldots]}
}}\\
{\scriptsize\color{brown}\itshape $\hookrightarrow$ Correct answer present but buried in a flat list spanning 25 years; year--role--name hierarchy flattened to uniform dashes.}\\[3pt]
\centerline{$\vdots$}\\[3pt]
\end{minipage}%
\hfill
\begin{minipage}[t]{0.48\textwidth}
\textbf{\sys{} screenshot top-3}\\[2pt]
\fbox{\parbox{\dimexpr\textwidth-2\fboxsep-2\fboxrule}{\centering\scriptsize
\textbf{rank 1, score 0.629} --- \textit{Morgan Prize} (tile 0)\\[2pt]
\includegraphics[width=0.93\textwidth]{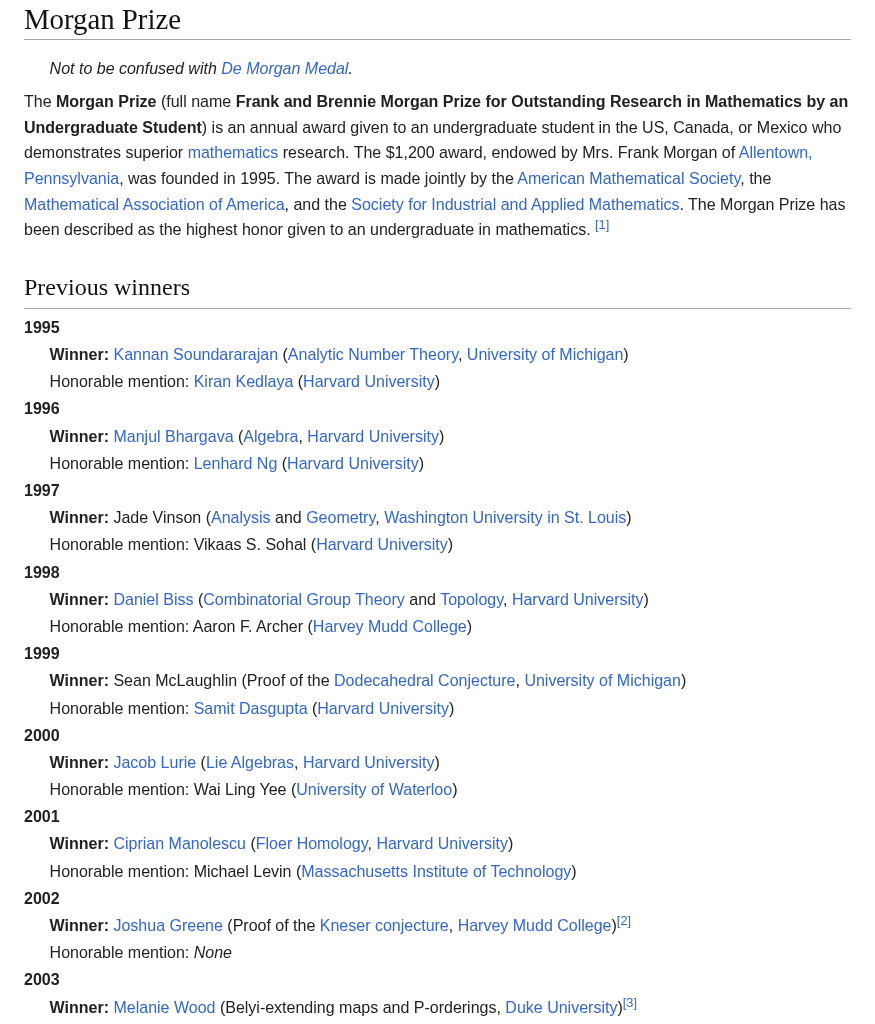}
}}\\
{\scriptsize\color{brown}\itshape $\hookrightarrow$ Visual grouping preserves year--role--name hierarchy; 1996 entry is unambiguous.}\\[2pt]
\centerline{$\vdots$}
\end{minipage}

\vspace{4pt}
\noindent
\begin{minipage}[t]{0.50\textwidth}\vspace{0pt}
\fbox{\parbox{\dimexpr\textwidth-2\fboxsep-2\fboxrule}{\scriptsize
\textbf{Text reader:} ``\textbf{Kiran Kedlaya} (Harvard University) received an honorable mention at the 1996 \ldots Morgan Prize.'' \quad{\color{red}$\times$}
}}\\
{\scriptsize\color{brown}\itshape $\hookrightarrow$ Picks the 1995 honorable mention instead of 1996 --- adjacent-entry confusion.}
\end{minipage}%
\hfill
\begin{minipage}[t]{0.48\textwidth}\vspace{0pt}
\fbox{\parbox{\dimexpr\textwidth-2\fboxsep-2\fboxrule}{\scriptsize
\textbf{Pixel reader:} ``\textbf{Lenhard Ng} (Harvard University)'' \quad{\color{green!50!black}$\checkmark$}
}}
\end{minipage}
\caption{\textbf{Reader loss.} Both modalities retrieve the same gold article at rank 1, and the answer appears verbatim in the text chunk. However, the linearized list flattens the year--role--name hierarchy into uniform dash-prefixed lines, and the text reader attributes the 1995 honorable mention to 1996. The rendered tile preserves the visual grouping by year, allowing the VLM to locate the correct entry.}
\label{fig:fm3_morgan_prize}
\end{figure}

Figures~\ref{fig:fm1_inter}--\ref{fig:fm3_morgan_prize} show fully expanded case studies (one for parser loss, two for rank loss, one for reader loss), including the actual top-3 retrieved chunks and tiles for each modality and the resulting reader outputs.


\newcommand{\dashedbox}[1]{%
\begin{tikzpicture}
\node[draw, dashed, inner sep=\fboxsep, line width=\fboxrule] {#1};
\end{tikzpicture}}

\begin{figure}[htbp]
\centering
\small
\fbox{\parbox{\dimexpr\textwidth-2\fboxsep-2\fboxrule}{\small
\emph{Q:} ``How many shots did Inter attempt on target in the Champions League Final \ldots May 23, 2010?'' \hfill \emph{A:} \textbf{7}
}}

\vspace{4pt}
\begin{minipage}[t]{0.50\textwidth}
\textbf{Trafilatura text top-3}\\[2pt]
\fbox{\parbox{\dimexpr\textwidth-2\fboxsep-2\fboxrule}{\raggedright\scriptsize
\textbf{rank 1, score 0.694} --- \textit{2010 UEFA Champions League final} (chunk 0)\\
\texttt{2010 UEFA Champions League final | Date | 22 May 2010 | Venue | Santiago Bernab\'eu, Madrid | \ldots}
}}\\
{\scriptsize\color{brown}\itshape $\hookrightarrow$ Infobox metadata; no shots-on-target stats.}\\[2pt]
\fbox{\parbox{\dimexpr\textwidth-2\fboxsep-2\fboxrule}{\raggedright\scriptsize
\textbf{rank 2, score 0.670} --- \textit{2009--10 UEFA Champions League} (chunk 0)\\
\texttt{2009--10 UEFA Champions League | Dates | 30 June \ldots{} 22 May 2010 | Champion | Inter Milan | \ldots}
}}\\
{\scriptsize\color{brown}\itshape $\hookrightarrow$ Season overview infobox, not match stats.}\\[2pt]
\fbox{\parbox{\dimexpr\textwidth-2\fboxsep-2\fboxrule}{\raggedright\scriptsize
\textbf{rank 3, score 0.667} --- \textit{Wesley Sneijder} (chunk 13)\\
\texttt{"Bayern Munich 0--2 Internazionale". ESPN Soccernet. 22 May 2010. Archived from \ldots}
}}\\
{\scriptsize\color{brown}\itshape $\hookrightarrow$ Bibliography, not data.}\\[3pt]
\centerline{$\vdots$}\\[3pt]
\dashedbox{\parbox{\dimexpr\textwidth-2\fboxsep-2\fboxrule}{\raggedright\scriptsize
\textbf{rank 26, score 0.614} --- \textit{2010 UEFA Champions League final} (chunk 4)\\
\texttt{Statistics | | | | Post-match As a result of Inter's victory, Italy held onto its position \ldots}
}}\\
{\scriptsize\color{brown}\itshape $\hookrightarrow$ Statistics table destroyed by linearization; header survived, table body reduced to empty pipes.}\\[2pt]
\end{minipage}%
\hfill
\begin{minipage}[t]{0.48\textwidth}
\textbf{\sys{} screenshot top-3}\\[2pt]
\fbox{\parbox{\dimexpr\textwidth-2\fboxsep-2\fboxrule}{\centering\scriptsize
\textbf{rank 1, score 0.601} --- \textit{2010 UEFA CL final} (tile 7)\\[2pt]
\includegraphics[width=0.93\textwidth]{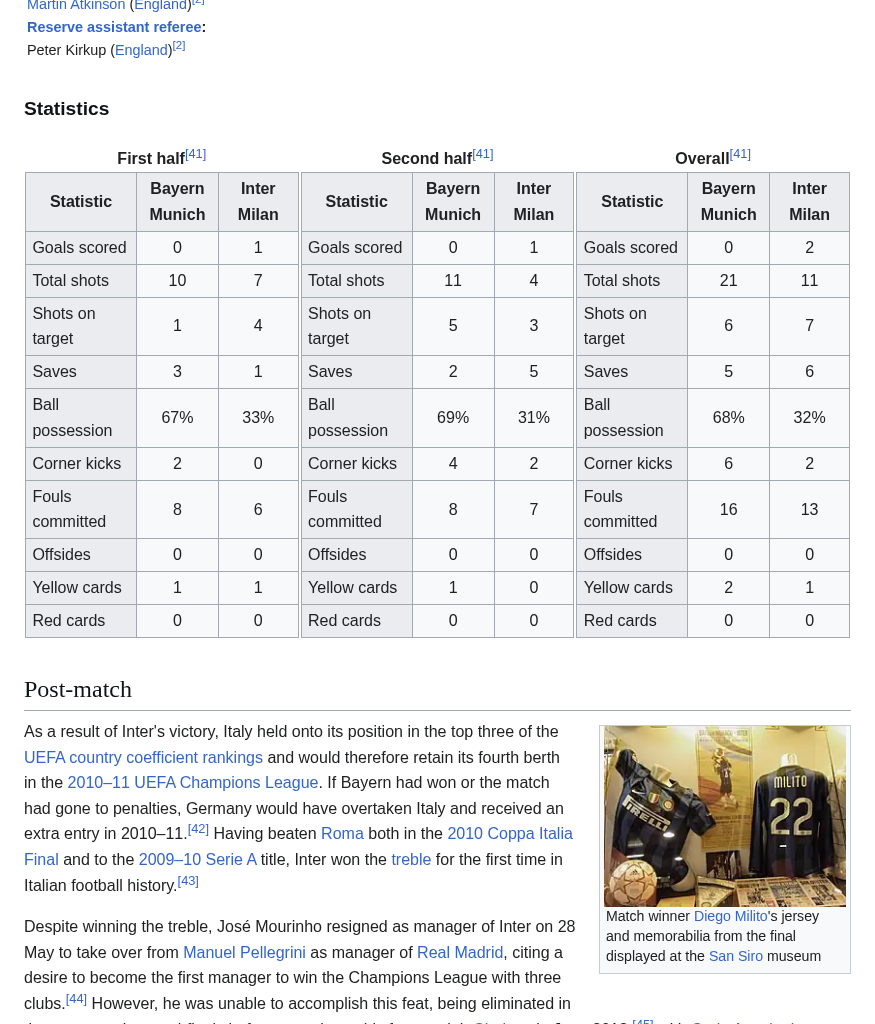}
}}\\[2pt]
\centerline{$\vdots$}
\end{minipage}

\vspace{4pt}
\noindent
\begin{minipage}[t]{0.50\textwidth}\vspace{0pt}
\fbox{\parbox{\dimexpr\textwidth-2\fboxsep-2\fboxrule}{\scriptsize
\textbf{Text reader:} ``\ldots the number of shots Inter attempted on target is not mentioned. \ldots'' \quad{\color{red}$\times$}
}}
\end{minipage}%
\hfill
\begin{minipage}[t]{0.48\textwidth}\vspace{0pt}
\fbox{\parbox{\dimexpr\textwidth-2\fboxsep-2\fboxrule}{\scriptsize
\textbf{Pixel reader:} ``\ldots under the `Overall' column for `Inter Milan', the number of `Shots on target' is \textbf{7}.'' \quad{\color{green!50!black}$\checkmark$}
}}
\end{minipage}
\caption{\textbf{Parser loss.} The 2010 Champions League Final article's match-statistics table is destroyed by HTML-to-text linearization, so no text chunk in the corpus contains the answer. The pixel retriever surfaces the rendered statistics table as the top tile. This is the same example shown in Figure~\ref{fig:pipeline}; here we expand the full retrieval lists for both modalities.}
\label{fig:fm1_inter}
\end{figure}


\begin{figure}[htbp]
\centering
\small
\fbox{\parbox{\dimexpr\textwidth-2\fboxsep-2\fboxrule}{\small
\emph{Q:} ``What day, month, and year did Salvador Dal\'i's mother pass away?'' \hfill \emph{A:} \textbf{6 February 1921}
}}

\vspace{4pt}
\begin{minipage}[t]{0.50\textwidth}
\textbf{Trafilatura text top-3}\\[2pt]
\fbox{\parbox{\dimexpr\textwidth-2\fboxsep-2\fboxrule}{\raggedright\scriptsize
\textbf{rank 1, score 0.662} --- \textit{Salvador Dal\'i} (chunk 0)\\
\texttt{Salvador Dal\'i | Born | 11 May 1904 | Died | 23 January 1989 | Education | San Fernando \ldots}
}}\\
{\scriptsize\color{brown}\itshape $\hookrightarrow$ Infobox; lists Dal\'i's own dates, not his mother's.}\\[2pt]
\fbox{\parbox{\dimexpr\textwidth-2\fboxsep-2\fboxrule}{\raggedright\scriptsize
\textbf{rank 2, score 0.649} --- \textit{Gala Dal\'i} (chunk 1)\\
\texttt{\ldots Death. Gala died in Port Lligat in Catalonia, Spain, earl\ldots}
}}\\
{\scriptsize\color{brown}\itshape $\hookrightarrow$ Wrong person (Dal\'i's wife, not his mother).}\\[2pt]
\fbox{\parbox{\dimexpr\textwidth-2\fboxsep-2\fboxrule}{\raggedright\scriptsize
\textbf{rank 3, score 0.639} --- \textit{Salvador Dal\'i} (chunk 20)\\
\texttt{at archive.today --- Boletin Oficial del Estado \ldots{} Gibson, Ian (1997) pp.\ 603--604 \ldots}
}}\\
{\scriptsize\color{brown}\itshape $\hookrightarrow$ References section, no biographical content.}\\[3pt]
\centerline{$\vdots$}\\[3pt]
\dashedbox{\parbox{\dimexpr\textwidth-2\fboxsep-2\fboxrule}{\raggedright\scriptsize
\textbf{rank 12} --- \textit{Salvador Dal\'i} (chunk 1)\\
\texttt{\ldots On 6 February 1921, Dal\'i's mother died of uterine cancer. Dal\'i was 16 years old and later said his mother's death ``was the greatest blow I had experienced in my life'' \ldots}
}}\\[2pt]
\end{minipage}%
\hfill
\begin{minipage}[t]{0.48\textwidth}
\textbf{\sys{} screenshot top-3}\\[2pt]
\fbox{\parbox{\dimexpr\textwidth-2\fboxsep-2\fboxrule}{\centering\scriptsize
\textbf{rank 1, score 0.566} --- \textit{Salvador Dal\'i} (tile 0) \hfill \textbf{rank 2, score 0.565} --- \textit{Salvador Dal\'i} (tile 1)\\[2pt]
\includegraphics[width=0.46\textwidth]{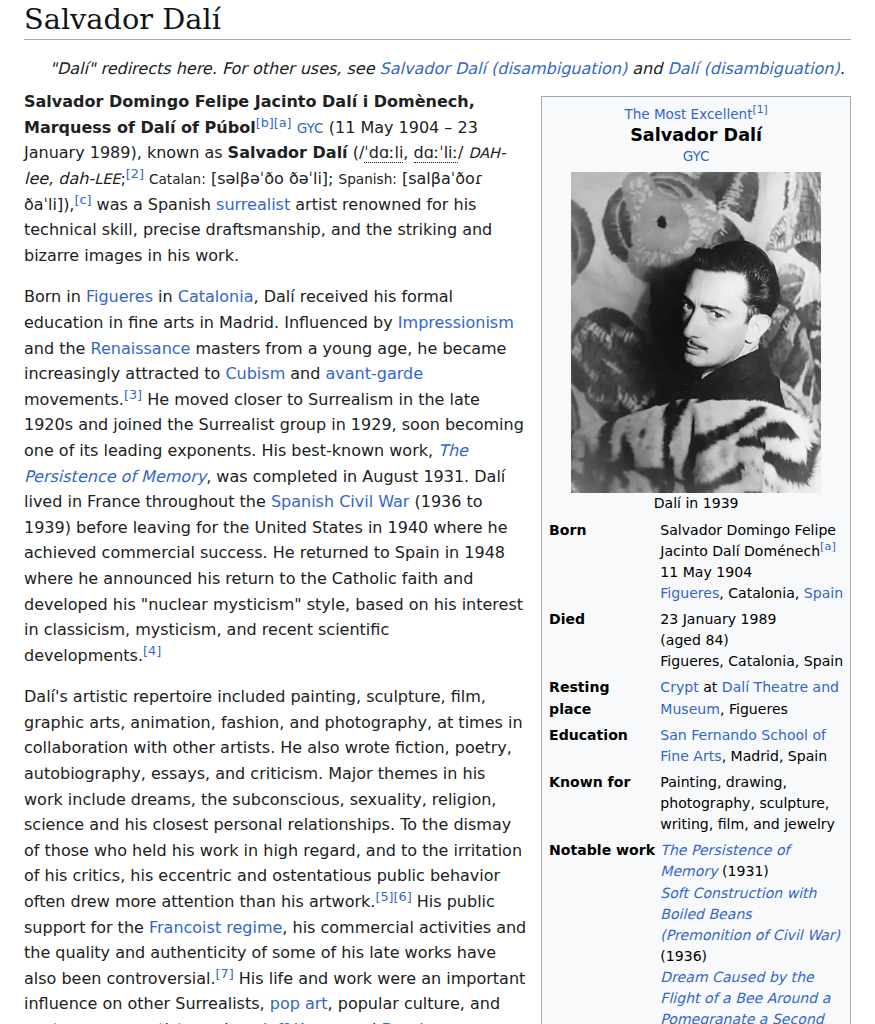}\hfill
\includegraphics[width=0.46\textwidth]{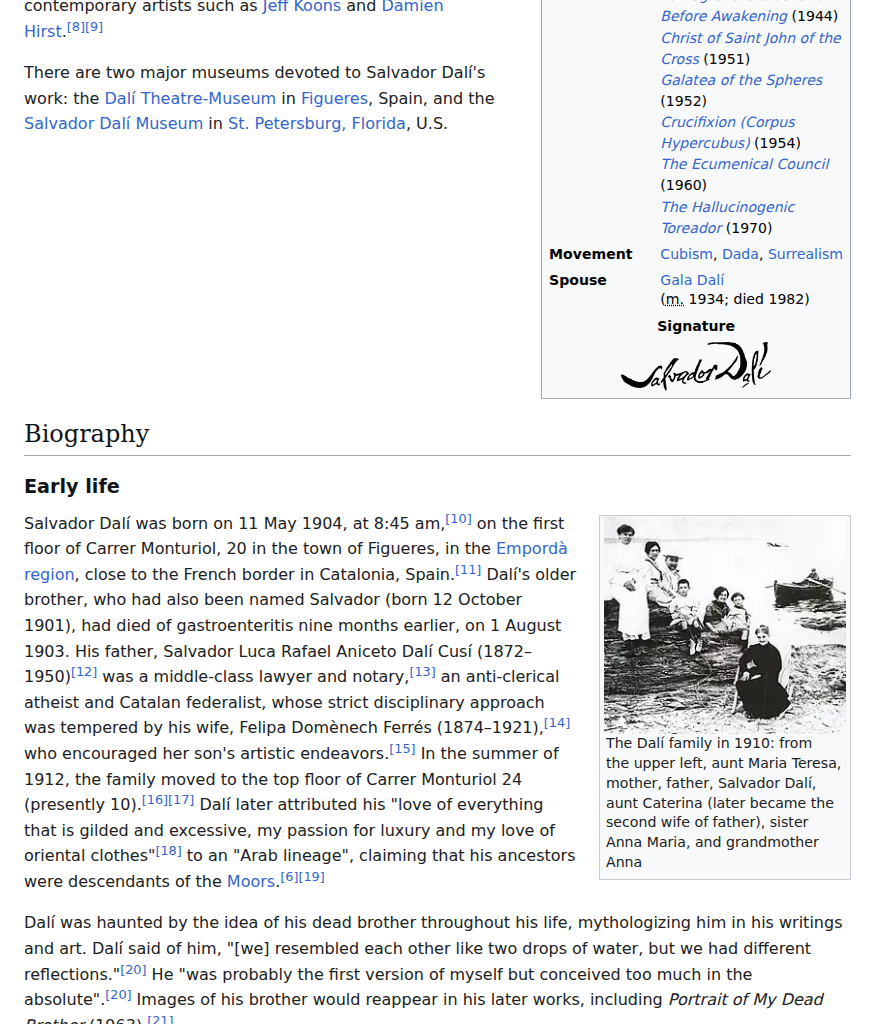}
}}\\[2pt]
\fbox{\parbox{\dimexpr\textwidth-2\fboxsep-2\fboxrule}{\centering\scriptsize
\textbf{rank 3, score 0.563} --- \textit{Salvador Dal\'i} (tile 2: contains ``\textbf{6 February 1921}'')\\[2pt]
\includegraphics[width=0.93\textwidth]{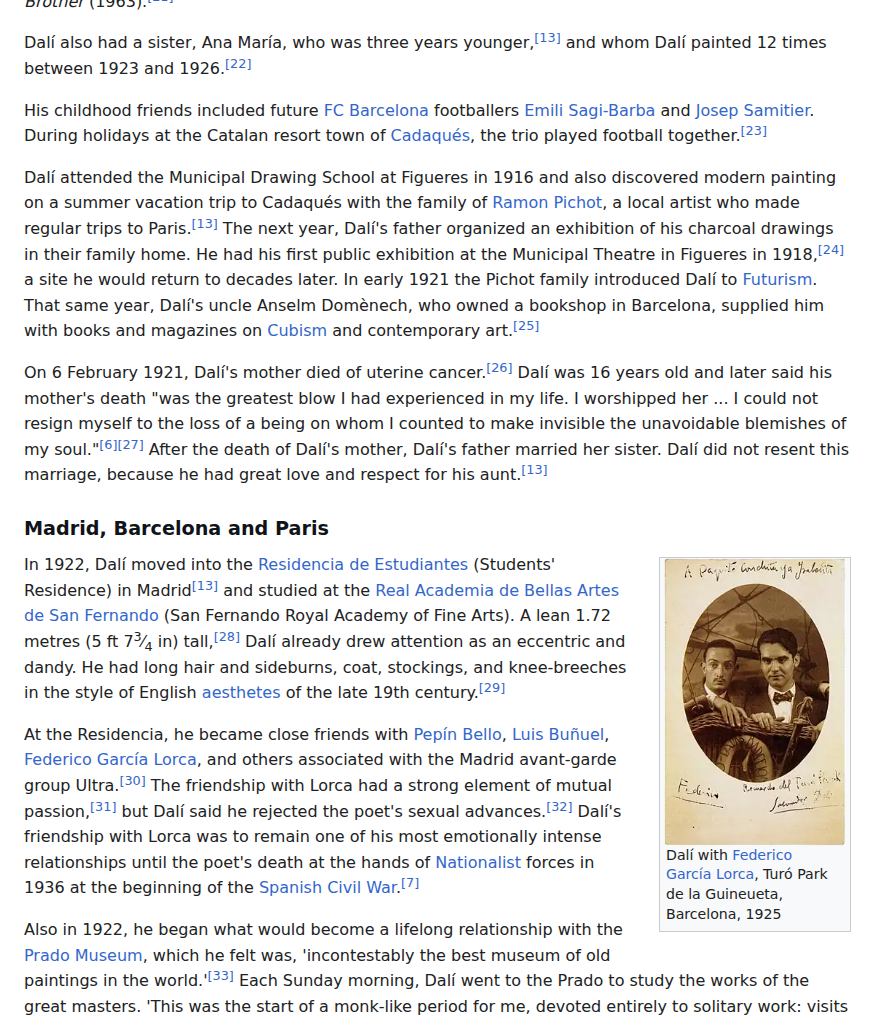}
}}
\end{minipage}

\vspace{4pt}
\noindent
\begin{minipage}[t]{0.50\textwidth}\vspace{0pt}
\fbox{\parbox{\dimexpr\textwidth-2\fboxsep-2\fboxrule}{\scriptsize
\textbf{Text reader:} ``\ldots the specific day and month of Salvador Dal\'i's mother's death are not mentioned; only the year 1921 is given.'' \quad{\color{red}$\times$}
}}
\end{minipage}%
\hfill
\begin{minipage}[t]{0.48\textwidth}\vspace{0pt}
\fbox{\parbox{\dimexpr\textwidth-2\fboxsep-2\fboxrule}{\scriptsize
\textbf{Pixel reader:} ``\ldots Salvador Dal\'i's mother passed away on \textbf{6 February 1921}.'' \quad{\color{green!50!black}$\checkmark$}
}}
\end{minipage}
\caption{\textbf{Rank loss (paragraph evidence).} Once the infobox is linearized, its flattened key--value text out-ranks the answer-bearing body paragraph (which falls to rank 12) --- the infobox lists Dal\'i's own birth/death, not his mother's, yet matches the query on the entity name. The visual embedding keeps the infobox sidebar structurally distinct from the body section, surfacing the relevant tile in the top-3.}
\label{fig:fm2_dali}
\end{figure}


\begin{figure}[htbp]
\centering
\small
\fbox{\parbox{\dimexpr\textwidth-2\fboxsep-2\fboxrule}{\small
\emph{Q:} ``What President nominated Elliott Fitch Shepard as U.S.\ Attorney for the Southern District of New York?'' \hfill \emph{A:} \textbf{Rutherford B.\ Hayes}
}}

\vspace{4pt}
\begin{minipage}[t]{0.50\textwidth}
\textbf{Trafilatura text top-3}\\[2pt]
\fbox{\parbox{\dimexpr\textwidth-2\fboxsep-2\fboxrule}{\raggedright\scriptsize
\textbf{rank 1, score 0.689} --- \textit{Elliott Fitch Shepard} (chunk 0)\\
\texttt{Elliott Fitch Shepard | Died | March 24, 1893 New York City | Occupation | lawyer, banker \ldots}
}}\\
{\scriptsize\color{brown}\itshape $\hookrightarrow$ Infobox; does not list the nominating President.}\\[2pt]
\fbox{\parbox{\dimexpr\textwidth-2\fboxsep-2\fboxrule}{\raggedright\scriptsize
\textbf{rank 2, score 0.668} --- \textit{U.S.\ Attorney for the District of New York} (chunk 0)\\
\texttt{The U.S.\ Attorney for the District of New York was from 1789 to 1815 the chief federal law \ldots}
}}\\
{\scriptsize\color{brown}\itshape $\hookrightarrow$ Office history page, no per-appointee details.\yichuan{can we change to some similar structure like before, some structural problem, and we should add some bounding box to the right}}\\[2pt]
\fbox{\parbox{\dimexpr\textwidth-2\fboxsep-2\fboxrule}{\raggedright\scriptsize
\textbf{rank 3, score 0.658} --- \textit{U.S.\ Attorney for the Southern District of New York} (chunk 0)\\
\texttt{Formed | September 24, 1789 Judiciary Act of 1789 | Jurisdiction | Southern District \ldots}
}}\\
{\scriptsize\color{brown}\itshape $\hookrightarrow$ Office overview infobox, no list of nominees.}\\[3pt]
\centerline{$\vdots$}\\[3pt]
\dashedbox{\parbox{\dimexpr\textwidth-2\fboxsep-2\fboxrule}{\raggedright\scriptsize
\textbf{rank 66} --- \textit{Elliott Fitch Shepard} (chunk 1)\\
\texttt{\ldots In 1881, US President Rutherford B.\ Hayes nominated him for United States Attorney for the Southern District of New York \ldots}
}}\\[2pt]
\end{minipage}%
\hfill
\begin{minipage}[t]{0.48\textwidth}
\textbf{\sys{} screenshot top-3}\\[2pt]
\fbox{\parbox{\dimexpr\textwidth-2\fboxsep-2\fboxrule}{\centering\scriptsize
\textbf{rank 1, score 0.576} --- \textit{Elliott Fitch Shepard} (tile 0) \hfill \textbf{rank 2, score 0.512} --- \textit{Elliott Shepard} (tile 0)\\[2pt]
\includegraphics[width=0.46\textwidth]{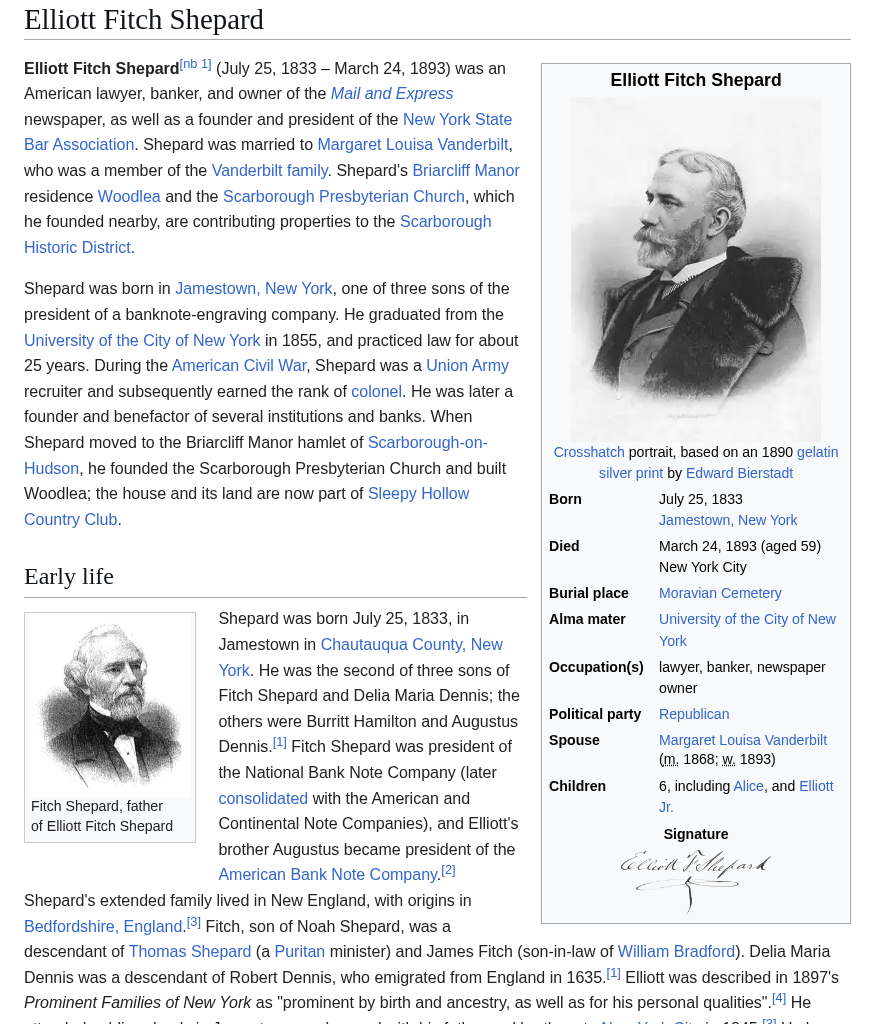}\hfill
\includegraphics[width=0.46\textwidth]{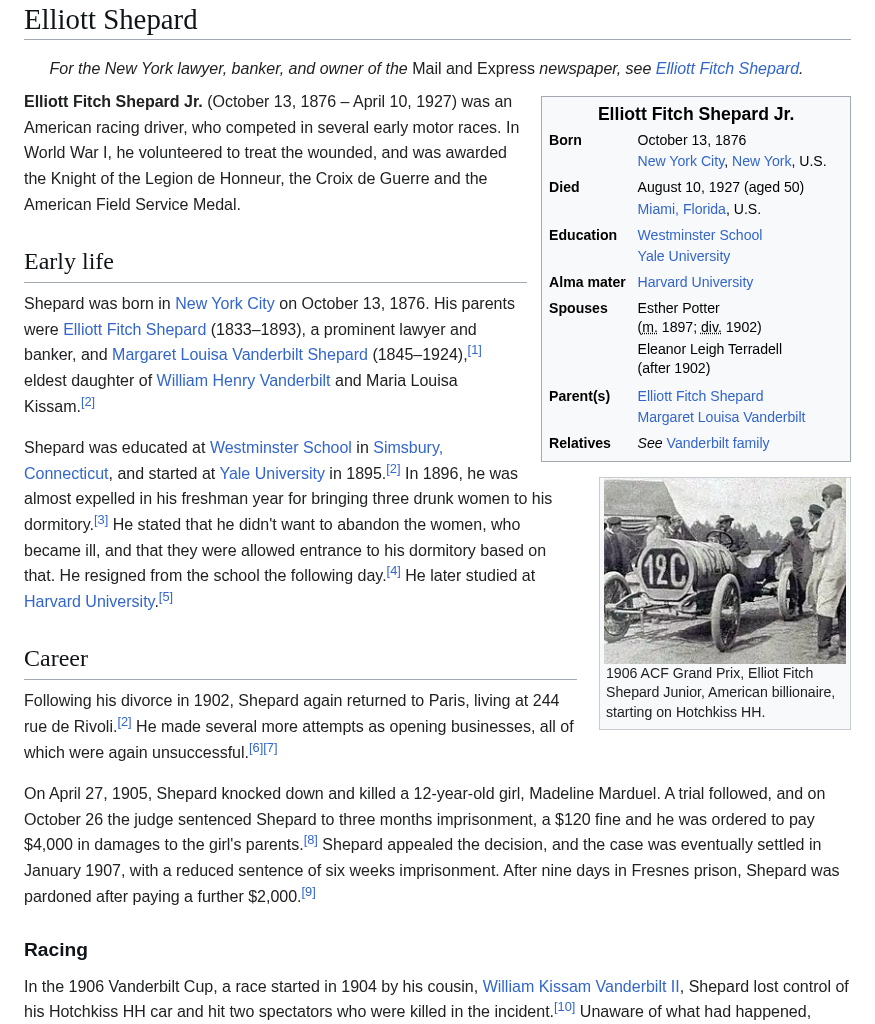}
}}\\[2pt]
\fbox{\parbox{\dimexpr\textwidth-2\fboxsep-2\fboxrule}{\centering\scriptsize
\textbf{rank 3, score 0.509} --- \textit{Elliott Fitch Shepard} (tile 2; answer region boxed)\\[2pt]
\begin{tikzpicture}
\node[inner sep=0pt] (img) {\includegraphics[width=0.93\textwidth]{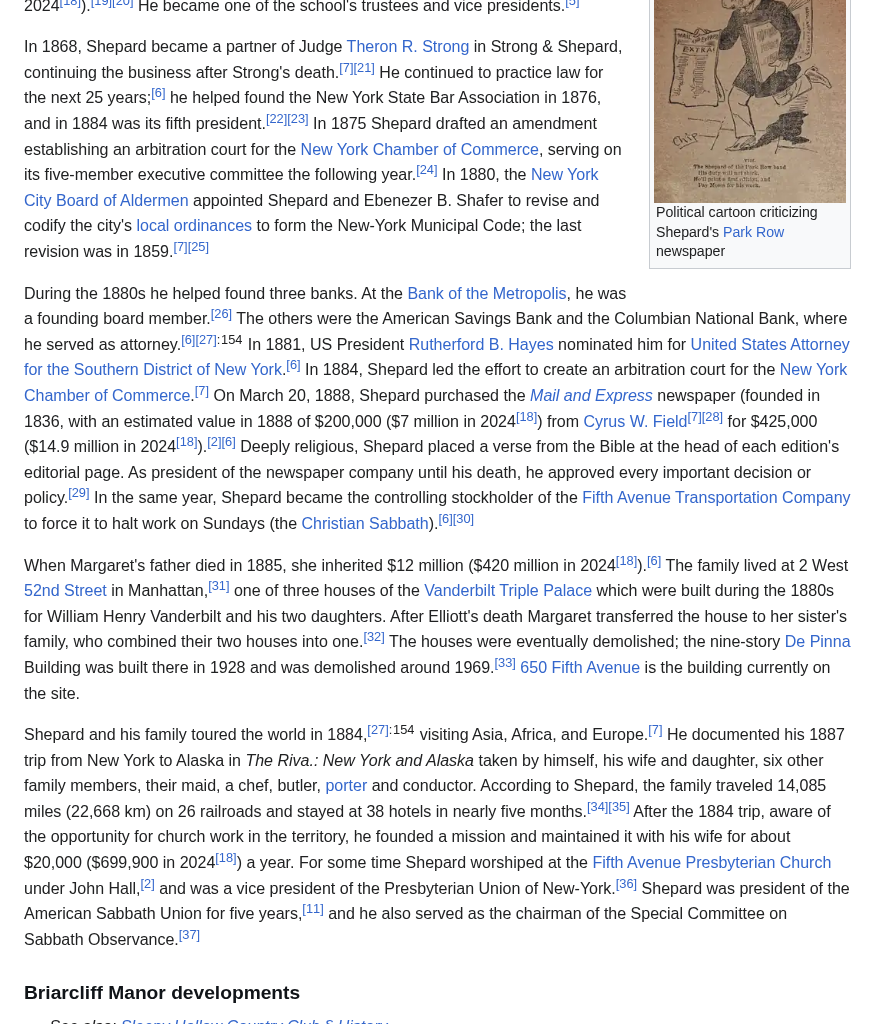}};
\draw[red, line width=1.2pt]
  ($(img.north west)!0.328!(img.south west)$)
  rectangle
  ($(img.north east)!0.382!(img.south east)$);
\end{tikzpicture}
}}
\end{minipage}

\vspace{4pt}
\noindent
\begin{minipage}[t]{0.50\textwidth}\vspace{0pt}
\fbox{\parbox{\dimexpr\textwidth-2\fboxsep-2\fboxrule}{\scriptsize
\textbf{Text reader:} ``\ldots there is no information stating that Elliott Fitch Shepard was nominated as United States Attorney \ldots'' \quad{\color{red}$\times$}
}}
\end{minipage}%
\hfill
\begin{minipage}[t]{0.48\textwidth}\vspace{0pt}
\fbox{\parbox{\dimexpr\textwidth-2\fboxsep-2\fboxrule}{\scriptsize
\textbf{Pixel reader:} ``\ldots US President \textbf{Rutherford B.\ Hayes} nominated Elliott Fitch Shepard \ldots for the Southern District of New York in 1881.'' \quad{\color{green!50!black}$\checkmark$}
}}
\end{minipage}
\caption{\textbf{Rank loss (extreme rank gap).} Text retrieval places the Shepard infobox chunk at rank 1 --- the correct article, but the infobox does not list the nominating President. The body paragraph that does contain the answer falls all the way to rank 66. The pixel retriever surfaces the answer-bearing tile at rank 3.}
\label{fig:fm2_shepard}
\end{figure}


\paragraph{Infobox rank displacement.}
Among the 91 rank-loss cases in Table~\ref{tab:failure_decomposition}, we find that 44\% share a common pattern: text retrieval ranks the article's infobox chunk (chunk~0) above the answer-bearing content despite the infobox not containing the answer.
Wikipedia infoboxes contain the entity name, key attributes, and category labels, producing high keyword overlap with factual queries.
Text retrieval systematically places chunk~0 at rank~1 for 75.9\% of queries (vs.\ 67.1\% for \sys{} retrieval), because the linearized infobox is a dense text-similarity target.
\sys{} retrieval is less susceptible: the visual embedding captures the distinct layout of an infobox (bordered sidebar with key-value pairs) versus a body paragraph (flowing text under a section heading), allowing the retriever to distinguish content type even when keyword overlap is similar.

The effect is counter-intuitive because it is worst on the \emph{correct article}: the retriever finds the right Wikipedia page and places its infobox at rank~1, yet the answer-bearing paragraph or table falls to rank~20+ because the infobox already saturates the top positions.

Examples include:
\begin{itemize}[leftmargin=1.2em]
    \item \textbf{``What President nominated Elliott Fitch Shepard as U.S.\ Attorney?''}
    (answer: Rutherford B.\ Hayes).  
    Text retrieval ranks the Shepard infobox chunk first (correct article, wrong section), while the answer-bearing paragraph falls to rank~66; \sys{} retrieves it at rank~3.

    \item \textbf{``What day did Dalí's mother die?''}
    (answer: 6 February 1921).  
    The Dalí infobox chunk is ranked first by text retrieval, but the answer appears in a body paragraph at rank~12; \sys{} retrieves it at rank~3.

    \item \textbf{``From which Israeli university did Judith Hemmendinger receive her master's degree?''}
    (answer: Bar-Ilan University).  
    Text retrieval places the infobox at rank~1, while the answer paragraph falls to rank~44; \sys{} retrieves it at rank~2.
\end{itemize}

The displacement effect is worst for paragraph evidence, where the mean rank of rank-loss cases is 22.5 (vs.\ 16--18 for other evidence types).
In these cases, \sys{} retrieval finds the answer at mean rank 1.8 on the same questions.
On the 42 questions where paragraph evidence ranks outside text retrieval's top-3 \emph{and} text places an infobox chunk at rank~1, \sys{} still recovers valid evidence in top-3 for 37 of 42 (88\%, paragraph evidence specifically for 20 of 42), directly confirming its immunity to the displacement described above.

Table~\ref{tab:failure_stage_text_vs_screenshot} reports, for each modality, the share of its failures falling into each of the three modes (\emph{parser loss}: no answer-bearing item anywhere in top-100; \emph{rank loss}: retrieved but outside top-3; \emph{reader loss}: in top-3 but the reader still errs). The headline text percentages in \S\ref{sec:qualitative_analysis} ($36.6\%/55.2\%/8.2\%$) are the question-level aggregate of these columns over all evidence types.
Table~\ref{tab:evidence_rank_margin} reports the rank distribution of valid type-$\tau$ evidence in top-100 retrieval, indicating how many evidence misses at $k{=}3$ are near misses recoverable at larger $k$.

\begin{table}[htbp]
\centering
\caption{SimpleQA failure causes by evidence type for text vs.\ \sys{} retrieval (Qwen3.5-4B reader, top-$k{=}3$). Parser/Rank/Reader give the \% of each modality's failures from parser, rank, and reader loss. Rows overlap.}
\label{tab:failure_stage_text_vs_screenshot}
\footnotesize
\begin{tabular*}{\textwidth}{@{\extracolsep{\fill}}llcccccc@{}}
\toprule
Modality & Evidence & $n$ & Failures & Parser & Rank & Reader & Ev. Recall@3 \\
\midrule
\sys{} & Table     & 282 &  78 &  6.4 & 56.4 & 37.2 & 81.2 \\
\sys{} & List      & 290 &  73 &  8.2 & 50.7 & 41.1 & 83.8 \\
\sys{} & Infobox   & 503 &  92 &  5.4 & 55.4 & 39.1 & 88.1 \\
\sys{} & Paragraph & 571 & 121 &  4.1 & 65.3 & 30.6 & 84.6 \\
\midrule
Text       & Table     & 282 &  95 & 35.8 & 49.5 & 14.7 & 68.1 \\
Text       & List      & 290 & 70  & 31.4 & 58.6 & 10.0 & 75.9 \\
Text       & Infobox   & 503 & 111 & 28.8 & 57.7 & 13.5 & 78.9 \\
Text       & Paragraph & 571 & 167 & 29.3 & 59.9 & 10.8 & 71.1 \\
\bottomrule
\end{tabular*}
\end{table}

\begin{table}[htbp]
\centering
\caption{SimpleQA evidence-tile rank distribution by evidence type (top-100 retrieval, Qwen3.5-4B reader). Rows overlap.}
\label{tab:evidence_rank_margin}
\footnotesize
\begin{tabular*}{\textwidth}{@{\extracolsep{\fill}}llrrrrr@{}}
\toprule
Modality & Evidence type & $n$ & Recall@3 & Recall@10 & Recall@50 & Median rank \\
\midrule
\sys{} & Table     & 282 & 34.8 & 51.8 & 77.7 & 7 \\
\sys{} & List      & 290 & 36.9 & 52.1 & 76.9 & 6 \\
\sys{} & Infobox   & 503 & 63.0 & 69.2 & 77.7 & 1 \\
\sys{} & Paragraph & 571 & 63.9 & 76.9 & 87.2 & 2 \\
\midrule
Text       & Table     & 282 & 23.8 & 34.0 & 45.7 & 4 \\
Text       & List      & 290 & 28.6 & 33.4 & 44.5 & 2 \\
Text       & Infobox   & 503 & 65.2 & 71.6 & 77.9 & 1 \\
Text       & Paragraph & 571 & 44.1 & 54.6 & 66.4 & 2 \\
\bottomrule
\end{tabular*}
\end{table}


\subsection{HTML DOM Lookup Baseline: Setup and Analysis}
\label{app:html_dom_lookup}


The \textbf{Text $\to$ HTML} row in Table~\ref{tab:ablation_modality} tests whether preserving the original DOM structure of Wikipedia articles can close the gap between text-based RAG and pixel-based RAG.
Standard text-based RAG linearizes HTML into flat strings, destroying table and list structure; this baseline instead feeds the reader \emph{raw HTML} with intact \texttt{<table>}, \texttt{<ul>}, and sectional markup.

\paragraph{Setup.}
We reuse the same Trafilatura text-chunk index and retrieval API (Qwen3-VL-Embedding-2B, 1024-token chunks, IVFFlat).
For each retrieved text chunk, a \emph{DOM lookup} step recovers the corresponding HTML from the original Wikipedia article:
\begin{enumerate}[leftmargin=*,itemsep=2pt]
    \item \textbf{Fetch HTML.} The article HTML is served from a local Kiwix ZIM archive~\cite{kiwixzim} via \texttt{kiwix-serve}, eliminating network latency.
    \item \textbf{Extract search keys.} Distinctive phrases are extracted from the text chunk: table cell values (e.g., codes like \texttt{B01AC06}, numbers with units) for table-heavy chunks, mid-line prose fragments for paragraph-heavy chunks. The first line (article title) is skipped to avoid matching the \texttt{<h1>} heading.
    \item \textbf{Locate in DOM.} Each key is searched within the \texttt{text\_content()} of every element under the article's \texttt{mw-parser-output} container. Both the key and element text are normalized (non-breaking spaces, dash variants, and diacritics are collapsed) to handle encoding mismatches between Trafilatura output and raw HTML. The tightest-matching element is selected.
    \item \textbf{Resolve to contiguous span.} Each matched element is walked up to its nearest direct-child ancestor of \texttt{mw-parser-output}. The final result is the contiguous range of direct children from the first matched child to the last---preserving all intermediate elements (tables, paragraphs, lists) that the original text chunk spanned.
    \item \textbf{Clean and return.} Inline \texttt{<style>}, \texttt{<script>}, and navigation-box (\texttt{navbox}) elements are stripped. The serialized HTML is returned to the reader. If no key matches in the DOM, the original flat text is used as fallback.
\end{enumerate}

\noindent The reader (Qwen3-VL-4B, \texttt{max\_model\_len}$=$65536) receives the concatenated HTML of all $k{=}3$ retrieved passages, separated by \texttt{<hr>} delimiters.

\paragraph{Results.}
HTML achieves 59.8\% QA accuracy vs.\ 71.6\% for flat text on SimpleQA, and 56.6\% vs.\ 59.0\% on LiveVQA (Table~\ref{tab:ablation_modality}).
Retrieval quality is identical (Recall@1 and Recall@any differ by $<$1\,pp), confirming that the gap is entirely in the reading stage.
Oversized HTML passages are truncated to 30k characters per passage to keep the total context within the reader's 65k-token window.

\paragraph{Why HTML hurts: tag dilution.}
HTML markup inflates the average context from 7{,}601 to 28{,}941 characters ($3.8\times$) on SimpleQA, consuming reader tokens on tags rather than content.
The reader sees the same factual content---but diluted by structural markup, it more frequently fails to locate the answer.
The gap is smaller on LiveVQA ($-2.4$\,pp vs.\ $-11.8$\,pp on SimpleQA) because news articles are shorter and have simpler DOM structure than Wikipedia pages.

\paragraph{Implications.}
Structured HTML is semantically richer than linearized text, yet this richness comes at a steep token cost that current context-window--limited readers cannot absorb.
Screenshots bypass both problems: a rendered tile encodes the same tabular and sectional structure in a fixed ${\sim}875$ visual tokens per tile regardless of article complexity, and no markup overhead is paid.
The HTML baseline thus supports the central claim of this paper: pixel-space retrieval preserves document structure without the linearization losses of text \emph{or} the token overhead of markup.


\subsection{Directly RAG on Raw HTML Data}
\label{app:html_rag_full}

One might expect that preserving the original HTML structure throughout the RAG pipeline would help, since linearization is lossy. We test this by building a fully HTML-native pipeline that indexes and reads raw HTML chunks directly, and find that it performs \emph{worse} than plain-text RAG on nearly every benchmark: retrieval quality is comparable, but HTML markup overwhelms the reader with tags, causing large accuracy drops.
We detail the setup and results below.

\paragraph{Setup.}
We extract raw HTML from the Kiwix ZIM archive and chunk at DOM boundaries: section headers (\texttt{<h2>}/\texttt{<h3>}) force chunk boundaries, prose elements accumulate up to 1{,}024 tokens, and tables are split at \texttt{<tr>} row boundaries with the header row prepended to each sub-chunk (preserving column context).
Navigation boxes, reference lists, and table-of-contents elements are filtered out.
Each chunk stores the raw HTML (with tags) and, for table sub-chunks, a \texttt{parent\_html} field containing the full original table.
The resulting corpus contains 25.7M chunks (vs.\ 15.7M for the Trafilatura 1{,}024-token text baseline).

Chunks are embedded with the same Qwen3-VL-Embedding-2B model used for all other baselines, with the raw HTML as input text (\texttt{max\_length}$=$1{,}024).
Short chunks are batched together via a dynamic token-budget scheduler to avoid padding waste.
A FAISS IVFFlat index (nlist$=$4{,}096, nprobe$=$128) is built over the 25.7M embeddings.
At query time, the search API returns the raw HTML chunk to the reader.

\paragraph{Results.}

\begin{table}[htbp]
\centering
\caption{HTML-RAG (full pipeline) vs.\ Trafilatura text baseline across Wikipedia benchmarks. Reader: Qwen3.5-4B, top-$k{=}3$, \texttt{--no-think}. Retrieval Recall@3 is computed over examples with ground-truth article annotations.}
\label{tab:html_rag_full}
\footnotesize
\begin{tabular}{l cc cc cc c cc}
\toprule
 & \multicolumn{2}{c}{NQ} & \multicolumn{2}{c}{NQ-Tables} & \multicolumn{2}{c}{SimpleQA} & MMSearch & \multicolumn{2}{c}{EVQA} \\
\cmidrule(lr){2-3} \cmidrule(lr){4-5} \cmidrule(lr){6-7} \cmidrule(l{6pt}r{6pt}){8-8} \cmidrule(lr){9-10}
Method & R@3 & Acc & R@3 & Acc & R@3 & Acc & Acc & R@3 & Acc \\
\midrule
Trafilatura (text) & 45.8 & 55.9 & 37.2 & 42.5 & 76.2 & 69.2 & 24.7 & 6.4 & 29.6 \\
HTML-RAG (ours)    & 47.9 & 26.5 & 38.2 & 18.4 & 69.6 & 59.8 & 22.3 & 1.5 & 34.7 \\
\midrule
$\Delta$           & {+2.1} & {$-$29.4} & {+1.0} & {$-$24.1} & {$-$6.6} & {$-$9.4} & {$-$2.4} & {$-$4.9} & {+5.1} \\
\bottomrule
\end{tabular}
\end{table}

Table~\ref{tab:html_rag_full} reveals a striking dissociation between retrieval and reading when HTML is used end-to-end:

\begin{itemize}[leftmargin=*,itemsep=2pt]
    \item \textbf{Retrieval quality is comparable or better.} HTML-RAG achieves higher Recall@3 than Trafilatura on NQ (+2.1\,pp) and NQ-Tables (+1.0\,pp). DOM-boundary chunking produces more semantically coherent units than fixed-window text splitting, and the embedding model handles HTML markup without difficulty.
    \item \textbf{Reading quality drops sharply.} QA accuracy falls on every benchmark except EVQA, with the largest gaps on NQ ($-$29.4\,pp) and NQ-Tables ($-$24.1\,pp). The cause is the same as in the DOM lookup baseline (Appendix~\ref{app:html_dom_lookup}): HTML tags inflate the reader's context, consuming tokens on markup rather than evidence. The effect is most severe on knowledge-intensive benchmarks where the reader must locate a specific fact within a dense passage.
    \item \textbf{EVQA is the sole exception} (+5.1\,pp). EVQA questions often target entity attributes found in structured infoboxes; the HTML preserves this structure, benefiting the reader even at higher token cost. However, the low retrieval recall (1.5\% vs.\ 6.4\%) suggests this gain is largely driven by the no-retrieval baseline rather than retrieved HTML content.
\end{itemize}

\paragraph{Implications.}
The HTML-RAG experiment reinforces the DOM lookup finding from a different angle: even when retrieval is built from scratch with HTML-native chunking and embedding, the reader bottleneck persists.
The fundamental issue is not retrieval quality but \emph{token efficiency}: HTML markup is a verbose encoding of structure that penalizes context-window--limited readers.
Screenshots encode the same structural information in a fixed number of visual tokens (${\sim}875$ per tile), avoiding both the linearization losses of text and the tag overhead of HTML.
Raw HTML is therefore not a viable alternative to either text-based or pixel-based RAG.
This underscores why text parsers exist in the first place: current LLM readers cannot consume raw HTML effectively, so careful extraction into clean text remains a necessary preprocessing step for text-based RAG, with all the associated information loss discussed in \S\ref{sec:background_motivation}.


\section{Limitations, Broader Impact, and Future Work}
\label{app:limitations}

\paragraph{Limitations.}
A key limitation of pixel-space retrieval is the loss of hyperlink structure.
In text-based retrieval, hyperlinks provide navigable connections between documents and serve as a rich signal for downstream tasks such as multi-hop reasoning and entity disambiguation.
In our screenshot-based representation, hyperlinks are visually rendered (e.g., as blue underlined text) but are not directly actionable; the system cannot follow a link to retrieve the target page.
One mitigation is to preserve hyperlink information as structured metadata alongside each tile: by extracting hyperlink URLs and their anchor text during the rendering stage and storing them as auxiliary fields, downstream models can access link targets without requiring the pixel representation itself to encode this information.

A second limitation is storage overhead: storing rendered screenshots requires substantially more disk space than raw text.
Our Wikipedia datastore alone occupies nearly 6\,TB of screenshot tiles.
A practical mitigation is a \emph{render-on-demand} strategy: once all tiles have been embedded, the raw screenshots can be deleted, retaining only the embeddings for retrieval.
At inference time, after retrieving the top-$K$ tiles, the system looks up the corresponding HTML source and re-renders the relevant screenshots on the fly before passing them to the VLM for generation.

A third limitation is language coverage: all datastores in this work are English-only (English Wikipedia and English-language news outlets), introducing a language bias. Extending to multilingual corpora is an important direction for future work.

\paragraph{Broader impact.}
By operating directly on rendered screenshots, \sys{} removes the dependency on HTML parsing and text extraction that systematically disadvantages visually rich webpages (e.g., infographics, styled tables, diagram-heavy articles).
This levels the playing field for content whose value is inseparable from its visual layout.
The same principle extends beyond webpages: any visually rich document, such as scanned PDFs, slide decks, or posters, can be ingested as screenshot tiles and retrieved without format-specific parsers.
Because pixel representations do not depend on language-specific tokenizers or extractors, this approach naturally extends to non-Latin-script and low-resource languages where text extraction pipelines are less mature.
On the risk side, screenshot-based retrieval faithfully preserves whatever appears on a rendered page, including potentially harmful, misleading, or private content.
Unlike text pipelines, where filtering can operate on extracted strings, pixel content is harder to moderate automatically.
Deploying \sys{} at scale therefore requires careful content filtering at the rendering or indexing stage to prevent surfacing inappropriate material.

\paragraph{Future work.}
Several directions can extend \sys{}.
First, our embedding model is trained on a single domain (English Wikipedia); adapting it to a given target datastore, or mixing datastores from diverse domains (e.g., scientific papers, e-commerce, forums) to train a cross-domain embedding model, is a natural next step enabled by our synthetic training pipeline.
Second, hybrid text and image retrieval, where text-based and pixel-based scores are combined, may capture complementary signals and further improve recall.
Third, pixel-space datastores open new possibilities for agentic workflows: an agent could retrieve relevant screenshot tiles, visually ground its reasoning, and iteratively refine its search, leveraging the rich visual context that static text pipelines discard.


\section{Prompt Listings}
\label{app:prompt_listings}

This section collects the full verbatim prompts referenced throughout the appendix.

\begin{figure}[htbp]
\begin{tcolorbox}[colback=chocolate!5!white,colframe=chocolate!75!black,
                  title=Synthetic query generation prompt]
\begin{tcolorbox}[colback=gray!15!white,colframe=gray!40!black,
                  title=Image input,fontupper=\small\ttfamily,
                  left=4pt,right=4pt,top=3pt,bottom=3pt]
\texttt{\{screenshot tile\}}
\end{tcolorbox}
\vspace{2pt}
\begin{VerbatimWrap}
You are generating a query-evidence pair for training a visual retrieval model over Wikipedia screenshot chunks.

TASK: Given this screenshot chunk, generate ONE factual question whose answer is explicitly and completely visible in this chunk.

STYLE -- write natural search-style questions, not templates.
Vary the phrasing: "how much", "in what year", "which", "who", "where", "what caused", "how long", etc. Examples of the target style:
  - "How much money, in euros, was the surgeon held responsible for paying in the Olivia Puls case?"
  - "In what city was the 2010 FIFA World Cup opening ceremony held?"
  - "How many days did the 1906 San Francisco earthquake fire burn?"
  - "Which award did Fullmetal Alchemist win at the American Anime Awards in 2007?"
  - "Who was the first Black female judge appointed to the Cook County Circuit Court?"

EVIDENCE: draw from any visible content -- prose, infobox fields, table cells, image captions, diagrams, or photographs. Pick whichever source yields the most natural question; do not default to infobox.

HARD RULES:
1. SELF-CONTAINED. The question must be understandable on its own; every entity must be named explicitly.
   BAD:  "Who composed the music for the film?"          (missing film name)
   BAD:  "On what date was Lerew awarded the DFC?"       (surname only + acronym)
   BAD:  "Which cyclist placed second in the Tempo race?" (missing event/year)
   BAD:  "Which mission is shown in the screenshot?"      (references layout)
   GOOD: "Who composed the music for Once Upon a Time in Hong Kong?"
   GOOD: "On what date was RAF pilot Arthur Lerew awarded the Distinguished Flying Cross in World War II?"
2. EVIDENCE COMPLETE. The answer must be fully visible in this chunk. The source span (S:) must be a complete, untruncated sentence.
3. DISTINCTIVE. Include enough specifics (names, dates, locations, titles) to distinguish this chunk from similar pages.

ANSWER: prefer a single concise entity -- name, date, place, number, title, or short phrase.

SKIP (write exactly: SKIP) if any of the following holds:
  - Content is a raw vote count, track listing, census table, or episode list.
  - The answer is not fully visible or requires external context.
  - You cannot write a self-contained question naming every entity.
  - The source sentence is truncated or a fragment.

OUTPUT -- exactly five lines, or the single literal word SKIP:
Q: <natural, self-contained question>
A: <concise answer>
S: <verbatim complete span from the chunk>
T: image | table | infobox | prose
C: science | medicine | history | geography | technology | education | culture | politics | economics | biology | sports | entertainment | other
\end{VerbatimWrap}
\end{tcolorbox}
\caption{Synthetic query generation prompt (Stage~1). The model is sent this text together with the rendered tile as an image in the same turn.}
\label{fig:prompt-query-gen}
\end{figure}

\begin{figure}[htbp]
\begin{tcolorbox}[colback=indigo!5!white,colframe=indigo!75!white,
                  title=Self-contained-query filter prompt]
\begin{VerbatimWrap}
For each numbered question, answer YES (self-contained) or NO (not self-contained).

A question is NOT self-contained (NO) if it requires knowing a specific Wikipedia page, table, or screenshot to understand WHAT is being asked. Specifically answer NO when:

1. The subject is a vague pronoun or generic noun without a proper name:
   NO: "What was the final score of the basketball game between THE TEAM and Marquette?"  ("the team" unnamed)
   NO: "Who directed the episode of THE TELEVISION SERIES titled 'X'?"  ("the television series" unnamed)
   NO: "In what year did THE SUBJECT OF THE ARTICLE move to Tokyo?"  ("the subject" unnamed)
   NO: "What is the running time of THE FILM DESCRIBED IN THE TEXT?"  (layout reference)

2. The question explicitly references document structure:
   NO: "Which item IS LISTED IN THE TABLE as X?"
   NO: "What is shown IN THE INFOBOX?"
   NO: "According to THE PROVIDED TABLE, which..."

3. A role/position question where no year or identifying event is given and the role has had many holders:
   NO: "Who was THE CAPTAIN of HMS Defence?"  (no year, hundreds of captains over centuries)

4. A geographic entity refers only to a category without naming which one:
   NO: "On what date did THE GOODS YARD at the London and North Eastern Railway station close?"  (LNER had hundreds of stations -- which one?)

Answer YES if all the key entities (people, places, works, teams, events) are explicitly named, even if the names are obscure. Proper names are always fine.
   YES: "Who did Sandefjord Fotball hire as manager after firing Arne Sandsto?"
   YES: "How many consonants does the Pesisir language have?"
   YES: "What 'fresh' rating did the film Our Man in Havana receive on Rotten Tomatoes?"
   YES: "In what city were the 2025 Special Olympics World Winter Games held?"
   YES: "Who did Emile Derlin Zinsou serve as assistant to in 1945?"

Output exactly one line per question, using the question number: "1: YES" or "1: NO"

Questions:
{questions}
\end{VerbatimWrap}
\end{tcolorbox}
\caption{Self-contained-query filter prompt (Stage~1, first false-positive filter). Queries labelled \texttt{NO} are dropped from the training set.}
\label{fig:prompt-selfcontained}
\end{figure}

\begin{figure}[htbp]
\begin{tcolorbox}[colback=lightblue!5!white,colframe=lightblue!75!white,
                  title=Hard-negative Stage~A: candidate-answer prompt (VLM)]
\begin{VerbatimWrap}
You are looking at {tile_count} screenshot tiles from Wikipedia pages.

Based ONLY on what you can see in these images, answer the following question.
If the answer is not visible in the images, reply "CANNOT_ANSWER".

Question: {question}

Give a short, direct answer (just the answer, no explanation).
\end{VerbatimWrap}
\end{tcolorbox}
\vspace{4pt}
\begin{tcolorbox}[colback=lightblue!5!white,colframe=lightblue!75!white,
                  title=Hard-negative Stage~B: judge prompt]
\begin{VerbatimWrap}
You are validating a candidate answer against screenshot tiles from Wikipedia pages.

Based ONLY on what you can see in these images, classify the candidate answer to the question as exactly one of:
- CORRECT: the candidate answer is visible in the images and is correct.
- WRONG: the images contain enough information to tell that the candidate answer is wrong.
- CANNOT_ANSWER: the images do not contain enough information to verify the candidate answer.

Question: {question}
Candidate answer: {candidate_answer}

Return exactly one token: CORRECT, WRONG, or CANNOT_ANSWER.
\end{VerbatimWrap}
\end{tcolorbox}
\vspace{6pt}
\begin{tcolorbox}[colback=gray!5!white,colframe=gray!50!black,
                  title=Evidence QA prompt (text-only query)]
\begin{VerbatimWrap}
You are a research assistant who answers questions based on provided evidence.
Use <think></think> tags to show your reasoning if needed.
Answer the question directly and concisely based ONLY on the provided evidence.
\end{VerbatimWrap}
\end{tcolorbox}
\vspace{4pt}
\begin{tcolorbox}[colback=gray!5!white,colframe=gray!50!black,
                  title=Evidence QA prompt (multimodal query)]
\begin{VerbatimWrap}
You are a research assistant who answers questions based on retrieved visual evidence.
You will receive: (1) a text question, (2) a query image, and (3) retrieved Wikipedia evidence images.
Use the query image and evidence images to answer the question.
Use <think></think> tags to show your reasoning if needed.
Answer the question directly and concisely.
\end{VerbatimWrap}
\end{tcolorbox}
\caption{Evaluation prompts. \textbf{Hard-negative Stage~A} (\emph{answer}, blue): the VLM sees only the candidate tile(s) and the query, returning a short answer or \texttt{CANNOT\_ANSWER}. \textbf{Stage~B} (\emph{judge}, blue): classifies the candidate as \texttt{CORRECT} (false negative, dropped), \texttt{WRONG}, or \texttt{CANNOT\_ANSWER} (hard negative, kept). \textbf{Evidence QA} (gray): reader system prompts for text-only (top) and multimodal (bottom) query benchmarks.}
\label{fig:prompt-hn-answer}
\label{fig:prompt-hn-judge}
\label{fig:prompt-evidence-qa}
\label{fig:prompt-multimodal-qa}
\end{figure}

\clearpage

\end{document}